\documentclass[aps,prd,reprint,superscriptaddress,nofootinbib,floatfix,showkeys,longbibliography]{revtex4-1}
\usepackage[T1]{fontenc}
\usepackage[utf8]{inputenc}
\usepackage{graphicx}
\usepackage{amsmath,amssymb,mathtools,bm,mathrsfs}
\usepackage{booktabs}
\usepackage{array}
\usepackage{dcolumn}
\usepackage{xcolor}
\usepackage{hyperref}
\usepackage{orcidlink}
\usepackage{microtype}
\usepackage{placeins}
\newif\ifshowchanges
\showchangestrue
\newcommand{\revcolor}{\ifshowchanges\color{black}\fi}
\newcommand{\rev}[1]{{\revcolor #1}}
\newenvironment{revision}{\par\begingroup\revcolor}{\par\endgroup}
\newcommand{\panel}[2]{\begin{minipage}[t]{#1\textwidth}\centering#2\end{minipage}}

\hypersetup{
  colorlinks=true,
  linkcolor=black,
  citecolor=black,
  urlcolor=black,
}

\graphicspath{{figures/}}
\newcommand{\dd}{\mathrm{d}}
\newcommand{\ii}{\mathrm{i}}
\newcommand{\e}{\mathrm{e}}
\newcommand{\order}{\mathcal{O}}
\newcommand{\Diss}[1]{\mathcal{D}[#1]}
\newcommand{\epsh}{\varepsilon}
\newcommand{\rh}{r_{\mathrm h}}
\newcommand{\OmH}{\Omega_{\mathrm H}}
\newcommand{\TempH}{T_{\mathrm H}}
\newcommand{\bH}{\beta_{\mathrm H}}
\newcommand{\kap}{\kappa}
\newcommand{\Mrem}{M_{\mathrm{em}}}

\begin{document}

\title{Acceleration radiation, quasinormal modes and quasi-bound states in generic rotating regular black holes}

\author{Uktamjon Uktamov\,\orcidlink{0009-0003-0423-2474}}
\email[Corresponding author:]{uktam.uktamov11@gmail.com}
\affiliation{School of Physics, Harbin Institute of Technology, Harbin 150001, People's Republic of China}
\affiliation{Institute for Advanced Studies, New Uzbekistan University, Movarounnahr Street 1, Tashkent 100000, Uzbekistan}

\author{Ali {\"O}vg{\"u}n\,\orcidlink{0000-0002-9889-342X}}
\email{ali.ovgun@emu.edu.tr}
\affiliation{Physics Department, Faculty of Arts and Sciences, Eastern Mediterranean University, Famagusta, 99628 North Cyprus via Mersin 10, T\"urkiye}

\author{Reggie C. Pantig\,\orcidlink{0000-0002-3101-8591}}
\email{rcpantig@mapua.edu.ph}
\affiliation{Physics Department, School of Foundational Studies and Education, Map\'ua University, 658 Muralla Street, Intramuros, Manila 1002, Philippines}

\author{Bobomurat Ahmedov\,\orcidlink{0000-0002-1232-610X}}
\email[Corresponding author:]{ahmedov@astrin.uz}
\affiliation{School of Physics, Harbin Institute of Technology, Harbin 150001, People's Republic of China}
\affiliation{Institute of Theoretical Physics, National University of Uzbekistan, Tashkent 100174, Uzbekistan}

\begin{abstract}
We study horizon-brightened acceleration radiation for a neutral scalar field in separable Kerr-like rotating metrics with nonlinear-electrodynamic mass profiles. The principal branch has $M=0$ and $\mu=3$, with finite but direction-dependent central curvature limits. An analytic extremal charge--spin parametrization fixes the black hole domain. At a simple outer horizon, the scalar equation reduces to local conformal quantum mechanics with corotating energy $\epsh=\omega-m\OmH$. For an isolated outgoing channel we derive the first finite-gap correction in proper time, including the radial Frobenius coefficient, angular variation, geodesic terms, and switching slope. This replaces a coordinate-dependent trajectory-phase multiplier by a protocol-defined detailed-balance affinity. Direct integration of the outgoing scalar equation and an equatorial infalling trajectory validates the expansion for an explicitly specified finite gate; a solvable logarithmic-mode limit separates switching from geometric contributions. We also analyze forced equatorial circular atoms and show that co-rotating and counter-rotating ultra-relativistic motion yields a mode-dependent orbital response rather than the infalling Hawking affinity. We identify the shrinking near-extremal radial scale and quantify the effect of an unselected second radial branch, without asserting a uniform extremal limit. Random atomic injection gives a selected-mode master equation and a nonnegative population entropy-production identity. A direct first-law integrability test shows why horizon-area bookkeeping remains conditional for this off-shell metric. Independently converged Chebyshev scalar quasinormal frequencies validate the WKB results and probe the exterior geometry beyond the leading horizon data. The calculation distinguishes branch-resolved horizon thermality, finite detector protocol, and exterior scalar dynamics.
\end{abstract}


\maketitle

\section{Introduction}\label{sec:intro}

The coexistence of the laws of black hole mechanics, the entropy--area relation, and Hawking radiation implies that horizon geometry carries thermodynamic information~\cite{Bardeen:1973gs,Bekenstein:1973ur,Hawking:1974rv,Hawking:1975vcx}.  In quantum field theory on curved spacetime, this conclusion is inseparable from the observer dependence of positive frequency and vacuum states~\cite{Birrell:1982ix,Wald:1995yp}.  The Fulling--Davies--Unruh effect provides the corresponding acceleration-induced response in flat spacetime~\cite{Fulling:1972md,Davies:1974th,Unruh:1976db}.  These results motivate operational probes in which a localized quantum system interrogates the near-horizon field rather than inferring thermality only from asymptotic particle production.

Horizon-brightened acceleration radiation is one such probe.  A ground-state two-level atom falls freely through a Boulware-type field state and can become excited while emitting a field quantum through the counter-rotating sector of a weak atom--field interaction~\cite{Boulware:1974dm,Scully:2017utk}.  The atom is locally inertial, but the exterior field modes are accelerated relative to it.  Near a nonextremal horizon, the transition amplitude is dominated by the logarithmic phase of the outgoing mode.  This phase is governed by an inverse-square radial problem, i.e. by conformal quantum mechanics (CQM)~\cite{Camblong:2020pme,deAlfaro:1976vlx,Camblong:2004ye}.  A dilute cloud with randomized injection times leads to a birth--death master equation and, under the usual mode-selection assumptions, to a thermal stationary state~\cite{Azizi:2021qcu,Azizi:2021yto,Ordonez:2025sqp}.

Rotation changes both the kinematics and the thermodynamic affinity.  The horizon is generated by
\begin{equation}
 \chi=\partial_t+\OmH\partial_\phi,
\end{equation}
and the relevant mode energy is not the coordinate frequency $\omega$ but
\begin{equation}\label{eq:eps-def-intro}
 \epsh=\omega-m\OmH.
\end{equation}
For Kerr, the near-horizon CQM derivation and the associated Planck kernel survive after transforming to the corotating frame~\cite{Azizi:2020gff}.  The positive-$\epsh$ sector behaves as a thermal bosonic sector, whereas $\epsh<0$ belongs to the superradiant gain regime and cannot be represented by a normalizable positive-temperature Gibbs state~\cite{Brito:2015oca}.

The present work combines this rotating HBAR construction with a broad family of regular metrics motivated by nonlinear electrodynamics (NLED).  Regular black holes replace the classical curvature singularity by a nonsingular core~\cite{Bardeen:1968,Hayward:2005gi,Verbin:2024ewl,Dymnikova:2004zc,Ayon-Beato:1999kuh,Bronnikov:2000vy,Ayon-Beato:2000mjt,Wang:2026sqr,Guo:2024svn,Zhang:2024hix}; the Fan--Wang class provides a systematic static realization within Einstein gravity coupled to NLED~\cite{Fan:2016hvf}.  An alternate Newman--Janis construction yields an analytically simple rotating Kerr-like family with a radial mass profile~\cite{Toshmatov:2017zpr}.  The rotating matter reconstruction is not exact in a strict NLED sense: a subsequent comment identified a mismatch between the reconstructed Lagrangian and its derivative~\cite{Rodrigues:2017tfm}, while a quantitative follow-up found that the discrepancy is small for representative parameters and advocated the metric as an effective analytic model~\cite{Toshmatov:2017note}.  We therefore formulate our results as geometric statements for the off-shell Kerr metric.  The local HBAR derivation requires its horizon structure and separability, not an exact microscopic NLED Lagrangian.

\begin{revision}
Detector calculations in generic static geometries and beyond the leading near-horizon approximation provide important comparisons~\cite{Sen:2022cdx,Sen:2023zfq}. Static NLED effects on acceleration radiation have been studied in model-specific and generic regular geometries~\cite{Ovgun:2025ehi,Uktamov:2026hsw,2026arXiv260809976X}, while massive-vector extensions separate the local thermal factor from polarization and propagation effects~\cite{Pantig:2025eew}. Finite switching and the order of large-gap limits require particular care~\cite{Fewster2016Waiting}. These results motivate a precise question: which part of an infalling detector response is fixed by the rotating horizon, and which part depends on the detector protocol and the exterior mode?

Our contribution is a rotating, general-angle proper-time expansion in which the radial mode and geodesic corrections are retained at the same order. We derive its first finite-gap affinity coefficient, identify cancellations that an isolated trajectory-phase correction misses, and test the expansion by integrating the exact outgoing scalar equation and an equatorial worldline with a specified exterior gate. The exact extremal parametrization provides a common admissible domain for the Maxwellian, Bardeen-like, and Hayward-like profiles. We distinguish a selected-mode entropy identity from an area law, since an explicit integrability check obstructs the naive fixed-work first law of the off-shell rotating family.

We also solve the neutral massless scalar quasinormal-mode (QNM) problem by independent WKB and Chebyshev methods. Rotating Bardeen/Hayward scalar spectra have already been studied~\cite{PengHuang2025}; links between quasinormal dynamics and acceleration radiation have also been explored~\cite{Ovgun2026QNMOptics,Pantig2025Ringdown,Lutfuoglu:2025mqa,Lutfuoglu:2026fpx,Lutfuoglu:2026boa,Lutfuoglu:2026etg,Wang:2026sqr}. Here the QNM calculation supplies a converged exterior-field comparison for the same separated operator used in the detector problem. It is neither a gravitational-wave calculation nor a prescription for replacing a real detector-mode frequency by a complex QNM frequency.

We use $G=c=\hbar=k_{\rm B}=1$ unless constants are restored, and signature $(-,+,+,+)$. Section~\ref{sec:geometry} defines the geometry and parameter domain; Sec.~\ref{sec:kg} gives the scalar separation and local CQM reduction. Section~\ref{sec:detector} develops the infalling trajectory and leading kernels. The proper-time finite-gap calculation and its operational assumptions are given in Sec.~\ref{sec:operational}. Sections~\ref{sec:master} and \ref{sec:numerics} describe selected-mode dynamics and numerical results. Section~\ref{sec:scalar-qnms} gives the independent scalar spectrum, followed by limitations and conclusions. The appendices provide derivative identities, geodesic and numerical coefficients, and charged-scalar separation for a prescribed magnetic background.
\end{revision}

\section{Generic rotating regular geometry}\label{sec:geometry}

\subsection{Metric, mass profile, and scaling}

We consider the Boyer--Lindquist form~\cite{Toshmatov:2017zpr}
\begin{align}\label{eq:metric}
 \dd s^2={}&-\frac{\Delta}{\Sigma}\left(\dd t-a\sin^2\theta\,\dd\phi\right)^2
 +\frac{\Sigma}{\Delta}\dd r^2+\Sigma\dd\theta^2 \nonumber\\
 &+\frac{\sin^2\theta}{\Sigma}
 \left[(r^2+a^2)\dd\phi-a\dd t\right]^2,
\end{align}
where
\begin{equation}\label{eq:SigmaDelta}
 \Sigma=r^2+a^2\cos^2\theta,
 \qquad
 \Delta=r^2-2r\rho(r)+a^2.
\end{equation}
Before nondimensionalization, the radial mass function is
\begin{equation}\label{eq:rho-dimensional}
 \rho_{\rm dim}(r)=M+\frac{q_{\rm dim}^3}{\alpha}
 \frac{r^\mu}{(r^\nu+q_{\rm dim}^\nu)^{\mu/\nu}},
\end{equation}
with $\alpha$ of dimension length squared and electromagnetic mass
\begin{equation}
 \Mrem=\frac{q_{\rm dim}^3}{\alpha}.
\end{equation}
Scaling every length by $\Mrem$ gives
\begin{equation}\label{eq:rho-dimensionless}
 \rho(r)=M+\frac{r^\mu}{(r^\nu+q^\nu)^{\mu/\nu}}.
\end{equation}
\rev{Unless dimensions are explicitly restored, all lengths, times, and frequencies below are expressed in units of $\Mrem$, $\Mrem$, and $\Mrem^{-1}$, respectively. Thus the dimensionless symbols $r$, $a$, and $q$ are not divided by $\Mrem$ again in the figures. The integration constant in Eq.~\eqref{eq:rho-dimensionless} is $M/\Mrem$, and $M_{\rm ADM}=M+1$. The profile parameter $q$ is not automatically the physical magnetic charge or the coefficient of a rotating gauge potential; these are distinguished in Appendix~\ref{app:charged}.}

\begin{figure*}[ht!]\centering
\includegraphics[width=0.4\textwidth]{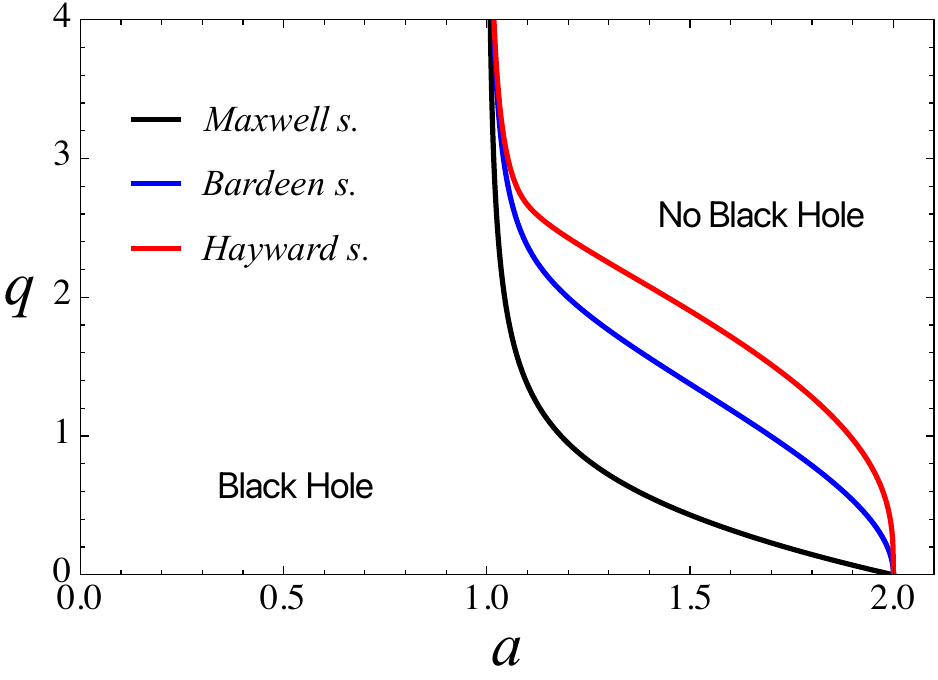}
\caption{The separation of the ($q,a$) parameter space into black hole regions and no-horizon (naked-singularity) regions.
\label{Fig.BH region}}
\end{figure*}

\begin{figure*}[ht!]\centering
\includegraphics[width=0.4\textwidth]{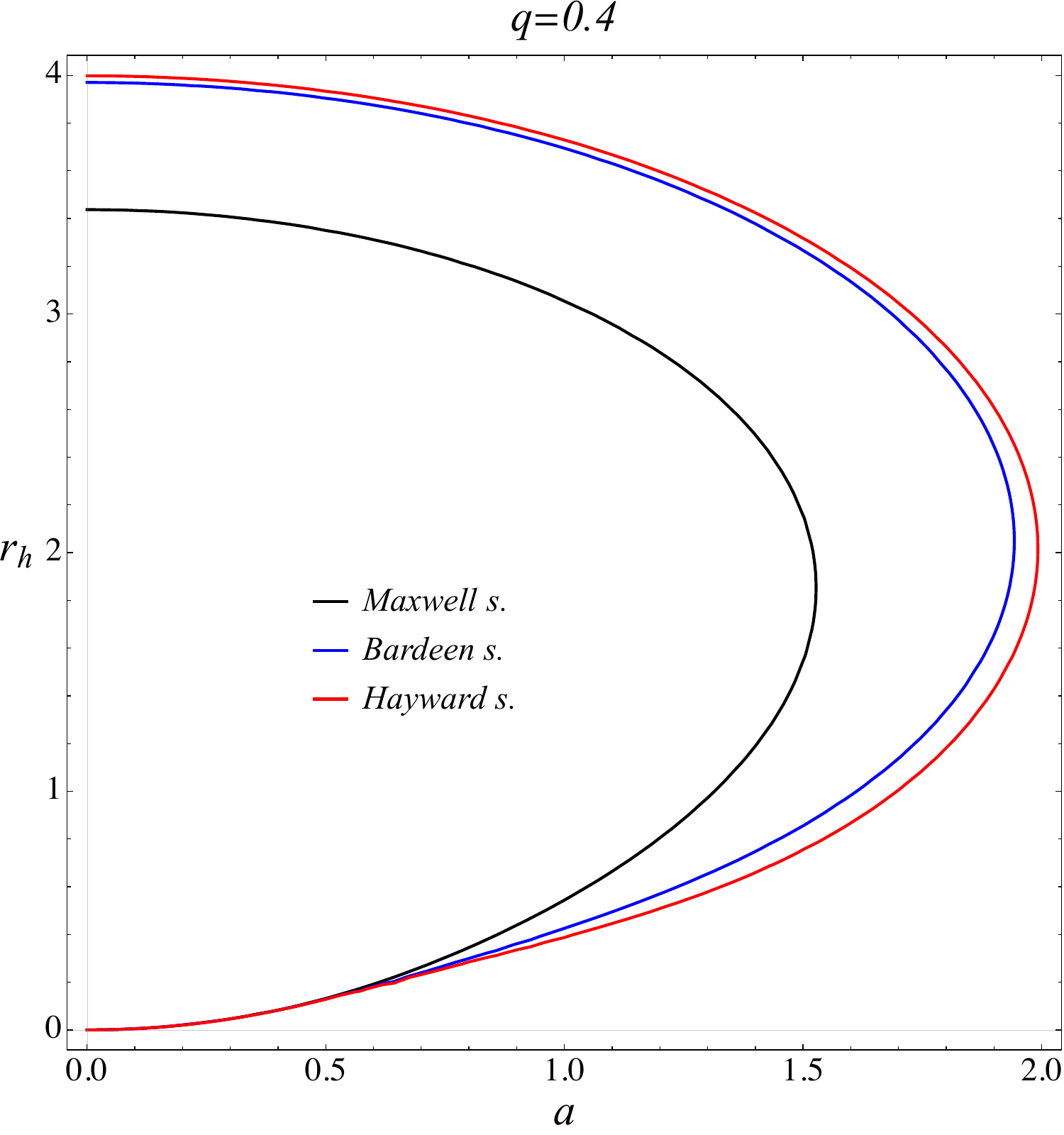}
\caption{Outer-horizon radius as a function of spin $r_h$ on the spin $a$ of the BH.
\label{Fig.rh}}
\end{figure*}

The allowable ranges of the black hole parameters $q$ and $a$ are determined by imposing the extremal condition:
\begin{eqnarray}
    \Delta=\frac{\partial\Delta}{\partial r}=0\,,
\end{eqnarray}  
which defines the boundary of the physical parameter space. Then we show these extreme values of the charge $q$ and the spin $a$ in Fig.(\ref{Fig.BH region}) for different rotating generic regular BH solutions. Fig.(\ref{Fig.rh}) shows the outer-horizon radius as a function of spin.

Finite curvature at the rotating center requires the branch $M=0$ with $\mu\geq3$; a nonzero Schwarzschild integration mass restores the Kerr-like central singularity~\cite{Toshmatov:2017zpr}.  We therefore use
\begin{equation}\label{eq:regular-sector}
 M=0,
 \qquad \mu=3,
\end{equation}
as the minimal finite-curvature branch.  For $\mu=3$, the curvature invariants remain finite as the center is approached, but their limiting values depend on the direction of approach; for $\mu>3$, the central limits are direction independent~\cite{Toshmatov:2017zpr}.  We consequently use ``regular'' in the standard finite-curvature sense and do not infer smooth extendibility through the rotating center from the $\mu=3$ metric alone.  The HBAR calculation is entirely exterior, $\rh>0$, and depends only on the local simple-horizon geometry.  The labels Maxwellian, Bardeen-like, and Hayward-like refer respectively to $\nu=1,2,3$.  The formal $q\rightarrow0$ limit of the dimensionless profile is $\rho\rightarrow1$ outside the origin, reproducing a Kerr metric of unit mass.

The profile derivatives needed below are
\begin{align}\label{eq:rho-derivatives}
 \rho'(r)&=\frac{\mu q^\nu r^{\mu-1}}
 {(r^\nu+q^\nu)^{\mu/\nu+1}},\\
 \rho''(r)&=\frac{\mu q^\nu r^{\mu-2}
 \left[(\mu-1)q^\nu-(\nu+1)r^\nu\right]}
 {(r^\nu+q^\nu)^{\mu/\nu+2}},
\end{align}
so that
\begin{align}\label{eq:Delta-derivatives}
 \Delta'&=2\left(r-\rho-r\rho'\right),\\
 \Delta''&=2-4\rho'-2r\rho''.
\end{align}
\rev{The explicit $\mu=3$ forms are listed in Appendix~\ref{app:derivatives}.}

\subsection{Restoration of dimensions and physical parameter paths}\label{subsec:dimensions}

Let a superscript ``geom'' denote a dimensionful quantity in geometrized units.  Because $\Mrem$ is the length unit, the conversion rules are
\begin{align}\label{eq:dimension-restoration}
 r_{\rm geom}&=\Mrem r,
 &a_{\rm geom}&=\Mrem a,
 &q_{\rm dim}&=\Mrem q,\\
 \kap_{\rm geom}&=\frac{\kap}{\Mrem},
 &\OmH^{\rm geom}&=\frac{\OmH}{\Mrem},
 &\TempH^{\rm geom}&=\frac{\TempH}{\Mrem},\\
 A_{\rm H}^{\rm geom}&=\Mrem^2 A_{\rm H}.
\end{align}
Restoring $G$, $c$, $\hbar$, and $k_{\rm B}$ gives
\begin{align}\label{eq:SI-restoration}
 M_{\rm ADM}^{\rm SI}
 &=\frac{c^2}{G}(M+1)\Mrem,\\
 T_{\rm H}^{\rm SI}
 &=\frac{\hbar c}{k_{\rm B}\Mrem}\,\TempH,\\
 \frac{S_{\rm BH}^{\rm SI}}{k_{\rm B}}
 &=\frac{c^3A_{\rm H}^{\rm geom}}{4G\hbar}
 =\frac{A_{\rm H}^{\rm geom}}{4\ell_{\rm P}^2},
 \qquad
 \ell_{\rm P}^2=\frac{G\hbar}{c^3}.
\end{align}
The corresponding geometrized ADM mass length is $M_{\rm ADM}^{\rm geom}=(M+1)\Mrem$.  The definitions
\begin{equation}
 \Mrem=\frac{q_{\rm dim}^3}{\alpha},
 \qquad
 q=\frac{q_{\rm dim}}{\Mrem}
\end{equation}
imply the correlated parameter path
\begin{equation}\label{eq:physical-parameter-path}
 q_{\rm dim}=q\Mrem,
 \qquad
 \alpha=q^3\Mrem^2.
\end{equation}
Thus, varying the dimensionless $q$ at fixed $\Mrem$ changes both the dimensional charge parameter and the NLED coupling; it is not a fixed-$\alpha$ scan within one microscopic theory.  Likewise, the formal Kerr limit $q\to0$ at fixed nonzero $\Mrem$ is the correlated limit $q_{\rm dim}\to0$ and $\alpha\to0$ with $q_{\rm dim}^3/\alpha=\Mrem$ held fixed.  All plots below state dimensionless quantities and should be interpreted along this path unless a different dimensional prescription is specified.

\subsection{Exact extremal charge--spin boundary}\label{subsec:extremal}

A nonextremal black hole has a simple outer root $\rh$ of $\Delta$.  Extremality occurs at $r=r_{\rm e}$ when
\begin{equation}\label{eq:extremality}
 \Delta(r_{\rm e})=0,
 \qquad
 \Delta'(r_{\rm e})=0.
\end{equation}
For $M=0$, introduce
\begin{equation}
 y=\left(\frac{q}{r_{\rm e}}\right)^\nu.
\end{equation}
\rev{For $\mu>1$ and $\nu>0$, Eqs.~\eqref{eq:extremality} admit the parametrization}
\begin{align}\label{eq:extremal-param}
 r_{\rm e}(y)&=\frac{1+(\mu+1)y}{(1+y)^{1+\mu/\nu}},\\
 q_{\rm e}(y)&=r_{\rm e}(y)y^{1/\nu},\\
 a_{\rm e}^2(y)&=\frac{2r_{\rm e}(y)}{(1+y)^{\mu/\nu}}-r_{\rm e}^2(y).
\end{align}
The condition $a_{\rm e}^2\geq0$ reduces to
\begin{equation}\label{eq:y-domain}
 0\leq y\leq\frac{1}{\mu-1}.
\end{equation}
The endpoint $y=0$ is extremal Kerr, $(q_{\rm e},a_{\rm e},r_{\rm e})=(0,1,1)$, while $y=1/(\mu-1)$ is the static extremal solution.  For $\mu=3$, the static critical charge and radius are
\begin{equation}\label{eq:static-critical}
 q_* =\left(\frac{2^{\nu+2}}{27}\right)^{1/\nu},
 \qquad
 r_*=2^{1/\nu}q_*.
\end{equation}
\rev{Table~\ref{tab:critical} and Fig.~\ref{fig:extremal} summarize the three subclasses. At fixed $0\leq a<1$ and $q>0$, the subextremal domain is $q<q_{\rm e}(a)$; crossing the curve removes the pair of positive roots. The $a=q=0$ Schwarzschild endpoint has only one positive horizon. On the $M=0$ branch, $\rho\leq1$ gives $\Delta\geq(r-1)^2+a^2-1$, so no black hole point with $a>1$ is admissible. A horizonless region is not, by that fact alone, a naked-singularity region.}

\subsection{Horizon quantities and ergoregion}

Let $\rh$ be the largest positive root of $\Delta$.  The null generator, angular velocity, surface gravity, Hawking temperature, area, and entropy are
\begin{align}\label{eq:horizon-data}
 \chi&=\partial_t+\OmH\partial_\phi,
 &\OmH&=\frac{a}{\rh^2+a^2},\\
 \kap&=\frac{\Delta'(\rh)}{2(\rh^2+a^2)},
 &\TempH&=\frac{\kap}{2\pi},\\
 A_{\rm H}&=4\pi(\rh^2+a^2),
 &S_{\rm BH}&=\frac{A_{\rm H}}{4}.
\end{align}
\rev{The nonextremal analysis requires $\Delta'(\rh)>0$. The extremal limit has $\kap=\TempH=0$ and cannot be substituted into the simple-horizon CQM formulas. The area is a geometric invariant; identifying $A_{\rm H}/4$ with the entropy in a dynamical first law requires the source and work terms discussed in Sec.~\ref{subsec:area-bookkeeping}.}

The stationary-limit surface follows from $g_{tt}=0$:
\begin{equation}\label{eq:static-limit}
 r^2-2r\rho(r)+a^2\cos^2\theta=0.
\end{equation}
Its outer root $r_{\rm s}(\theta)$ encloses the event horizon except on the rotation axis.  Equation~\eqref{eq:static-limit} is the horizon equation evaluated at the effective spin $a|\cos\theta|$, which gives a stable numerical construction of the ergosurface.

\begin{figure*}[ht!]\centering
\includegraphics[width=0.23\textwidth]{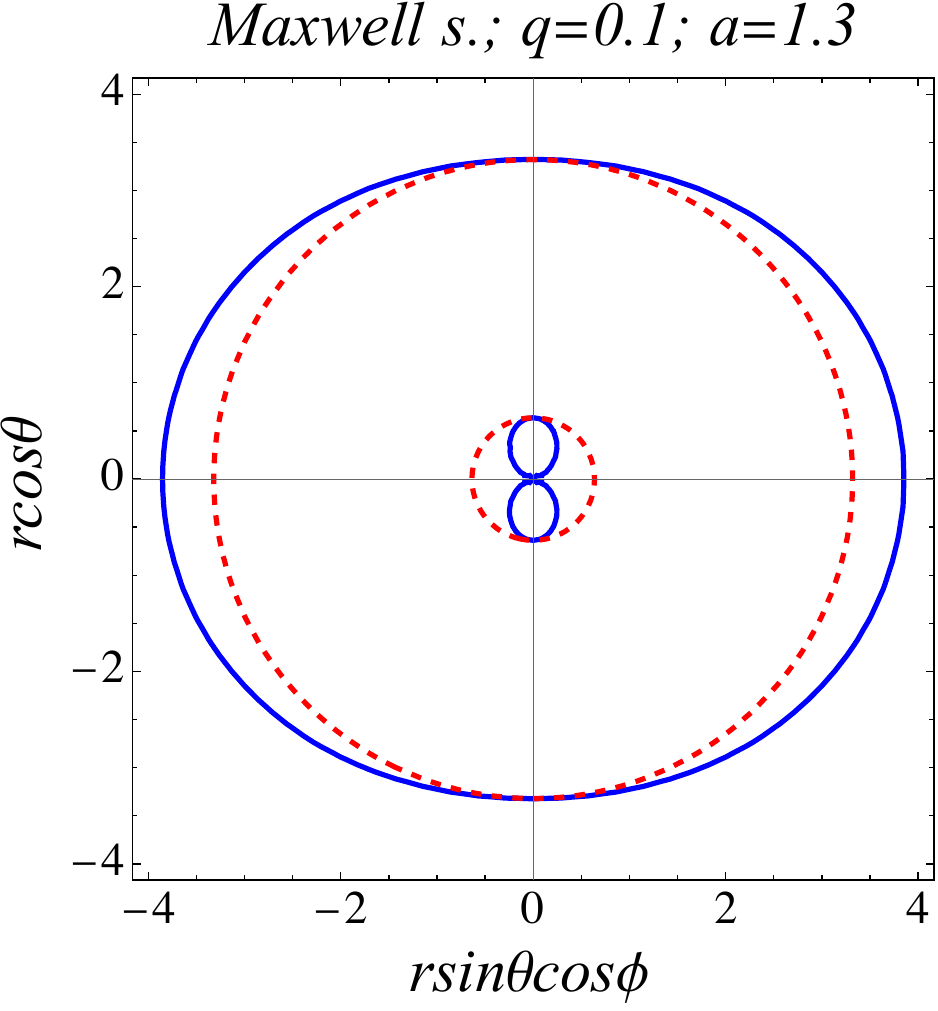}
\includegraphics[width=0.23\textwidth]{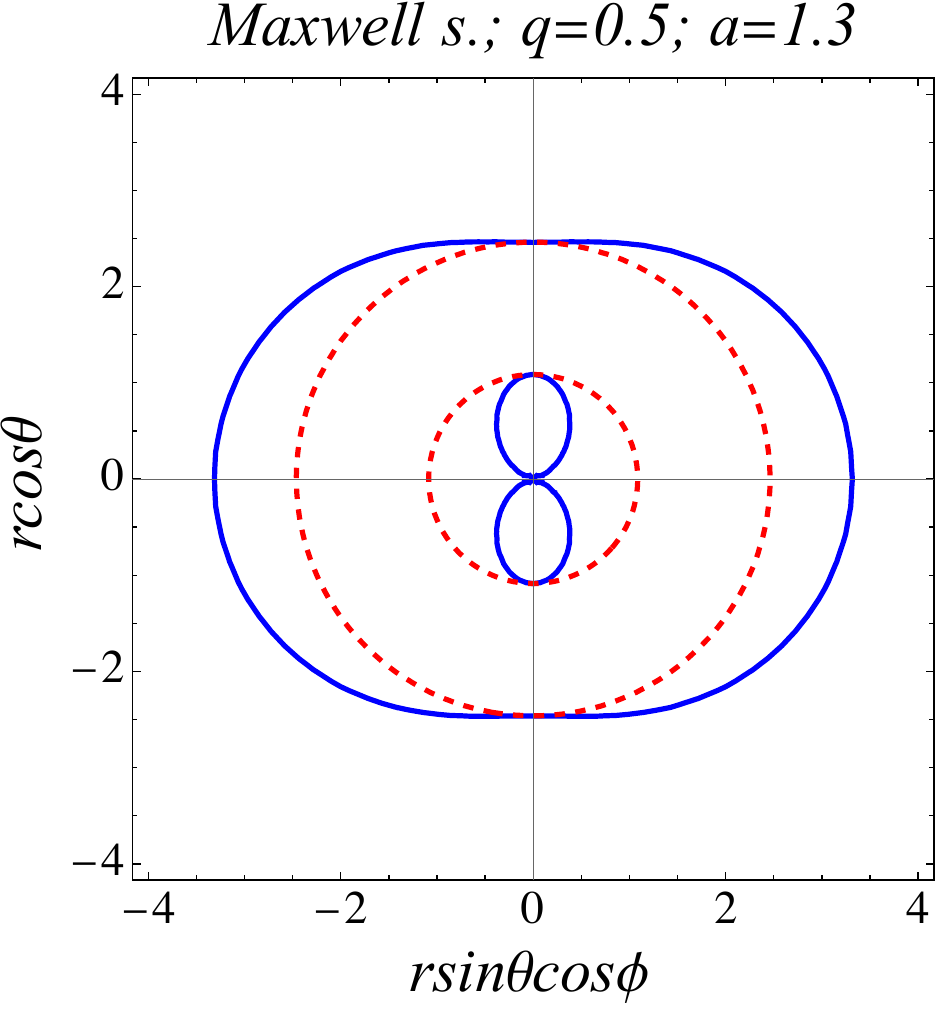}
\includegraphics[width=0.23\textwidth]{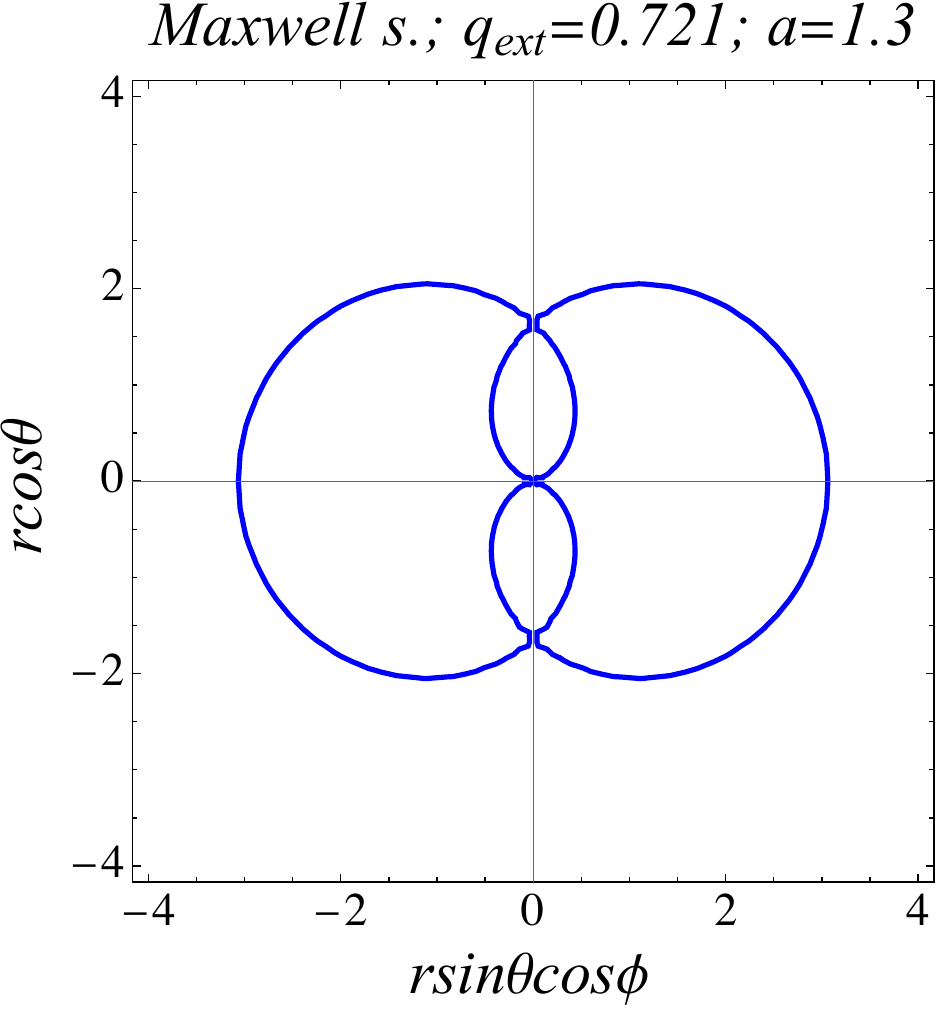}
\includegraphics[width=0.23\textwidth]{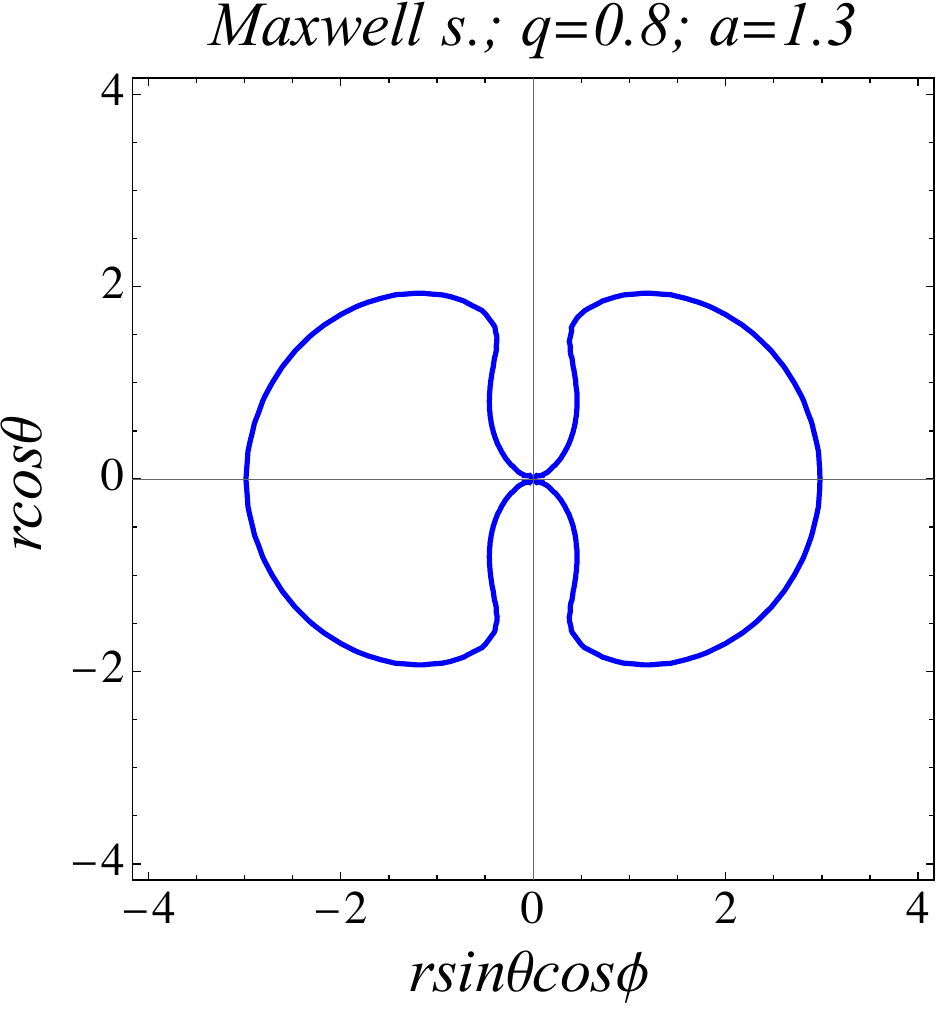}
\includegraphics[width=0.23\textwidth]{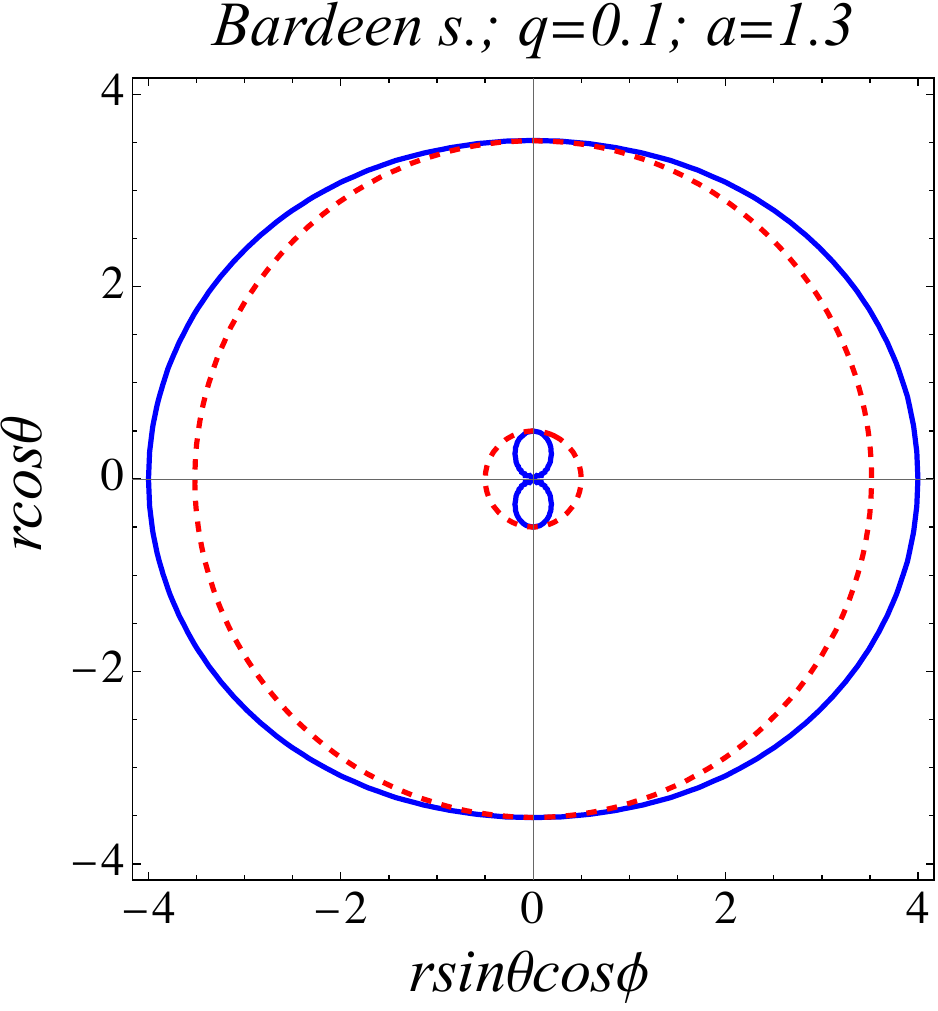}
\includegraphics[width=0.23\textwidth]{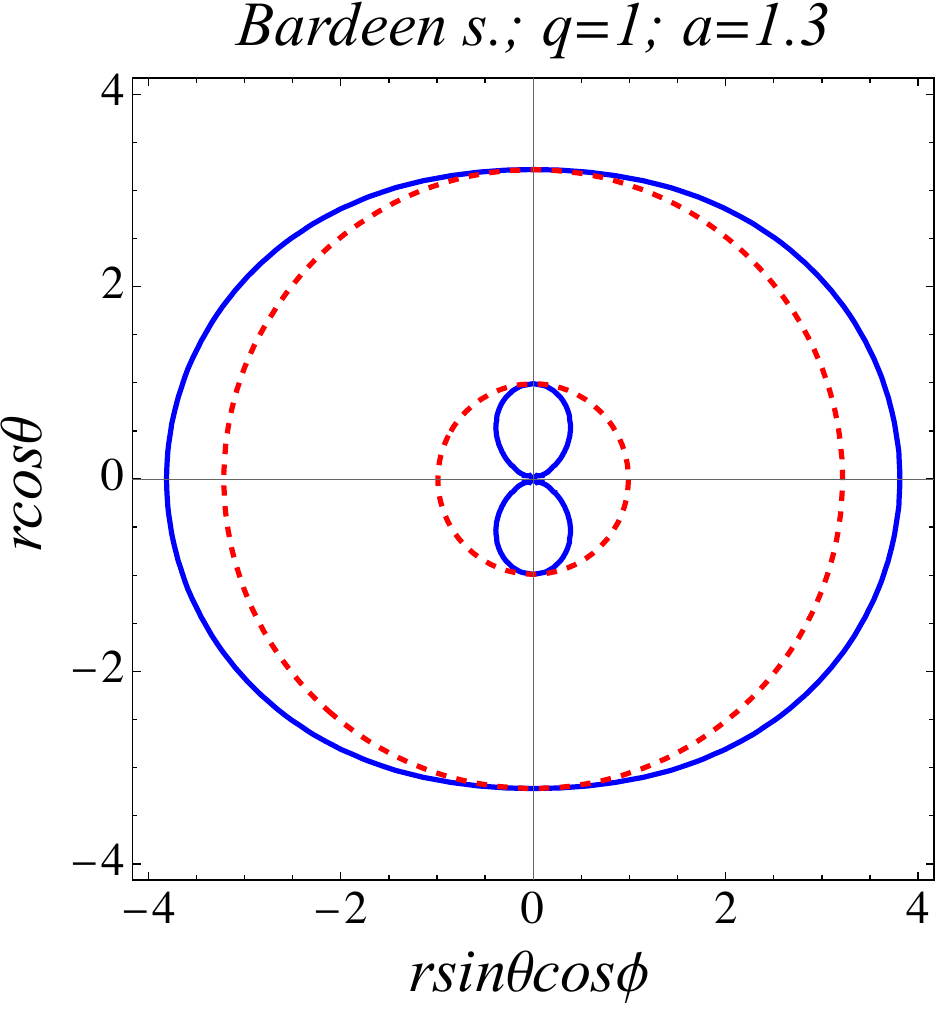}
\includegraphics[width=0.23\textwidth]{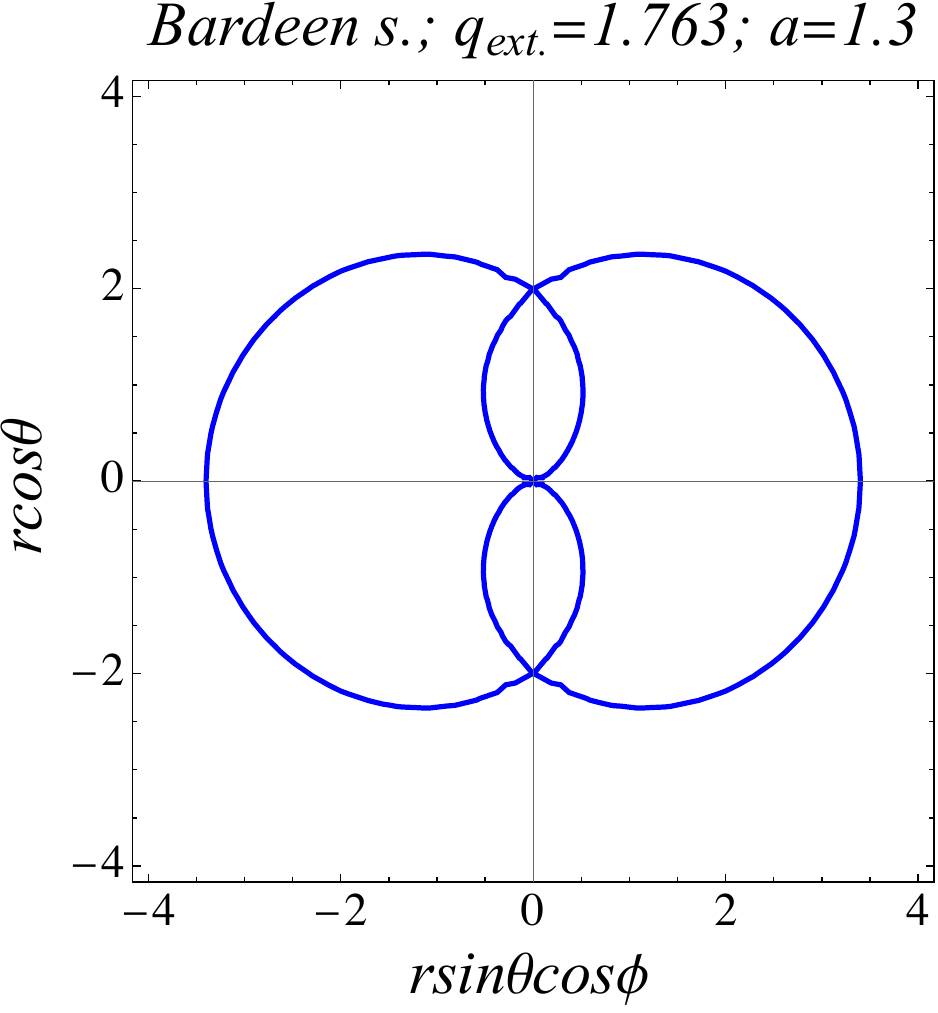}
\includegraphics[width=0.23\textwidth]{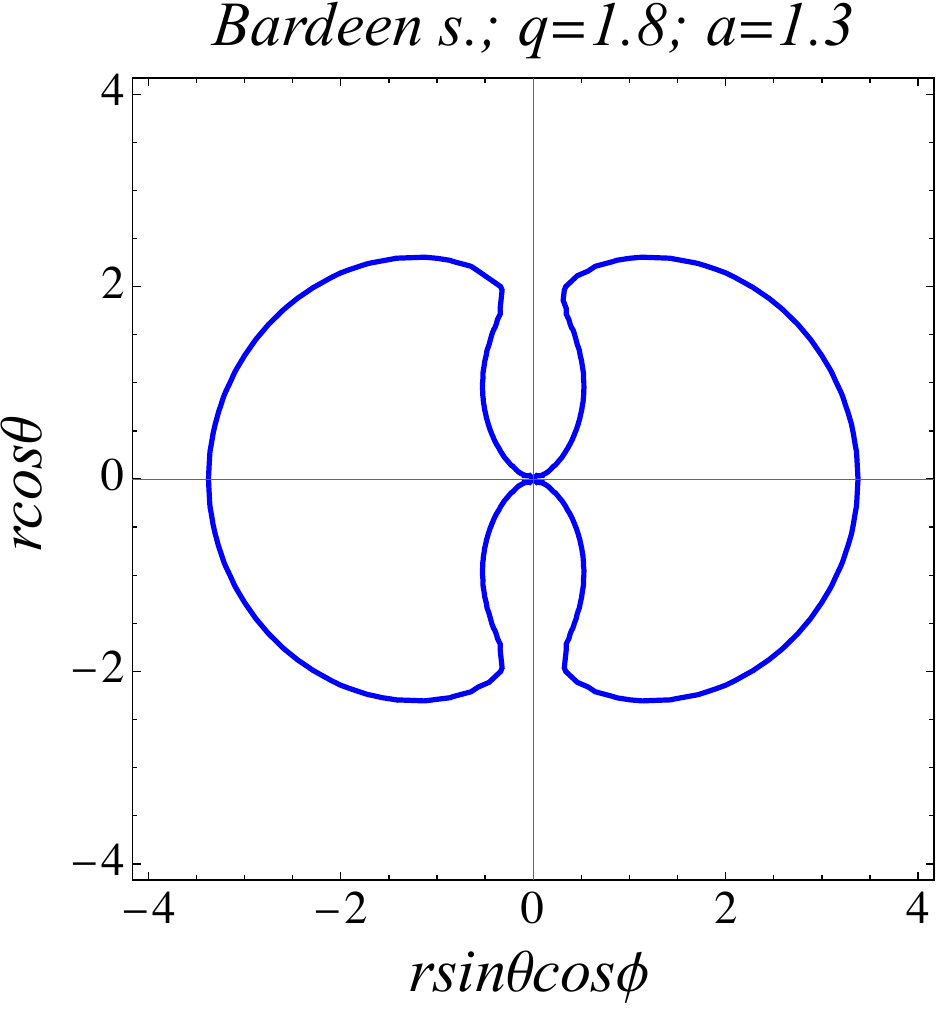}
\includegraphics[width=0.23\textwidth]{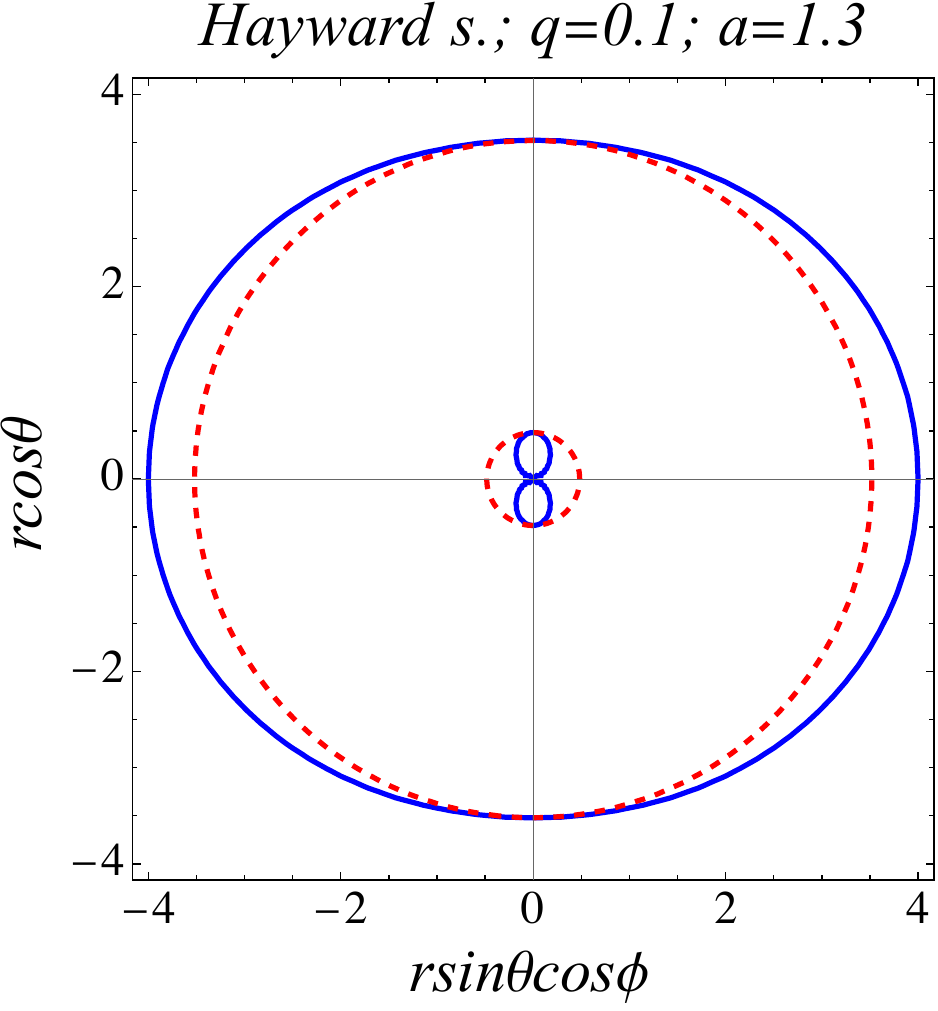}
\includegraphics[width=0.23\textwidth]{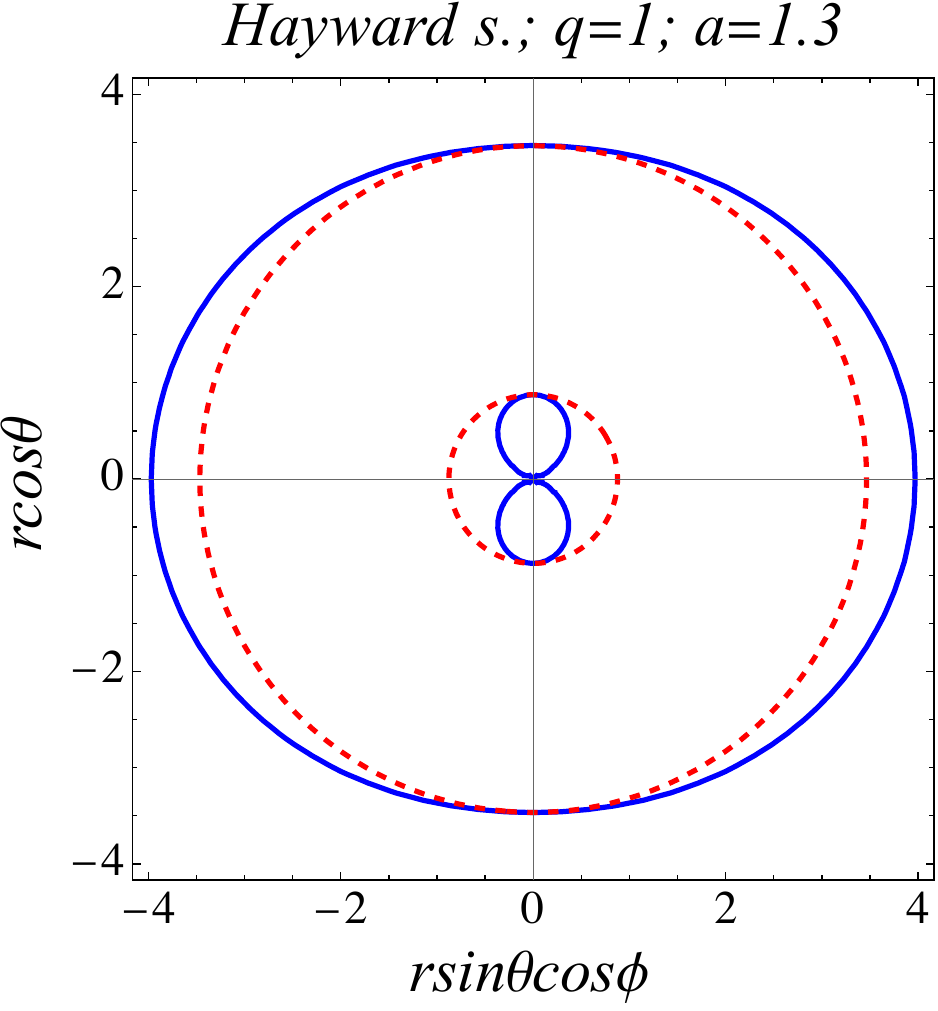}
\includegraphics[width=0.23\textwidth]{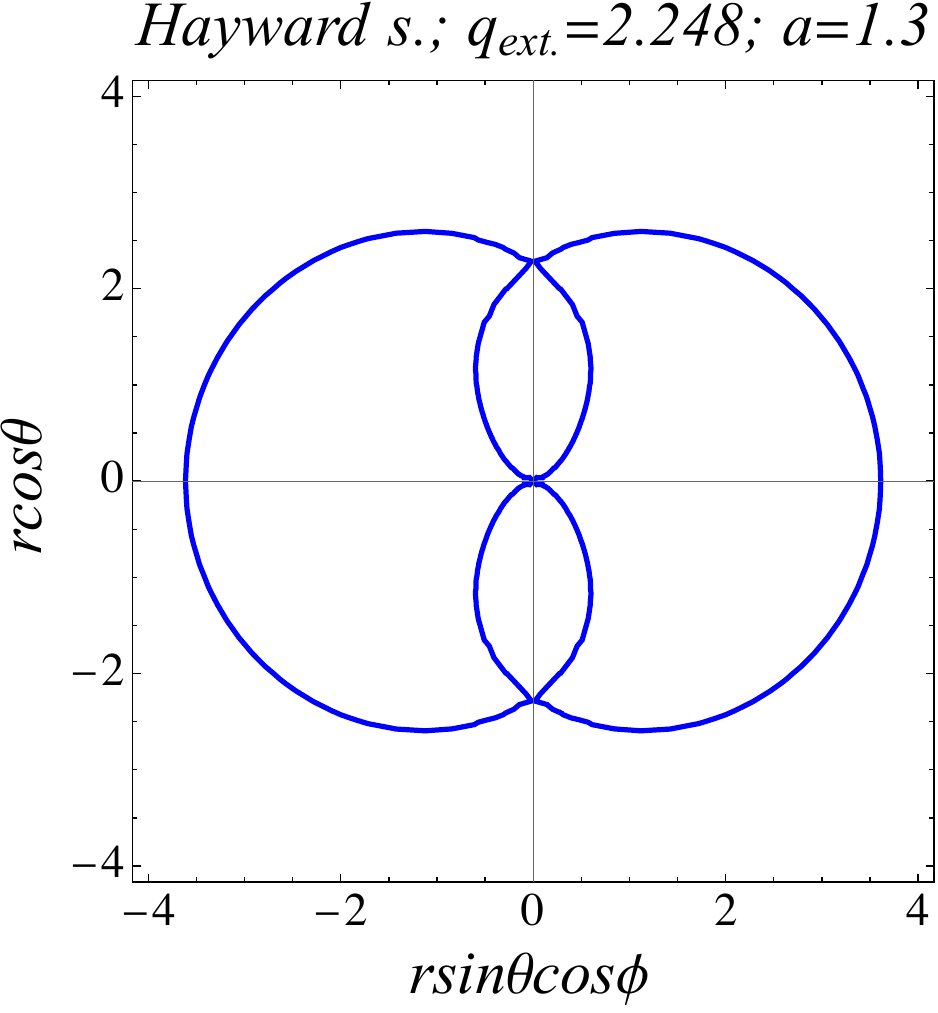}
\includegraphics[width=0.23\textwidth]{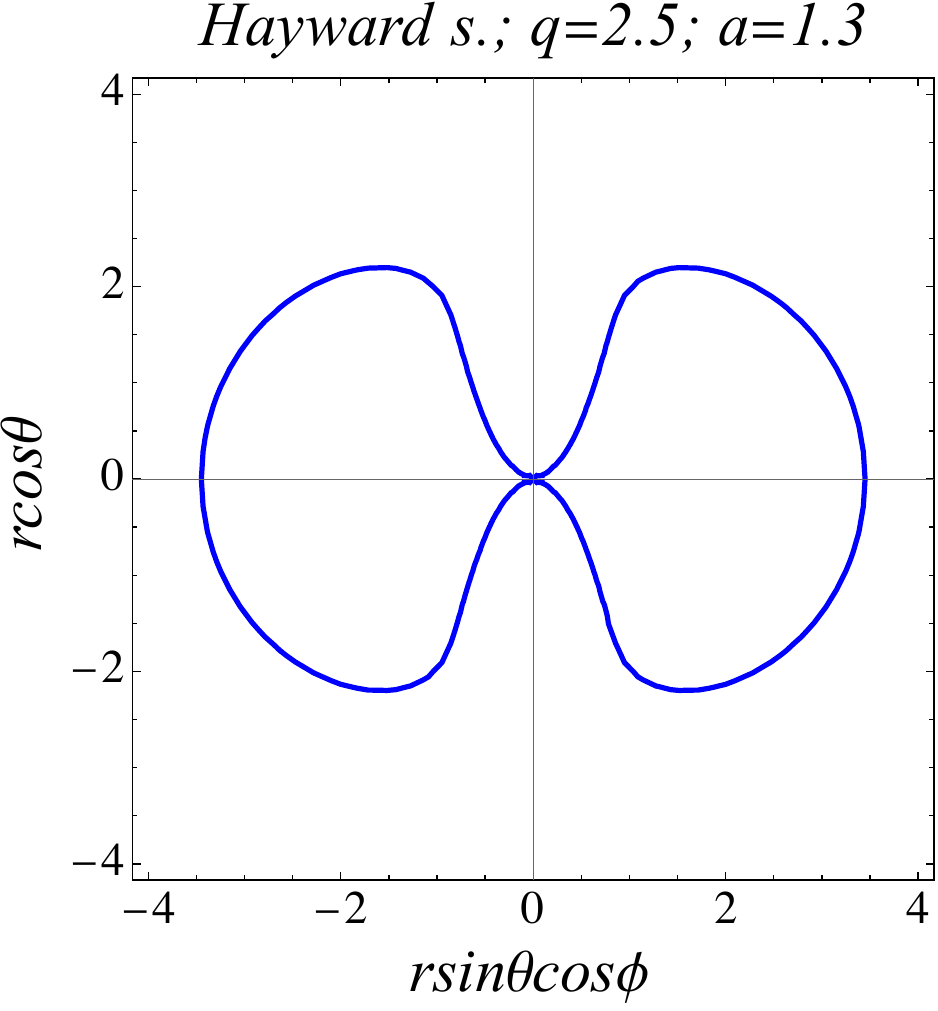}
\caption{The shape of the static limit (solid blue lines) and event horizon (dashed red curves) of the generic rotating BH in $xz$-plane for different cases.
\label{Fig.gtt}}
\end{figure*}

Then, employing the condition $g_{tt}=0$ to determine the ergoregion boundaries, we have displayed the ergosphere shapes for various cases in Fig.(\ref{Fig.gtt}). It is clear from first and second columns of  Fig.(\ref{Fig.gtt}) that increasing the value of the charge $q$ of the generic rotating BHs causes reducing the ergosphere of the generic regular BHs.

\begin{revision}
\subsection{Horizon data and the local response}\label{subsec:geometry-hbar-map}
At leading order, the normalized outgoing-channel kernel depends on the mass profile only through $\kap$ and $\OmH$:
\begin{equation}\label{eq:local-kernel-map}
 \mathcal K_{\rm HBAR}(\omega,m)
 =\frac{\epsh/\kap}{\exp(2\pi\epsh/\kap)-1},
 \qquad \epsh=\omega-m\OmH>0.
\end{equation}
A lower temperature does not by itself imply a smaller response at fixed $\omega$, because the corotating energy changes with $\OmH$. Geometries with the same two horizon parameters have the same leading normalized kernel, but need not have the same finite-gate response or scalar spectrum. The next local coefficient contains $\Delta''(\rh)$ through both the radial Frobenius series and the trajectory. Global normalization, angular weights, interference and transmission require information beyond this leading kernel. The formal $q\to0$ limit gives Kerr; setting $a=0$ gives the static regular family, and both limits give Schwarzschild.
\end{revision}

\section{Scalar field and near-horizon conformal quantum mechanics}\label{sec:kg}

\subsection{Exact Klein--Gordon separation}

A real scalar field of mass $\mu_\Phi$ obeys
\begin{equation}\label{eq:KG}
 (\Box-\mu_\Phi^2)\Phi=0.
\end{equation}
For the metric~\eqref{eq:metric}, $\sqrt{-g}=\Sigma\sin\theta$.  Multiplying Eq.~\eqref{eq:KG} by $\Sigma$ yields
\begin{align}\label{eq:KG-expanded}
 0={}&-\frac{(r^2+a^2)^2-a^2\Delta\sin^2\theta}{\Delta}
 \partial_t^2\Phi
 -\frac{4ar\rho(r)}{\Delta}\partial_t\partial_\phi\Phi
 \nonumber\\
 &+\left(\frac{1}{\sin^2\theta}-\frac{a^2}{\Delta}\right)
 \partial_\phi^2\Phi
 +\partial_r(\Delta\partial_r\Phi)
 \nonumber\\
 &+\frac{1}{\sin\theta}\partial_\theta
 (\sin\theta\partial_\theta\Phi)-\mu_\Phi^2\Sigma\Phi.
\end{align}
No slow-rotation expansion is needed.  With
\begin{equation}\label{eq:separation-ansatz}
 \Phi=R_{\omega lm}(r)S_{\omega lm}(\theta)
 \e^{\ii m\phi-\ii\omega t},
\end{equation}
we obtain
\begin{align}\label{eq:angular}
 \frac{1}{\sin\theta}\frac{\dd}{\dd\theta}
 \left(\sin\theta\frac{\dd S}{\dd\theta}\right)
 +\bigg[&a^2(\omega^2-\mu_\Phi^2)\cos^2\theta
 \nonumber\\
 &-\frac{m^2}{\sin^2\theta}+\Lambda_{\omega lm}\bigg]S=0
\end{align}
and
\begin{align}\label{eq:radial-exact}
 \frac{\dd}{\dd r}\left(\Delta\frac{\dd R}{\dd r}\right)
 +\bigg\{&\frac{\left[(r^2+a^2)\omega-am\right]^2}{\Delta}
 -\mu_\Phi^2r^2
 \nonumber\\
 &-a^2\omega^2+2am\omega-\Lambda_{\omega lm}\bigg\}R=0.
\end{align}
The numerator can equivalently be written as
\begin{equation}\label{eq:radial-expanded-numerator}
 (r^2+a^2)^2\omega^2-4ar\rho(r)m\omega+a^2m^2,
\end{equation}
provided the $2am\omega$ term in Eq.~\eqref{eq:radial-exact} is combined over the common denominator.  The square on $(r^2+a^2)$ is essential.  Equations~\eqref{eq:angular} and \eqref{eq:radial-exact} are the off-shell Kerr scalar equations and follow from the same hidden-symmetry structure that separates the Hamilton--Jacobi equation~\cite{Carter:1968rr,Frolov:2017kze}.

\subsection{Corotating frequency and CQM equation}

Introduce horizon-adapted coordinates
\begin{equation}\label{eq:corotating}
 \widetilde t=t,
 \qquad
 \widetilde\phi=\phi-\OmH t,
\end{equation}
so that
\begin{equation}
 \e^{\ii m\phi-\ii\omega t}
 =\e^{\ii m\widetilde\phi-\ii\epsh\widetilde t},
 \qquad
 \epsh=\omega-m\OmH.
\end{equation}
For a simple outer horizon set $x=r-\rh$.  Then
\begin{equation}\label{eq:Delta-NH}
 \Delta=\Delta_1x+\frac{1}{2}\Delta_2x^2+\order(x^3),
 \qquad
 \Delta_n=\left.\frac{\dd^n\Delta}{\dd r^n}\right|_{\rh}.
\end{equation}
The singular terms in Eq.~\eqref{eq:radial-exact} give
\begin{equation}\label{eq:CQM-R}
 \left[
 \frac{1}{x}\frac{\dd}{\dd x}
 \left(x\frac{\dd}{\dd x}\right)
 +\frac{\epsh^2}{4\kap^2x^2}
 \right]R=\order(x^{-1})R.
\end{equation}
After the Liouville transformation $R=x^{-1/2}u$,
\begin{equation}\label{eq:CQM-u}
 u''+\frac{\lambda_{\rm CQM}}{x^2}u=\order(x^{-1})u,
 \qquad
 \lambda_{\rm CQM}=\frac14+\left(\frac{\epsh}{2\kap}\right)^2.
\end{equation}
The leading solutions are
\begin{equation}\label{eq:CQM-modes}
 R_\pm\sim\left(\frac{x}{x_0}\right)^{\pm\ii\epsh/(2\kap)},
\end{equation}
where $x_0$ is an arbitrary reference length.  The $+$ and $-$ branches are outgoing and ingoing, respectively, with the convention used below.  Equivalently, the tortoise coordinate
\begin{equation}
 \frac{\dd r_*}{\dd r}=\frac{r^2+a^2}{\Delta}.
\end{equation}
Near the horizon, it satisfies
\begin{equation}
 r_*\sim\frac{1}{2\kap}\ln x,
\end{equation}
and the modes are local plane waves in $\widetilde t\mp r_*$.  Field mass, angular barrier, and separation constant are subleading at the horizon.

Here CQM denotes the local asymptotic $SO(2,1)$-type scaling sector of the radial wave equation.  For $\epsh\neq0$, Eq.~\eqref{eq:CQM-u} has $\lambda_{\rm CQM}>1/4$, so a separate global quantization of the inverse-square Hamiltonian would require a self-adjoint boundary prescription and can introduce a scale through dimensional transmutation~\cite{deAlfaro:1976vlx,Camblong:2004ye}.  No such global CQM Hamiltonian or bound-state spectrum is assumed in the HBAR calculation: only the local Frobenius/WKB phase of the spacetime field is used.

Equations~\eqref{eq:CQM-u}--\eqref{eq:CQM-modes} display a useful separation between universality and model dependence.  Any simple horizon within the canonical separable Kerr-like class considered here produces the same inverse-square operator after the replacement of the asymptotic frequency by the corotating energy.  The nonlinear-electrodynamic profile changes the leading result only through $\kap$ and $\OmH$. Higher radial derivatives are absent at leading order, but enter both the subleading radial mode and the freely falling trajectory.

\subsection{Vacuum and superradiant sector}

\begin{revision}
A globally regular, isometry-invariant Hartle--Hawking state for an asymptotically flat rotating bosonic black hole is obstructed by superradiance~\cite{Kay:1991}. Mirror constructions can remove the region beyond the speed-of-light surface in Kerr~\cite{Duffy:2002ss,Duffy2008Kerr}. They do not automatically select a single radial branch of a standing-wave mode. We therefore formulate the local result for an isolated outgoing channel with
\begin{equation}\label{eq:positive-eps}
 \epsh>0.
\end{equation}
The selected-mode open-system construction below uses the actual channel probabilities; its use of the outgoing-only kernels requires directional selection or a bound on the second-branch amplitude. The $\epsh<0$ sector requires a separate superradiant gain treatment. We do not assume that a Kerr mirror-state existence result proves the corresponding global state for every off-shell mass profile.

When the outgoing mode is evaluated on an infalling trajectory, its radial logarithmic phase and the logarithmic Boyer--Lindquist-time phase add. The full endpoint index is then $b_s=\epsh_s/\kap$, rather than the radial half-index $\epsh_s/(2\kap)$.
\end{revision}

\section{Freely falling atom and local HBAR transition kernels}\label{sec:detector}

\subsection{Atom--field interaction and normalization}

Let $|b\rangle$ and $|a\rangle$ be the ground and excited states of a two-level atom with proper transition frequency $\omega_0>0$.  To avoid confusion, $\omega_0$ is distinct from the NLED index $\nu$.  In the interaction picture, a scalar monopole/dipole analogue is
\begin{equation}\label{eq:interaction}
 V_I(\tau)=g\,\chi_s(\tau)
 \left(\sigma_-\e^{-\ii\omega_0\tau}+\sigma_+\e^{\ii\omega_0\tau}\right)
 \Phi[x^\mu(\tau)],
\end{equation}
\begin{revision}
where $g$ is weak, $\chi_s$ is a real switching function smooth on the exterior proper-time interval, and $\tau=0$ is the horizon crossing. Field emission and absorption by ground-state incident atoms are, respectively,
\begin{equation}\label{eq:atomic-processes}
 |b,0_s\rangle\longrightarrow|a,1_s\rangle,
 \qquad
 |b,1_s\rangle\longrightarrow|a,0_s\rangle.
\end{equation}
Both excite the atom; absorption is not the time reverse of emission. For a specified complex mode function $\Phi_s$,
\begin{align}\label{eq:Pe-def}
 P_{e,s}&=g^2\left|\int\dd\tau\,\chi_s(\tau)
          \e^{\ii\omega_0\tau}\Phi_s^*[x(\tau)]\right|^2,\\
 P_{a,s}&=g^2\left|\int\dd\tau\,\chi_s(\tau)
          \e^{\ii\omega_0\tau}\Phi_s[x(\tau)]\right|^2.
 \label{eq:Pa-def}
\end{align}
These define physical probabilities only after the field channel and its normalization are specified. For a branch-resolved outgoing solution with nonzero horizon amplitude, freezing the smooth factors gives
\begin{equation}\label{eq:physical-local-factorization}
 P_{e,s}^{\rm out}=\mathcal W_s\mathscr R_{e,s}^{\rm loc}
 (1+\mathcal E_{e,s}),
\end{equation}
where $\mathcal W_s$ contains the common Klein--Gordon normalization, angular and switching weights, and $\mathcal E_{e,s}$ is a relative error for this specified channel. There is an analogous absorption expression. A full scattering or cavity mode can contain two radial branches whose amplitudes interfere; Eq.~\eqref{eq:physical-local-factorization} is not a general factorization for such a mode. We assume $\chi_s(0)\neq0$ and $S_{\omega lm}(\theta_{\rm h})\neq0$. A node changes the leading endpoint power and requires a new expansion. In the numerical ratio tests we use a horizon-normalized outgoing solution; a common overall normalization cancels from $P_e/P_a$, but an omitted branch does not.
\end{revision}

\subsection{Generic off-shell Kerr geodesic}

For an atom of unit rest mass, define the specific energy $e$, axial angular momentum $\ell$, and Carter constant $\mathcal Q$.  The first-order equations retain the Kerr form with the replacement $\Delta_{\rm Kerr}\to\Delta(r)$~\cite{Toshmatov:2017zpr,Frolov:2017kze}:
\begin{align}\label{eq:geodesics}
 \Sigma\dot r&=-\sqrt{\mathcal R},
 &\mathcal R&=\mathcal P^2-\Delta\mathcal B,\\
 \Sigma\dot t&=-a(ae\sin^2\theta-\ell)
 +\frac{r^2+a^2}{\Delta}\mathcal P,\\
 \Sigma\dot\phi&=-\left(ae-\frac{\ell}{\sin^2\theta}\right)
 +\frac{a}{\Delta}\mathcal P.
\end{align}
The polar motion is
\begin{align}\label{eq:polar-geodesic}
 \Sigma\dot\theta&=\pm\sqrt{\Theta_\theta},\\
 \Theta_\theta&=\mathcal Q-\cos^2\theta
 \left[a^2(1-e^2)+\frac{\ell^2}{\sin^2\theta}\right].
\end{align}
where an overdot is $\dd/\dd\tau$ and
\begin{equation}\label{eq:PB}
 \mathcal P=e(r^2+a^2)-a\ell,
 \qquad
 \mathcal B=r^2+(\ell-ae)^2+\mathcal Q.
\end{equation}
A future-directed horizon crossing requires
\begin{equation}\label{eq:etilde}
 \widetilde e=e-\OmH\ell>0.
\end{equation}
If the trajectory crosses the horizon at polar angle $\theta_{\rm h}$, its near-horizon form is
\begin{align}\label{eq:geodesic-expansions}
 \tau&=\rev{-kx-\tfrac12 k_1x^2+\order(x^3)},\\
 t&=-\frac{1}{2\kap}\ln\left(\frac{x}{x_0}\right)-Cx+\order(x^2),\\
 \widetilde\phi&=\alpha_\phi x+\order(x^2).
\end{align}
The leading coefficient is
\begin{equation}\label{eq:kcoeff}
 k=\frac{\Sigma_{\rm h}}{(\rh^2+a^2)\widetilde e},
 \qquad
 \Sigma_{\rm h}=\rh^2+a^2\cos^2\theta_{\rm h},
\end{equation}
while $C$ and $\alpha_\phi$ depend on $\Delta_2$, the constants of motion, and $\theta_{\rm h}$.  Their exact generic-$\Delta$ expressions are given in Appendix~\ref{app:geodesics}.  In particular, $C$ contains a term $2\rh/\Delta_1$ generated by
\begin{equation}
 \left.\frac{\dd}{\dd r}(r^2+a^2)\right|_{\rh}=2\rh.
\end{equation}
The logarithmic divergences in $t$ and $\phi$ cancel in $\widetilde\phi$.

\subsection{Phase-improved endpoint integral and large-gap detailed balance}\label{subsec:endpoint}

The creation part of the field operator contributes the complex-conjugate outgoing mode.  Combining that mode with Eqs.~\eqref{eq:geodesic-expansions} gives the local endpoint amplitude
\begin{equation}\label{eq:Ie-endpoint}
 \mathscr I_e^{\rm loc}\simeq
 k\int_0^{x_f}\dd x\,x^{-\ii a_\epsh}
 \e^{-\ii s_e x},
 \qquad
 a_\epsh=\frac{\epsh}{\kap},
\end{equation}
with
\begin{equation}\label{eq:se-sa}
 s_e=k\omega_0+\delta,
 \qquad
 s_a=k\omega_0-\delta,
 \qquad
 \delta=C\epsh+\alpha_\phi m.
\end{equation}
We assume $s_e>0$ and $s_a>0$; if either scale crosses zero, the endpoint contour and the physical switching protocol must be treated separately.  Holding all smooth prefactors at their horizon values while retaining the complete phase through linear order in $x$ defines the phase-improved endpoint truncation.  With a regulator $\e^{-\epsilon_{\rm reg}x}$ and $\epsilon_{\rm reg}\to0^+$,
\begin{equation}\label{eq:gamma-integral}
 \mathscr I_e^{\rm loc}
 =k\,\Gamma(1-\ii a_\epsh)(\ii s_e)^{-1+\ii a_\epsh}
\end{equation}
\rev{for the Abel-regularized infinite endpoint model. Replacing a finite physical gate by this regulator requires separate control of its remote contribution. The resulting phase-only kernels are}
\begin{align}\label{eq:probabilities-exact}
 \mathscr R_{e,s}^{\rm loc}
 &=g^2k^2\frac{2\pi a_\epsh}
 {s_e^2\left(\e^{2\pi a_\epsh}-1\right)},\\
 \mathscr R_{a,s}^{\rm loc}
 &=g^2k^2\frac{2\pi a_\epsh\e^{2\pi a_\epsh}}
 {s_a^2\left(\e^{2\pi a_\epsh}-1\right)}.
\end{align}
Within this truncation,
\begin{equation}\label{eq:ratio-finite-gap}
 \frac{\mathscr R_{e,s}^{\rm loc}}{\mathscr R_{a,s}^{\rm loc}}
 =\e^{-\bH\epsh}
 \left(\frac{s_a}{s_e}\right)^2,
 \qquad
 \bH=\frac{2\pi}{\kap}.
\end{equation}
\rev{The first factor is the leading geometric Hawking ratio for this outgoing branch. The second isolates the Boyer--Lindquist trajectory phase. It is not the complete finite-gap correction, because the quadratic term in $\tau(x)$ and the smooth mode factors contribute at the same inverse-gap order.}

\rev{At fixed $b_s=\epsh_s/\kap$ with a controlled remote switching remainder, the large-gap outgoing-channel kernels are}
\begin{align}\label{eq:Pe-large-gap}
 \mathscr R_{e,s}^{\rm loc}
 &=\frac{2\pi g^2\epsh}{\kap\omega_0^2}
 \frac{1}{\e^{2\pi\epsh/\kap}-1},\\
 \mathscr R_{a,s}^{\rm loc}
 &=\frac{2\pi g^2\epsh}{\kap\omega_0^2}
 \frac{1}{1-\e^{-2\pi\epsh/\kap}}.
\end{align}
Their ratio is the Hawking detailed-balance factor.  The zero-charge limit gives the Kerr kernel, the nonrotating limit gives the static finite-curvature kernel, and taking both limits gives Schwarzschild.  The notebook checks all three reductions.

\subsection{Coordinate robustness and endpoint scales}\label{subsec:coordinate-robustness}

To determine which finite-gap terms are physical, retain a general smooth endpoint amplitude:
\begin{align}\label{eq:general-endpoint}
 I_e&=\int_0^{x_f}\dd x\,
 F_e(x)x^{-\ii a_\epsh}\e^{-\ii s_ex},\\
 F_e(x)&=F_{e0}(1+f_{e1}x+\cdots).
\end{align}
Consider a regular nonlinear reparametrization
\begin{equation}
 x=Ay+By^2+\order(y^3),
 \qquad A>0.
\end{equation}
The transformed integrand is
\begin{widetext}
\begin{align}\label{eq:coordinate-transform-integrand}
 \dd x\,F_e(x)x^{-\ii a_\epsh}\e^{-\ii s_ex}
 ={}&F_{e0}A^{1-\ii a_\epsh}\dd y\,
 y^{-\ii a_\epsh}\e^{-\ii As_ey}\nonumber\\
 &\times\left\{1+
 \left[Af_{e1}+\frac{(2-\ii a_\epsh)B}{A}\right]y
 -\ii s_eBy^2+\cdots\right\}.
\end{align}
\end{widetext}
\begin{revision}
At fixed $b_s=a_\epsh$, endpoint power counting gives $y\sim s_e^{-1}$. The amplitude derivative, Jacobian, logarithmic phase and $s_eBy^2$ term contribute at the same relative inverse-gap order. A nonlinear radial relabeling therefore changes the isolated factor $(s_a/s_e)^2$ while leaving the complete detector integral invariant. For a fixed, isolated outgoing channel and a sufficiently suppressed remote remainder, the asymptotic statement is
\begin{equation}\label{eq:invariant-detailed-balance}
 -\ln\frac{P_{e,s}^{\rm out}}{P_{a,s}^{\rm out}}
 =\bH\epsh_s+\order(\omega_0^{-1}).
\end{equation}
The coefficient implicit in the remainder depends on the geometry, the mode and the protocol. This is not a uniform claim as $b_s\to\infty$.

For example, the first endpoint moment obeys
\begin{equation}\label{eq:endpoint-moment}
 \frac{\int_0^\infty x^{1-\ii b_s}\e^{-\ii sx}\dd x}
 {\int_0^\infty x^{-\ii b_s}\e^{-\ii sx}\dd x}
 =\frac{1-\ii b_s}{\ii s},
\end{equation}
with the same regulator in numerator and denominator. A smooth derivative $f_1$ is therefore enhanced by $\sqrt{1+b_s^2}/s$, not merely by $1/s$. The characteristic radial scale and the quadratic horizon scale are
\begin{equation}\label{eq:large-gap}
 x_{\rm char}\sim\frac{\max(1,b_s)}{k\omega_0},
 \qquad L_\Delta=\frac{2\Delta_1}{|\Delta_2|},
\end{equation}
when $\Delta_2\neq0$. Local endpoint control requires
\begin{equation}\label{eq:uniform-cutoff}
 x_{\rm char}\ll x_f,\qquad x_{\rm char}\ll L_\Delta,
\end{equation}
and corresponding conditions for higher derivatives and all smooth factors. These are scale checks, not a sufficient uniform error bound. If $\Delta_2=0$, the next nonzero term in $\Delta$ defines the relevant radial scale.

A hard outer cutoff adds a boundary term of order $x_f^{-\ii b_s}\e^{-\ii sx_f}/s$, the same inverse-gap order as the horizon term. It cannot be discarded solely because $sx_f\gg1$. A physical switching remainder must be bounded relative to the emission amplitude, which is exponentially small at large $b_s$. At fixed $\epsh>0$, the formal nonextremal kernel tends to zero as $\kap\to0^+$, but a fixed atom and gate need not follow that limit. The proper-time formulation below states what is calculated without asserting an unproved uniform hierarchy.
\end{revision}

\begin{revision}
\section{Operational finite-gap response}
\label{sec:operational}

\subsection{A proper-time coefficient for the outgoing channel}
\label{subsec:proper-time-finite-gap}

Let $z=-\tau>0$ be the proper-time distance to crossing and set $b_s=\epsh_s/\kap>0$. For a specified outgoing solution with a nonzero leading amplitude, write
\begin{equation}\label{eq:proper-time-mode}
 \chi_s(-z)\Phi_s[x(-z)]
 =\mathcal B_s(z/z_0)^{\ii b_s}
 \left[1+d_sz+\order(z^2)\right].
\end{equation}
The constant reference $z_0$ affects only an overall phase. The coefficient $d_s$ depends on the worldline, mode and gate, but not on a relabeling of the radial coordinate with these physical choices held fixed. We assume a real, differentiable exterior switching function and exclude angular or switching nodes at crossing.

The radial contribution follows from Eq.~\eqref{eq:radial-exact}. Define
\begin{align}
 K_0&=(\rh^2+a^2)\epsh_s, &K_1&=2\rh\omega,\nonumber\\
 p_s&=\frac{\ii K_0}{\Delta_1}=\frac{\ii b_s}{2}, &
 W_0&=\mu_\Phi^2\rh^2+a^2\omega^2-2am\omega+\Lambda_{\omega lm}.
 \label{eq:Frobenius-data}
\end{align}
For $R_s=x^{p_s}(1+c_{R,s}x+\order(x^2))$, the first Frobenius coefficient is
\begin{align}\label{eq:radial-first-Frobenius}
 c_{R,s}=-\frac{1}{1+2p_s}\bigg[
 &\frac{\Delta_2p_s}{2\Delta_1}
 +\frac{2K_0K_1}{\Delta_1^2}\nonumber\\
 &-\frac{K_0^2\Delta_2}{\Delta_1^3}
 -\frac{W_0}{\Delta_1}\bigg].
\end{align}
Using $z=kx+k_1x^2/2+\order(x^3)$ and $\theta=\theta_{\rm h}+\theta_1x+\order(x^2)$ gives
\begin{align}\label{eq:proper-time-d-coefficient}
 d_s={}&\frac{c_{R,s}+\ii\delta_s
       +\theta_1(\partial_\theta\ln S_{\omega lm})_{\rm h}}{k}
 \nonumber\\
 &-\frac{\ii b_sk_1}{2k^2}
 -\frac{\chi_s'(0)}{\chi_s(0)},
 \qquad \delta_s=C\epsh_s+m\alpha_\phi.
\end{align}
Here the prime on $\chi_s$ is a proper-time derivative; $k_1$ and $\theta_1$ are given in Appendix~\ref{app:geodesics}. Equation~\eqref{eq:proper-time-d-coefficient} keeps the radial, angular, trajectory and switching contributions at the same order.

At fixed $b_s$, the regulated endpoint integrals yield
\begin{align}\label{eq:proper-time-probabilities}
 P_{e,s}^{\rm out}&=P_{e,s}^{(0)}
 \left[1-\frac{2E_s}{\omega_0}+\order(\omega_0^{-2})\right],\\
 P_{a,s}^{\rm out}&=P_{a,s}^{(0)}
 \left[1+\frac{2E_s}{\omega_0}+\order(\omega_0^{-2})\right],\nonumber\\
 E_s&=b_s\operatorname{Re}d_s+\operatorname{Im}d_s.\nonumber
\end{align}
The leading probabilities contain the same channel weight and obey $P_{e,s}^{(0)}/P_{a,s}^{(0)}=\e^{-2\pi b_s}$. Thus the outgoing-channel detailed-balance affinity is
\begin{equation}\label{eq:proper-time-affinity}
 \mathcal A_s^{\rm out}
 =2\pi b_s+\frac{4E_s}{\omega_0}+\order(\omega_0^{-2}).
\end{equation}
The expansion presupposes a remote switching remainder small relative to both endpoint amplitudes. Its error constants depend on $b_s$ and the protocol. Equations~\eqref{eq:proper-time-probabilities} and \eqref{eq:proper-time-affinity} are therefore fixed-$b_s$ asymptotic statements, not a theorem uniform in the extremal limit.

There is a useful cancellation for real $\omega$, real $\Lambda_{\omega lm}$, and a real background. Equation~\eqref{eq:radial-first-Frobenius} implies
\begin{equation}\label{eq:radial-affinity-cancellation}
 b_s\operatorname{Re}c_{R,s}+\operatorname{Im}c_{R,s}
 =-\frac{b_s\Delta_2}{4\Delta_1}.
\end{equation}
The explicit real $W_0$ term cancels from this combination. In an equatorial trajectory, the direct angular-variation term also vanishes. This does not remove mass, angular normalization or propagation effects from the full channel; it isolates the dependence of the first local affinity coefficient. As a limiting check, radial Schwarzschild infall with $e=1$, $\ell=\mathcal Q=0$ has $k=1$, $k_1=1/4$, $C=3/2$, $b_s=4\omega$, and
\begin{equation}\label{eq:Schwarzschild-cancellation}
 E_s^{\rm geom}=0
 \quad\text{when}\quad \chi_s'(0)=0.
\end{equation}
An isolated Boyer--Lindquist phase would instead give a nonzero inverse-gap term. This cancellation provides a direct check on the necessity of the complete proper-time coefficient.

\subsection{Finite gate and direct outgoing-branch integration}
\label{subsec:finite-gate}

To specify an operational test, choose the one-sided exterior gate
\begin{equation}\label{eq:exponential-gate}
 \chi_s(-z)=\e^{-z/T},\qquad z\geq0,\quad T>0.
\end{equation}
It is smooth on the exterior half-line and has nonzero value at crossing. It is not a smooth switch-off across the horizon; no interior interaction is included. For the pure logarithmic channel $\Phi_s=\mathcal B_s(z/z_0)^{\ii b_s}$, the integrals are elementary. With $Z=T^{-1}+\ii\omega_0$, and suppressing common constants and reference phases,
\begin{align}
 I_{e,s}&=\Gamma(1-\ii b_s)Z^{-1+\ii b_s},\nonumber\\
 I_{a,s}&=\Gamma(1+\ii b_s)Z^{-1-\ii b_s}.
 \label{eq:finite-gate-integrals}
\end{align}
Consequently,
\begin{align}\label{eq:finite-switch-affinity}
 \frac{P_{e,s}}{P_{a,s}}
 &=\exp[-4b_s\arctan(\omega_0T)],\\
 \mathcal A_s^{\rm gate}&=4b_s\arctan(\omega_0T).\nonumber
\end{align}
This exact local-model result gives
\begin{equation}\label{eq:gate-large-gap}
 \mathcal A_s^{\rm gate}-2\pi b_s
 =-\frac{4b_s}{\omega_0T}
 +\order\!\left(\frac{b_s}{(\omega_0T)^3}\right).
\end{equation}
An absolute affinity error much smaller than unity requires $\omega_0T\gg b_s$, not just $\omega_0T\gg1$. The pure-power channel extended over the gate is a check of switching effects, not a global spacetime solution. Finite-duration detector effects and finite-gate branch interference have prior treatments~\cite{Fewster2016Waiting,Sen:2023zfq,LobosRodulfo2026TwoBranch}.

We go beyond this pure-power test by numerically integrating the exact separated radial equation on the prescribed geometry together with the equatorial geodesic. We take
\begin{equation}\label{eq:finite-gate-numerical-parameters}
 \begin{gathered}
 a=0.6,\qquad q=q_{\rm ext}(a)/2,\\
 \omega=0.55,\qquad l=m=1,\qquad\mu_\Phi=0,
 \end{gathered}
\end{equation}
with $e=1$, $\ell=\mathcal Q=0$ and $T=0.03$. An outgoing Frobenius series initializes $R=x^{\ii b_s/2}V(x)$, with $V(0)=1$, and the exact radial and Carter equations are evolved in $\ln x$. The regular proper-time factor
\begin{equation}\label{eq:full-branch-factor}
 \Phi_s[x(-z)]=\mathcal B_s(z/z_0)^{\ii b_s}F_s(z),
 \qquad F_s(0)=1,
\end{equation}
is then inserted into Eqs.~\eqref{eq:Pe-def} and \eqref{eq:Pa-def}. Unlike the pure logarithmic test, $F_s(z)$ includes the radial amplitude and complete exterior trajectory across the gate. Its derivative at crossing agrees with $d_s^{\rm geom}$, obtained from Eq.~\eqref{eq:proper-time-d-coefficient} without the switching term.

The numerical integrals subtract the analytically integrable constant and linear terms before oscillatory quadrature. We vary the differential-equation tolerance, Frobenius order, interpolation resolution and gate-tail truncation; the settings and errors are recorded in the notebook. This is a full finite-gate test of the selected outgoing branch in the prescribed exterior metric. It is not a calculation of the Klein--Gordon normalization, the second branch of a specified scattering state, or propagation to an asymptotic detector. Section~\ref{subsec:numerical-response} reports its agreement with Eq.~\eqref{eq:proper-time-affinity}.

\subsection{What radial branch selection must control}
\label{subsec:branch-selection}

For a mode containing outgoing and ingoing radial pieces, write the two transition amplitudes as
\begin{equation}
 I_e=I_e^{\rm out}(1+\zeta_e),\qquad
 I_a=I_a^{\rm out}(1+\zeta_a).
\end{equation}
The complex ratios $\zeta_e$ and $\zeta_a$ include the branch weights and the same physical gate; they need not be equal. The exact change in affinity is
\begin{equation}\label{eq:branch-affinity}
 \mathcal A_s-\mathcal A_s^{\rm out}
 =-\ln\frac{|1+\zeta_e|^2}{|1+\zeta_a|^2}.
\end{equation}
If $|\zeta_e|,|\zeta_a|\leq\zeta<1$, then
\begin{equation}\label{eq:branch-bound}
 |\mathcal A_s-\mathcal A_s^{\rm out}|
 \leq2\ln\frac{1+\zeta}{1-\zeta}.
\end{equation}
This supplies a quantitative branch-isolation criterion. At large $b_s$, the suppressed outgoing emission amplitude makes $\zeta_e$ particularly sensitive to even a small unselected component. Frequency selection alone does not impose this bound. Likewise, a small absolute quadrature error is insufficient unless it is small relative to $I_e^{\rm out}$.

A physical atomic implementation must also retain weak coupling, negligible recoil and point-detector resolution. In particular, increasing the proper gap cannot violate $\hbar\omega_0\ll m_Ac^2$ for a fixed rest mass $m_A$. The numerical two-level detector treats its internal gap as a prescribed parameter; it does not establish an astrophysical implementation or an unbounded-gap limit for a fixed atom.
\end{revision}

\section{Circular orbital response of a two-level atom}\label{sec6}

The freely falling calculation probes the logarithmic field phase along a horizon-crossing geodesic. We now consider a different motion in the same geometry. We place the two-level atom on a circular equatorial orbit at fixed radius and ask how rotation, local orbital speed, and proper acceleration modify the selected-mode response. We distinguish circular geodesics from externally maintained circular trajectories. The distinction is essential because a rapidly moving geodesic atom still has vanishing four-acceleration, whereas an atom held on a prescribed nongeodesic orbit can experience an arbitrarily large proper acceleration as its trajectory approaches a local null direction.

\subsection{Stationary equatorial circular worldlines}

We set
\begin{equation}
 r=r_0,
 \qquad
 \theta=\frac{\pi}{2},
 \qquad
 \phi=\Omega t,
\label{eq:orbital-worldline}
\end{equation}
where \(r_0>\rh\) and \(\Omega\) is constant along one orbit. On the equatorial plane, the metric components following from Eq.~\eqref{eq:metric} are
\begin{align}
 g_{tt}
 &=
 -1+\frac{2\rho(r)}{r},
 \nonumber\\
 g_{t\phi}
 &=
 -\frac{2a\rho(r)}{r},
 \nonumber\\
 g_{\phi\phi}
 &=
 \frac{\mathscr G(r)}{r^2},
\label{eq:orbital-metric-components}
\end{align}
where we define
\begin{equation}
 \mathscr G(r)
 =
 (r^2+a^2)^2-a^2\Delta(r).
\label{eq:G-orbital}
\end{equation}
The four-velocity takes the form (\cite{UktamjonUktamov:2025aqz,UktamjonUktamov:2025cso,UktamjonUktamov:2025qts,Uktamov:2025bth})
\begin{equation}
 u^\mu
 =
 u^t
 \left(
 1,0,0,\Omega
 \right).
\label{eq:orbital-four-velocity}
\end{equation}
Normalization gives
\begin{equation}
 u^t
 =
 \frac{1}{\mathcal N_\Omega},
\qquad
 \mathcal N_\Omega^2
 =
 -
 \left(
 g_{tt}
 +
 2\Omega g_{t\phi}
 +
 \Omega^2g_{\phi\phi}
 \right).
\label{eq:orbital-normalization}
\end{equation}
A timelike circular trajectory requires
\begin{equation}
 \mathcal N_\Omega^2>0.
\label{eq:orbital-timelike}
\end{equation}

We introduce the angular velocity of the locally nonrotating frame
\begin{equation}
 \omega_{\rm Z}
 =
 -\frac{g_{t\phi}}{g_{\phi\phi}}
 =
 \frac{2a\rho(r)r}{\mathscr G(r)}
\label{eq:zamo-angular-velocity}
\end{equation}
and the corresponding lapse
\begin{equation}
 N_{\rm Z}^2
 =
 \frac{
 g_{t\phi}^2-g_{tt}g_{\phi\phi}
 }{
 g_{\phi\phi}
 }
 =
 \frac{r^2\Delta}{\mathscr G(r)}.
\label{eq:zamo-lapse}
\end{equation}
Equation~\eqref{eq:orbital-normalization} then becomes
\begin{equation}
 \mathcal N_\Omega^2
 =
 N_{\rm Z}^2
 -
 g_{\phi\phi}
 \left(
 \Omega-\omega_{\rm Z}
 \right)^2.
\label{eq:orbital-normalization-zamo}
\end{equation}
The physical azimuthal speed measured by the locally nonrotating observer is
\begin{equation}
 v
 =
 \frac{\sqrt{g_{\phi\phi}}}{N_{\rm Z}}
 \left(
 \Omega-\omega_{\rm Z}
 \right),
\label{eq:local-orbital-speed}
\end{equation}
so that
\begin{equation}
 \gamma_{\rm L}
 =
 \frac{1}{\sqrt{1-v^2}},
 \qquad
 \mathcal N_\Omega
 =
 \frac{N_{\rm Z}}{\gamma_{\rm L}},
 \qquad
 u^t
 =
 \frac{\gamma_{\rm L}}{N_{\rm Z}}.
\label{eq:orbital-lorentz-factor}
\end{equation}

The limiting circular null directions follow from
\(\mathcal N_\Omega^2=0\). We obtain
\begin{align}
 \Omega_\pm^{\rm null}
 &=
 \omega_{\rm Z}
 \pm
 \frac{N_{\rm Z}}{\sqrt{g_{\phi\phi}}}
 \nonumber\\
 &=
 \frac{
 2a\rho(r)r
 \pm
 r^2\sqrt{\Delta}
 }{
 \mathscr G(r)
 }.
\label{eq:null-angular-velocities}
\end{align}
Every stationary timelike circular trajectory satisfies
\begin{equation}
 \Omega_-^{\rm null}
 <
 \Omega
 <
 \Omega_+^{\rm null}.
\label{eq:timelike-angular-window}
\end{equation}

We define local co-rotation and counter-rotation through the sign of \(v\), rather than through the sign of \(\Omega\). We write
\begin{equation}
 v_\varsigma
 =
 \varsigma v,
 \qquad
 0<v<1,
 \qquad
 \varsigma=\pm1,
\label{eq:orbital-branch-sign}
\end{equation}
where \(\varsigma=+1\) denotes local co-rotation and \(\varsigma=-1\) denotes local counter-rotation. Their angular velocities are
\begin{equation}
 \Omega_\varsigma
 =
 \omega_{\rm Z}
 +
 \varsigma
 \frac{N_{\rm Z}}{\sqrt{g_{\phi\phi}}}
 v.
\label{eq:orbital-branches}
\end{equation}
This definition remains meaningful inside the ergoregion, where frame dragging can make both coordinate angular velocities have the same sign.

\subsection{Circular geodesics and accelerated circular motion}

We first identify the geodesic subset of Eq.~\eqref{eq:orbital-worldline}. For a stationary circular trajectory generated by
\[
 K_\Omega
 =
 \partial_t+\Omega\partial_\phi,
\]
we have
\[
 u^\mu
 =
 \frac{K_\Omega^\mu}{\mathcal N_\Omega}.
\]
The four-acceleration satisfies
\begin{equation}
 \mathfrak a_\mu
 \equiv
 u^\nu\nabla_\nu u_\mu
 =
 \partial_\mu\ln\mathcal N_\Omega.
\label{eq:orbital-four-acceleration}
\end{equation}
Reflection symmetry removes the polar component on the equatorial plane. A circular geodesic therefore obeys
\begin{equation}
 \left.
 \partial_r\mathcal N_\Omega
 \right|_{r_0}
 =
 0,
\label{eq:circular-geodesic-condition}
\end{equation}
with the derivative evaluated at fixed \(\Omega\).

Solving Eq.~\eqref{eq:circular-geodesic-condition} gives the two equatorial circular geodesic angular velocities
\begin{equation}
 \Omega_{\rm geo}^{\pm}
 =
 \frac{
 -\partial_r g_{t\phi}
 \pm
 \sqrt{
 (\partial_r g_{t\phi})^2
 -
 (\partial_r g_{tt})
 (\partial_r g_{\phi\phi})
 }
 }{
 \partial_r g_{\phi\phi}
 }.
\label{eq:geo-omega-general}
\end{equation}
For the radial mass profile used here, we can write this result in the compact form
\begin{equation}
 \Omega_{\rm geo}^{\pm}
 =
 \frac{
 \pm\Upsilon(r)
 }{
 1\pm a\Upsilon(r)
 },
\qquad
 \Upsilon(r)
 =
 \sqrt{
 \frac{
 \rho(r)-r\rho'(r)
 }{
 r^3
 }
 }.
\label{eq:geo-omega-profile}
\end{equation}
We require the quantity inside the square root to be positive and the resulting worldline to remain timelike. Orbital stability would impose an additional radial condition, which we do not require for the finite-duration response considered here.

The Kerr limit \(\rho(r)=M\) gives
\begin{equation}
 \Omega_{\rm geo}^{\pm}
 =
 \frac{
 \pm\sqrt{M}
 }{
 r^{3/2}
 \pm
 a\sqrt{M}
 },
\label{eq:kerr-circular-omega}
\end{equation}
as expected. For either circular geodesic,
\begin{equation}
 \mathfrak a^\mu=0.
\label{eq:geodesic-zero-acceleration}
\end{equation}
A large orbital speed therefore does not by itself imply a large invariant acceleration.

We obtain the accelerated orbital case by choosing an angular velocity different from
\(\Omega_{\rm geo}^{\pm}\). The invariant acceleration magnitude is
\begin{equation}
 \mathfrak a_\Omega
 =
 \sqrt{
 \mathfrak a_\mu\mathfrak a^\mu
 }
 =
 \left.
 \frac{\sqrt{\Delta}}{r}
 \left|
 \partial_r\ln\mathcal N_\Omega
 \right|
 \right|_{r=r_0},
\label{eq:orbital-acceleration}
\end{equation}
where the derivative is again evaluated at fixed \(\Omega\). Equation~\eqref{eq:orbital-acceleration} gives the external acceleration required to maintain the prescribed circular trajectory.

The ultra-relativistic regime corresponds to
\begin{equation}
 \gamma_{\rm L}\gg1,
\qquad
 |v|\rightarrow1.
\label{eq:ultrarelativistic-orbit}
\end{equation}
The two branches then approach the circular null directions,
\begin{equation}
 \Omega_\varsigma
 =
 \Omega_\varsigma^{\rm null}
 -
 \varsigma
 \frac{N_{\rm Z}}
 {2\sqrt{g_{\phi\phi}}}
 \gamma_{\rm L}^{-2}
 +
 \order(\gamma_{\rm L}^{-4}).
\label{eq:null-approach}
\end{equation}

We can also examine the simple-horizon limit. Let
\begin{equation}
 x_{\rm orb}
 =
 r_0-\rh,
 \qquad
 0<x_{\rm orb}\ll\rh.
\label{eq:orbital-horizon-distance}
\end{equation}
Using
\(
\Delta=\Delta_1x_{\rm orb}+\order(x_{\rm orb}^2)
\),
we obtain
\begin{align}
 N_{\rm Z}
 &=
 \frac{
 \rh\sqrt{\Delta_1}
 }{
 \rh^2+a^2
 }
 \sqrt{x_{\rm orb}}
 +
 \order(x_{\rm orb}^{3/2}),
 \nonumber\\
 \omega_{\rm Z}
 &=
 \OmH
 +
 \order(x_{\rm orb}),
 \nonumber\\
 \sqrt{g_{\phi\phi}}
 &=
 \frac{
 \rh^2+a^2
 }{
 \rh
 }
 +
 \order(x_{\rm orb}).
\label{eq:orbital-horizon-expansions}
\end{align}
Hence
\begin{equation}
 \Omega_\varsigma
 =
 \OmH
 +
 \varsigma
 \frac{
 \rh^2\sqrt{\Delta_1}
 }{
 (\rh^2+a^2)^2
 }
 v\sqrt{x_{\rm orb}}
 +
 \order(x_{\rm orb}).
\label{eq:orbital-omega-near-horizon}
\end{equation}
Both local directions therefore approach the horizon angular velocity.

For fixed finite \(\gamma_{\rm L}\) in the simple-horizon limit, Eq.~\eqref{eq:orbital-acceleration} gives
\begin{equation}
 \mathfrak a_\varsigma
 =
 \frac{
 \gamma_{\rm L}^2\kap
 }{
 N_{\rm Z}
 }
 \left[
 1+\order(\sqrt{x_{\rm orb}})
 \right].
\label{eq:orbital-acceleration-near-horizon}
\end{equation}
Equivalently,
\begin{equation}
 \mathfrak a_\varsigma
 =
 \frac{
 \gamma_{\rm L}^2\sqrt{\Delta_1}
 }{
 2\rh\sqrt{x_{\rm orb}}
 }
 \left[
 1+\order(\sqrt{x_{\rm orb}})
 \right].
\label{eq:orbital-acceleration-divergence}
\end{equation}
We therefore obtain two independent enhancements of the proper acceleration. The horizon redshift produces the factor \(x_{\rm orb}^{-1/2}\), while ultra-relativistic motion produces the factor \(\gamma_{\rm L}^2\). The two orbital directions have the same leading acceleration magnitude, although they do not see the same field frequency.

\subsection{Selected-mode response along the orbit}

We now evaluate the same two-level interaction introduced in Eq.~\eqref{eq:interaction} along the circular worldline. For a selected stationary field channel with coordinate dependence
\[
 \Phi_s
 \sim
 R_{\omega lm}(r)
 S_{\omega lm}(\theta)
 \e^{\ii m\phi-\ii\omega t},
\]
the fixed orbital radius makes the radial and angular factors constant along the worldline. We therefore write
\begin{equation}
 \Phi_s[x(\tau)]
 =
 \Phi_{s0}
 \e^{-\ii\varpi_{s,\varsigma}\tau},
\label{eq:orbital-mode-proper-time}
\end{equation}
where \(\Phi_{s0}\) contains the selected-channel amplitude at \(r_0\). The proper-time frequency is
\begin{align}
 \varpi_{s,\varsigma}
 &=
 u^t_\varsigma
 \left(
 \omega-m\Omega_\varsigma
 \right)
 \nonumber\\
 &=
 \frac{\gamma_{\rm L}}{N_{\rm Z}}
 \left(
 \omega-m\Omega_\varsigma
 \right).
\label{eq:orbital-proper-frequency}
\end{align}
Using Eq.~\eqref{eq:orbital-branches}, we obtain
\begin{equation}
 \varpi_{s,\varsigma}
 =
 \frac{\gamma_{\rm L}}{N_{\rm Z}}
 \left(
 \omega-m\omega_{\rm Z}
 \right)
 -
 \varsigma
 \frac{
 \gamma_{\rm L}vm
 }{
 \sqrt{g_{\phi\phi}}
 }.
\label{eq:orbital-frequency-zamo}
\end{equation}
The first term contains the gravitational frequency shift, while the second separates local co-rotation from local counter-rotation.

Near the horizon, Eq.~\eqref{eq:orbital-horizon-expansions} gives
\begin{equation}
 \varpi_{s,\varsigma}
 =
 \frac{\gamma_{\rm L}}{N_{\rm Z}}
 \epsh_s
 -
 \varsigma
 \gamma_{\rm L}v
 \frac{
 m\rh
 }{
 \rh^2+a^2
 }
 +
 \order(
 \gamma_{\rm L}\sqrt{x_{\rm orb}}
 ),
\qquad
 \epsh_s=\omega-m\OmH.
\label{eq:orbital-frequency-near-horizon}
\end{equation}
For fixed positive \(\epsh_s\), the first term becomes large as \(r_0\) approaches the horizon. The second term remains sensitive to the direction of the atomic orbit.

Substitution of Eq.~\eqref{eq:orbital-mode-proper-time} into Eqs.~\eqref{eq:Pe-def} and \eqref{eq:Pa-def} gives
\begin{align}
 P_{e,s}^{\rm orb,\varsigma}
 &=
 g^2
 |\Phi_{s0}|^2
 \left|
 \int\dd\tau\,
 \chi_s(\tau)
 \e^{
 \ii(
 \omega_0+\varpi_{s,\varsigma}
 )\tau
 }
 \right|^2,
 \nonumber\\
 P_{a,s}^{\rm orb,\varsigma}
 &=
 g^2
 |\Phi_{s0}|^2
 \left|
 \int\dd\tau\,
 \chi_s(\tau)
 \e^{
 \ii(
 \omega_0-\varpi_{s,\varsigma}
 )\tau
 }
 \right|^2.
\label{eq:orbital-transition-probabilities}
\end{align}
It is useful to introduce the Fourier transform of the physical switching function
\begin{equation}
 \widehat\chi_s(\zeta)
 =
 \int\dd\tau\,
 \chi_s(\tau)
 \e^{\ii\zeta\tau}.
\label{eq:switch-fourier-orbit}
\end{equation}
We then obtain
\begin{align}
 P_{e,s}^{\rm orb,\varsigma}
 &=
 g^2
 |\Phi_{s0}|^2
 \left|
 \widehat\chi_s
 \left(
 \omega_0+\varpi_{s,\varsigma}
 \right)
 \right|^2,
 \nonumber\\
 P_{a,s}^{\rm orb,\varsigma}
 &=
 g^2
 |\Phi_{s0}|^2
 \left|
 \widehat\chi_s
 \left(
 \omega_0-\varpi_{s,\varsigma}
 \right)
 \right|^2.
\label{eq:orbital-probabilities-fourier}
\end{align}
Their ratio becomes
\begin{equation}
 \frac{
 P_{e,s}^{\rm orb,\varsigma}
 }{
 P_{a,s}^{\rm orb,\varsigma}
 }
 =
 \frac{
 \left|
 \widehat\chi_s
 (
 \omega_0+\varpi_{s,\varsigma}
 )
 \right|^2
 }{
 \left|
 \widehat\chi_s
 (
 \omega_0-\varpi_{s,\varsigma}
 )
 \right|^2
 }.
\label{eq:orbital-rate-ratio-general}
\end{equation}
Unlike the horizon-crossing result, Eq.~\eqref{eq:orbital-rate-ratio-general} does not reduce in general to
\(
\e^{-\bH\epsh_s}
\).
The orbital atom remains at one radius and samples a harmonic proper-time phase. It does not generate the logarithmic endpoint integral responsible for the freely falling Hawking ratio.

For direct comparison with the one-sided exponential gate used in Sec.~V, we choose
\begin{equation}
 \chi_s^{\rm orb}(\tau)
 =
 \e^{-\tau/T},
 \qquad
 \tau\geq0.
\label{eq:orbital-exponential-gate}
\end{equation}
Its Fourier factor satisfies
\begin{equation}
 \left|
 \widehat\chi_s^{\rm orb}(\zeta)
 \right|^2
 =
 \frac{1}{
 T^{-2}+\zeta^2
 }.
\label{eq:orbital-switch-transform}
\end{equation}
We then find
\begin{align}
 P_{e,s}^{\rm orb,\varsigma}
 &=
 \frac{
 g^2|\Phi_{s0}|^2
 }{
 T^{-2}
 +
 (
 \omega_0+\varpi_{s,\varsigma}
 )^2
 },
 \nonumber\\
 P_{a,s}^{\rm orb,\varsigma}
 &=
 \frac{
 g^2|\Phi_{s0}|^2
 }{
 T^{-2}
 +
 (
 \omega_0-\varpi_{s,\varsigma}
 )^2
 }.
\label{eq:orbital-exponential-probabilities}
\end{align}
The corresponding orbital affinity is
\begin{align}
 \mathcal A_{s,\varsigma}^{\rm orb}
 &\equiv
 -\ln
 \left(
 \frac{
 P_{e,s}^{\rm orb,\varsigma}
 }{
 P_{a,s}^{\rm orb,\varsigma}
 }
 \right)
 \nonumber\\
 &=
 \ln
 \left[
 \frac{
 T^{-2}
 +
 (
 \omega_0+\varpi_{s,\varsigma}
 )^2
 }{
 T^{-2}
 +
 (
 \omega_0-\varpi_{s,\varsigma}
 )^2
 }
 \right].
\label{eq:orbital-affinity}
\end{align}
Away from resonance, the long-gate limit gives
\begin{equation}
 \mathcal A_{s,\varsigma}^{\rm orb}
 =
 2
 \ln
 \left|
 \frac{
 \omega_0+\varpi_{s,\varsigma}
 }{
 \omega_0-\varpi_{s,\varsigma}
 }
 \right|
 +
 \order(T^{-2}).
\label{eq:orbital-affinity-long-gate}
\end{equation}
We therefore obtain a mode-dependent orbital affinity rather than a universal Hawking affinity.

For
\begin{equation}
 \varpi_{s,\varsigma}>0,
\label{eq:positive-orbital-frequency}
\end{equation}
the absorption channel becomes resonant near
\begin{equation}
 \varpi_{s,\varsigma}
 =
 \omega_0.
\label{eq:orbital-absorption-resonance}
\end{equation}
The excitation-emission channel instead becomes resonant near
\begin{equation}
 \varpi_{s,\varsigma}
 =
 -\omega_0.
\label{eq:orbital-emission-resonance}
\end{equation}
The latter condition requires a negative proper-time field frequency,
\begin{equation}
 \omega-m\Omega_\varsigma<0.
\label{eq:negative-orbital-frequency}
\end{equation}
At finite radius, this condition differs from the horizon superradiant condition
\(
\epsh_s<0
\).
Only as the allowed angular velocities collapse toward \(\OmH\) near the horizon do the two frequency criteria approach one another.

The circular calculation therefore separates proper acceleration from horizon-crossing thermality. A circular geodesic atom has
\(
\mathfrak a^\mu=0
\)
and responds through the stationary frequency
\(
\varpi_{s,\varsigma}
\).
A forced ultra-relativistic atom can have
\(
\mathfrak a_\varsigma\gg\kap
\),
yet its selected-mode response remains governed by Eqs.~\eqref{eq:orbital-transition-probabilities} and \eqref{eq:orbital-affinity}. The large proper acceleration does not by itself reproduce the freely falling HBAR factor. In the present rotating geometry, co-rotation and counter-rotation instead enter through the local Doppler term in Eq.~\eqref{eq:orbital-frequency-zamo}, while the horizon-crossing HBAR response remains tied to the logarithmic phase derived for the infalling atom.

\begin{figure*}[t]\revcolor\centering
 \panel{0.48}{\textbf{(a)}\\\includegraphics[width=\linewidth]{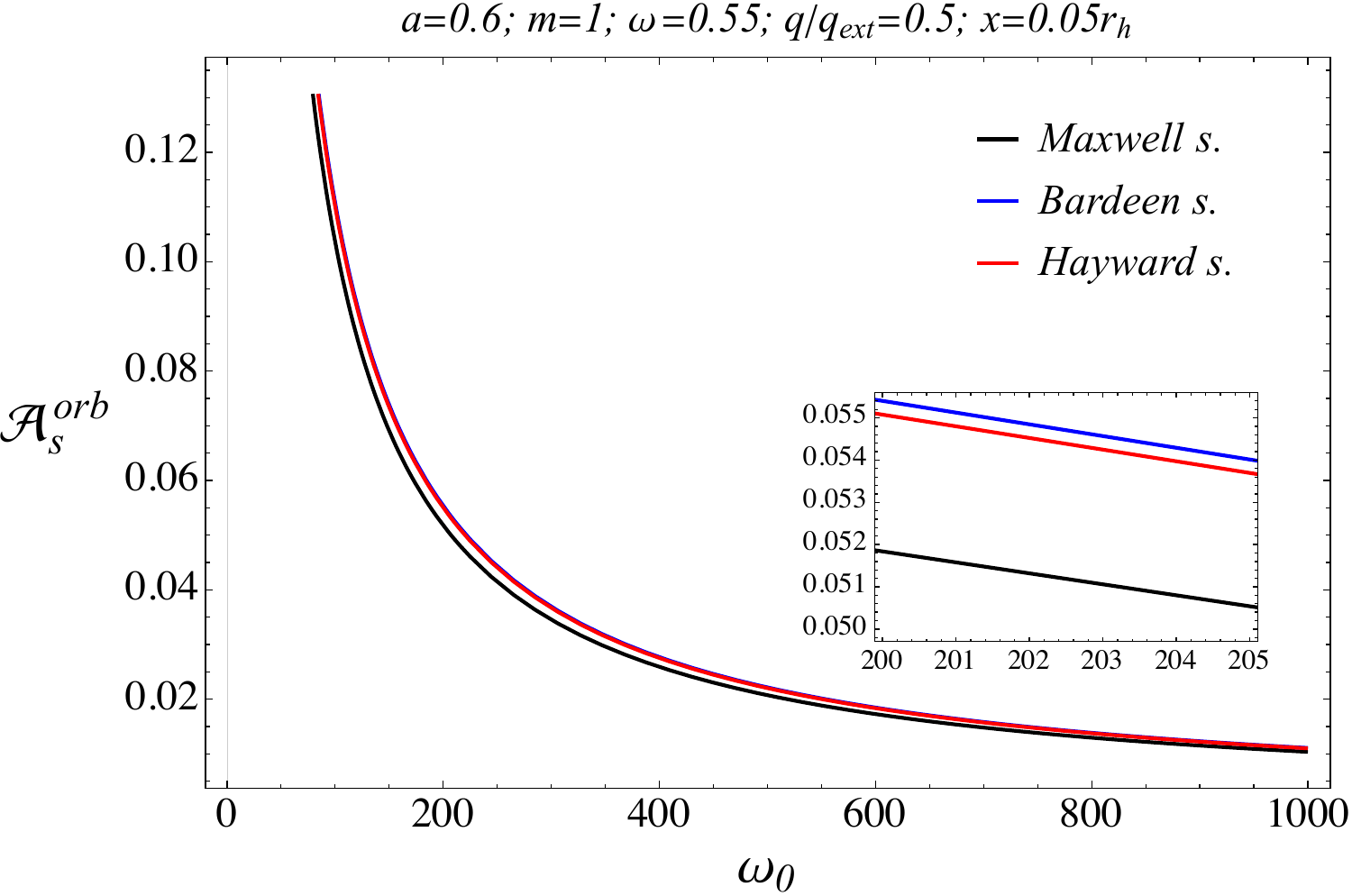}}\hfill
 \panel{0.48}{\textbf{(b)}\\\includegraphics[width=\linewidth]{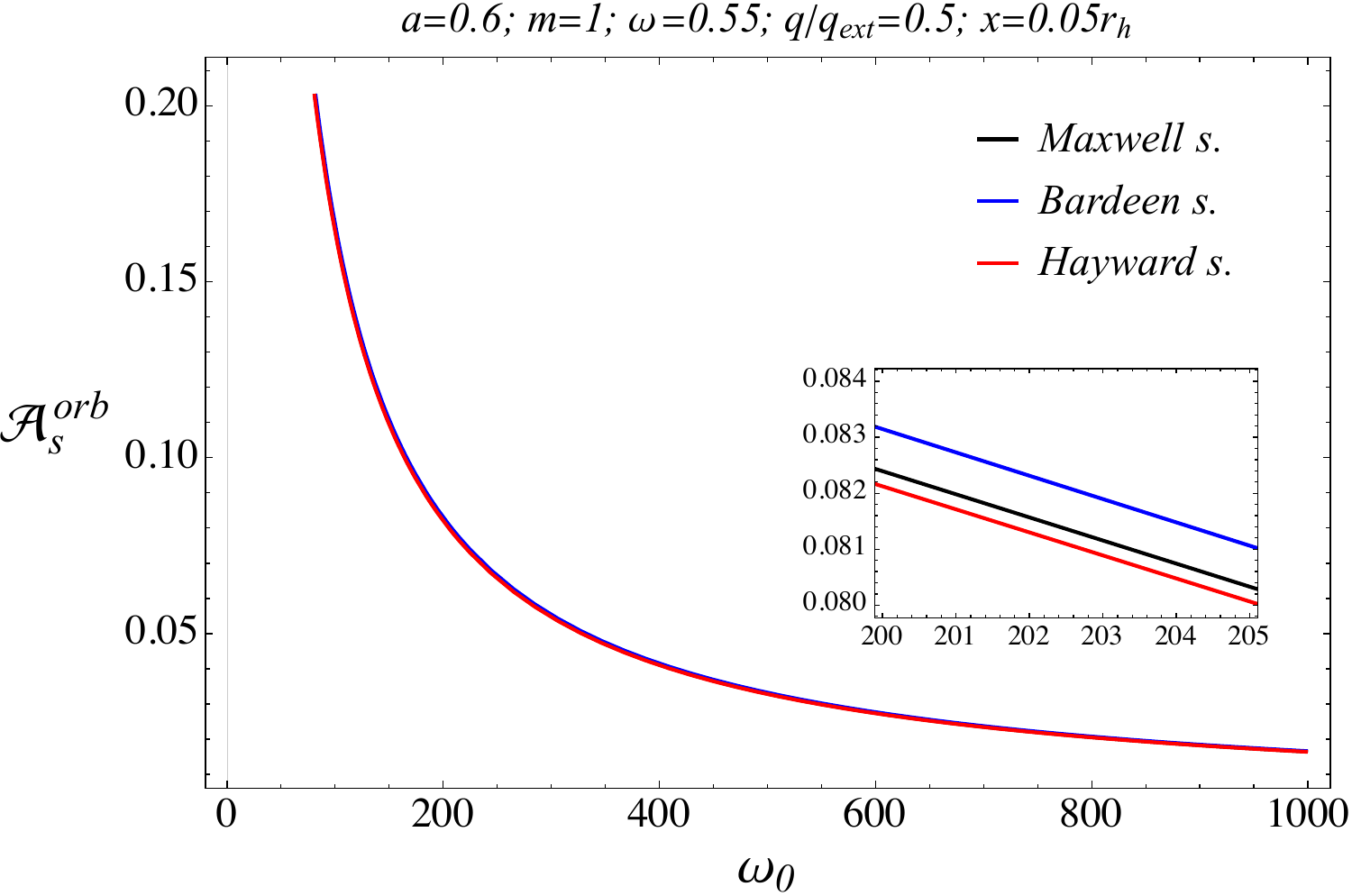}}
 \caption{Dependence of the orbital affinity $\mathcal{A}_s^{orb}$ on the $\omega_0$ for co-rotating (the left panel) and counter-rotating atoms (the right panel) for different solutions.}
 \label{fig:A}
\end{figure*}

Then we show how orbital affinity $\mathcal{A}_s^{orb}$ depends on $\omega_0$ for different solutions in Fig.(\ref{fig:A}).

\section{Atomic cloud, stationary state, and entropy balance}\label{sec:master}

\subsection{Single-mode Lindblad dynamics and detailed-balance affinity}

Consider independent atoms injected at Poisson rate $r_{\rm inj}$, with random injection times and sufficiently weak, nonoverlapping interactions.  \rev{For a selected nondegenerate bosonic mode with independently specified boundary conditions, define the complete channel rates}
\begin{equation}
 R_{e,s}=r_{\rm inj}P_{e,s},
 \qquad
 R_{a,s}=r_{\rm inj}P_{a,s}.
\end{equation}
Suppressing the mode label, the reduced field state obeys the phase-insensitive generator~\cite{Azizi:2021qcu,Breuer:2002}
\begin{equation}\label{eq:Lindblad}
 \dot\rho=R_a\Diss{a}\rho+R_e\Diss{a^\dagger}\rho,
\end{equation}
where
\begin{equation}
 \Diss{L}\rho=L\rho L^\dagger-\frac{1}{2}\{L^\dagger L,\rho\}.
\end{equation}
For number populations $p_n=\langle n|\rho|n\rangle$,
\begin{align}\label{eq:population-master}
 \dot p_n={}&-R_e\left[(n+1)p_n-np_{n-1}\right]\nonumber\\
 &-R_a\left[np_n-(n+1)p_{n+1}\right].
\end{align}
When
\begin{equation}\label{eq:qstable}
 q_s=\frac{R_{e,s}}{R_{a,s}}<1,
\end{equation}
the unique stationary distribution is
\begin{equation}\label{eq:stationary}
 p_{n,s}^{\rm ss}=(1-q_s)q_s^n,
 \qquad
 \bar n_s^{\rm ss}=\frac{q_s}{1-q_s}.
\end{equation}
The natural nonequilibrium control variable is the mode-resolved detailed-balance affinity
\begin{equation}\label{eq:affinity}
 \mathcal A_s\equiv-\ln q_s
 =-\ln\left(\frac{R_{e,s}}{R_{a,s}}\right).
\end{equation}
Within the Boyer--Lindquist phase-only truncation,
\begin{equation}\label{eq:affinity-phase}
 \mathcal A_s^{(\rm ph)}
 =\bH\epsh_s-2\ln\left(\frac{s_{a,s}}{s_{e,s}}\right).
\end{equation}
A single geometric distribution can always be assigned a formal mode temperature $\epsh_s/\mathcal A_s$, but a multimode field is thermal only if
\begin{equation}\label{eq:global-thermal-condition}
 \mathcal A_s=\beta\,\epsh_s
\end{equation}
with the same $\beta$ for every populated mode.  The finite-gap phase correction is generally mode dependent and therefore does not define a universal effective temperature.  \rev{When the actual selected-channel probabilities approach the outgoing Hawking ratio,}
\begin{equation}
 \mathcal A_s\longrightarrow\bH\epsh_s,
 \qquad
 \bar n_s^{\rm ss}\longrightarrow
 \frac{1}{\e^{\bH\epsh_s}-1}.
\end{equation}

The mean occupation evolves as
\begin{equation}\label{eq:nbar-evolution}
 \dot{\bar n}=R_e(\bar n+1)-R_a\bar n
 =-(R_a-R_e)(\bar n-\bar n_{\rm ss}).
\end{equation}
If all leading horizon weights in Eq.~\eqref{eq:physical-local-factorization} are absorbed into a common mode factor $\mathcal W_s$, then in the large-gap limit
\begin{equation}\label{eq:relax-rate}
 R_{a,s}-R_{e,s}
 =\mathcal W_s R_0\frac{\epsh_s}{\kap},
 \qquad
 R_0=\frac{2\pi r_{\rm inj}g^2}{\omega_0^2}.
\end{equation}

Random injection times secularize unequal coordinate frequencies but do not, by themselves, diagonalize a frequency-degenerate angular sector.  Equation~\eqref{eq:Lindblad} therefore assumes mode selection or additional angular averaging.  A fully degenerate multimode problem requires positive Kossakowski matrices and matrix-valued detailed balance; this issue is outside the present single-mode calculation.

\subsection{Entropy production}

Define the probability current between adjacent number states,
\begin{equation}\label{eq:Jn}
 J_n=(n+1)\left(R_ep_n-R_ap_{n+1}\right).
\end{equation}
Then $\dot{\bar n}=\sum_nJ_n$.  \rev{For a number-diagonal state, the population entropy equals the von Neumann entropy, $S_\Phi=-\sum_np_n\ln p_n$, and satisfies}
\begin{equation}\label{eq:entropy-identity}
 \dot S_\Phi=\sigma+\mathcal A_s\dot{\bar n},
\end{equation}
where
\begin{equation}\label{eq:sigma}
 \sigma=\sum_{n=0}^\infty J_n
 \ln\left(\frac{R_ep_n}{R_ap_{n+1}}\right)\geq0.
\end{equation}
\rev{The inequality follows term by term from $(x-y)\ln(x/y)\geq0$ and is the population form of the Spohn theorem~\cite{Spohn:1978}. Number-diagonal initial data remain diagonal under the phase-insensitive generator. For a state with coherences, the corresponding quantum-relative-entropy expression must be used rather than identifying the diagonal Shannon entropy with the full von Neumann entropy.}  Relative to the Hawking affinity, define
\begin{equation}\label{eq:delta-affinity}
 \Delta\mathcal A_s=\mathcal A_s-\bH\epsh_s.
\end{equation}
Equation~\eqref{eq:entropy-identity} becomes
\begin{equation}\label{eq:entropy-finite-gap}
 \dot S_\Phi=\bH\epsh_s\dot{\bar n}
 +\sigma+\Delta\mathcal A_s\dot{\bar n}.
\end{equation}
Within the phase-only truncation,
\begin{equation}
 \Delta\mathcal A_s^{(\rm ph)}
 =-2\ln\left(\frac{s_{a,s}}{s_{e,s}}\right).
\end{equation}
This term measures the mismatch between the actual stationary reference and the Hawking reference; it is not negative entropy production.  A complete next-to-leading detector model must compute $\mathcal A_s$ from the full physical rate ratio, including switching, normalization, and mode-amplitude derivatives.

\begin{revision}
\subsection{Mode leakage and conditional horizon bookkeeping}
\label{subsec:area-bookkeeping}

The energy transferred into a selected mode is not automatically the energy escaping to infinity. For a cavity coupled to a vacuum output channel at rate $\gamma_s$, the generator acquires $\gamma_s\Diss{a}$ and
\begin{align}\label{eq:lossy-mode}
 \dot{\bar n}_s&=R_{e,s}(\bar n_s+1)
 -(R_{a,s}+\gamma_s)\bar n_s,\\
 \dot N_s^{\rm out}&=\gamma_s\bar n_s.\nonumber
\end{align}
When $R_{a,s}+\gamma_s>R_{e,s}$, the stationary occupation is $R_{e,s}/(R_{a,s}+\gamma_s-R_{e,s})$. It can support a nonzero output even when $\dot{\bar n}_s=0$. The entropy identity above applies to the stated lossless atomic generator; with leakage, the additional reservoir contribution must be included. We therefore do not identify $\epsh_s\dot{\bar n}_s$ with an asymptotic outgoing flux without a boundary and energy-accounting prescription.

There is a separate geometric obstruction to assuming the Kerr first law. In a stationary Einstein--matter completion one would seek a relation of the schematic form
\begin{align}\label{eq:first-law-full}
 \dd M_{\rm ADM}={}&T_{\rm H}\dd S_{\rm BH}+\OmH\dd J\nonumber\\
 &+\Phi_m\dd Q_m+\Pi_\alpha\dd\alpha+\delta W_{\rm other}.
\end{align}
Its existence and conjugates are not supplied by the off-shell rotating metric. To test the fixed-work projection, keep the dimensional profile parameter and coupling fixed, so that $\Mrem$ is fixed, and extend the family locally to $\rho(r)=M+f(r)$. With $M_{\rm ADM}=M+1$ and $J=aM_{\rm ADM}$, differentiation of $\Delta(\rh)=0$ gives
\begin{equation}\label{eq:horizon-variation}
 \dd M=\frac{\Delta_1}{2\rh}\dd\rh+\frac{a}{\rh}\dd a.
\end{equation}
Substitution of the geometric quantities in Eq.~\eqref{eq:horizon-data} yields
\begin{align}\label{eq:first-law-defect}
 &\dd M_{\rm ADM}-T_{\rm H}\dd(A_{\rm H}/4)-\OmH\dd J\nonumber\\
 &\quad=\frac{a}{\rh^2+a^2}
 \big[\rho(\rh)+\rh\rho'(\rh)-M_{\rm ADM}\big]\dd a.
\end{align}
The right-hand side is generally nonzero, although it vanishes in the Kerr limit. Thus holding charge and coupling fixed does not by itself make the naive area first law an identity of this rotating family. An exact source completion and any additional work or boundary terms must be specified~\cite{Rodrigues:2017tfm,Toshmatov:2017note}. The number-state entropy-production identity is independent of this obstruction.

Only conditionally, in a completion with the appropriate fixed-work first law and quasistatic flux balance, could one write
\begin{equation}\label{eq:total-area}
 \dot S_{\rm BH}=\bH\mathcal F_{\rm H},\qquad
 \mathcal F_{\rm H}=\dot E_{\rm H}-\OmH\dot J_{\rm H},
\end{equation}
where $\mathcal F_{\rm H}$ is the actual net generalized energy crossing the horizon. A negative radiation component $-\bH\sum_s\epsh_s\dot N_s^{\rm out}$ requires the corresponding global conservation prescription; it is not implied by the local response kernels. The incident and post-interaction atomic energies must not be counted twice. Moreover, generic accretion need not preserve the $M=0$ slice. No unconditional generalized-second-law or dynamical area-change result is asserted for the present off-shell geometry.
\end{revision}

\begin{revision}
\section{Numerical results}\label{sec:numerics}

All figures and numerical tables are regenerated by the accompanying notebook. Unless stated otherwise, $M=0$, $\mu=3$, and lengths are dimensionless in $\Mrem$. The outer root is bracketed using the analytic extremal radius at fixed spin, avoiding an unconstrained search for nearly coincident roots. Exact double roots are obtained from the parametrization, not identified by a small sampled value of $\Delta$. The static bounds and representative horizon data are given in Tables~\ref{tab:critical} and \ref{tab:representative}.

\begin{table}[t]\revcolor
 \caption{Static extremal parameters for $M=0$, $\mu=3$.}
 \label{tab:critical}
 \begin{ruledtabular}
 \begin{tabular}{lcc}
 Profile & $q_*$ & $r_*$\\\hline
 Maxwellian & 0.296296 & 0.592593\\
 Bardeen-like & 0.769800 & 1.088662\\
 Hayward-like & 1.058267 & 1.333333
 \end{tabular}
 \end{ruledtabular}
\end{table}
\begin{figure}[t]\revcolor
 \centering\includegraphics[width=\columnwidth]{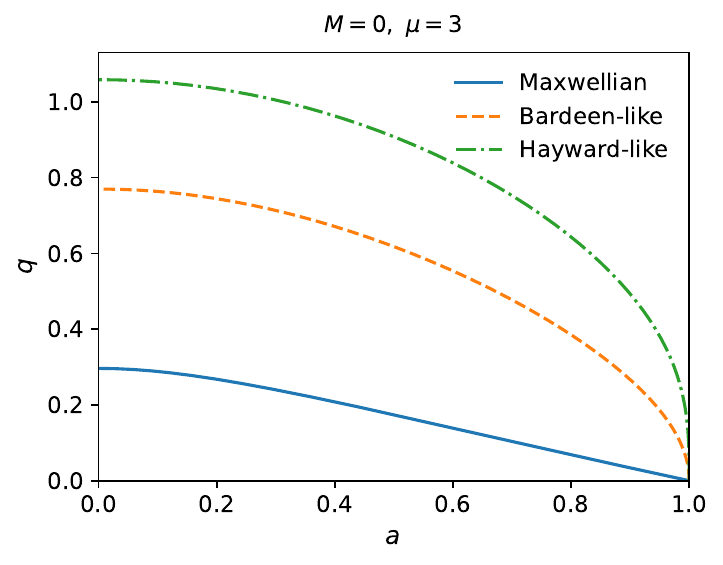}
 \caption{Exact extremal charge--spin boundaries for $M=0$, $\mu=3$. Positive-charge subextremal black holes lie below each curve, the curve itself is extremal, and the exterior region is horizonless. The $q=0$ boundary is the formal Kerr limit.}
 \label{fig:extremal}
\end{figure}

\begin{table*}[t]\revcolor
 \caption{Representative dimensionless horizon data at $a=0.6$, with $q=q_{\rm ext}(a)/2$ for each regular profile. The Kerr row is the formal $q=0$ reference.}
 \label{tab:representative}
 \begin{ruledtabular}
 \begin{tabular}{lrrrrrr}
 Profile & $q$ & $\rh$ & $\kap$ & $\TempH$ & $\OmH$ & $A_{\rm H}$\\\hline
 Kerr & 0 & 1.800000 & 0.222222 & 0.035368 & 0.166667 & 45.238934\\
 Maxwellian & 0.069507 & 1.508554 & 0.197107 & 0.031371 & 0.227641 & 33.121616\\
 Bardeen-like & 0.276804 & 1.714250 & 0.205774 & 0.032750 & 0.181892 & 41.452081\\
 Hayward-like & 0.419259 & 1.770450 & 0.213122 & 0.033919 & 0.171699 & 43.913104
 \end{tabular}
 \end{ruledtabular}
\end{table*}

\subsection{Horizon structure and exterior geometry}

Figure~\ref{fig:delta} shows subextremal, exactly extremal and supercritical radial functions at $a=0.6$. The double-root curve touches zero; the supercritical curve has no horizon. All plots use the same $M=0$ branch. Equal $q/q_{\rm ext}$ compares distances within model-dependent admissible intervals, not equal magnetic charge or equal deformation in a fixed microscopic theory.

\begin{figure*}[t]\revcolor\centering
 \includegraphics[width=0.48\textwidth]{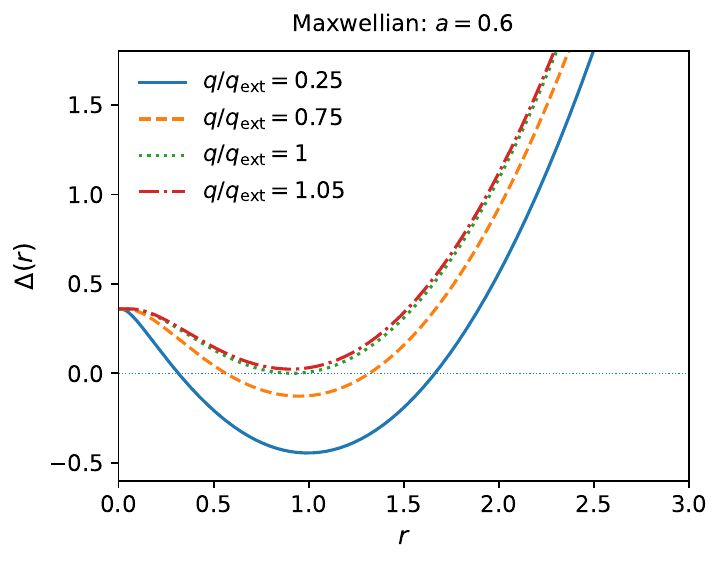}\hfill
 \includegraphics[width=0.48\textwidth]{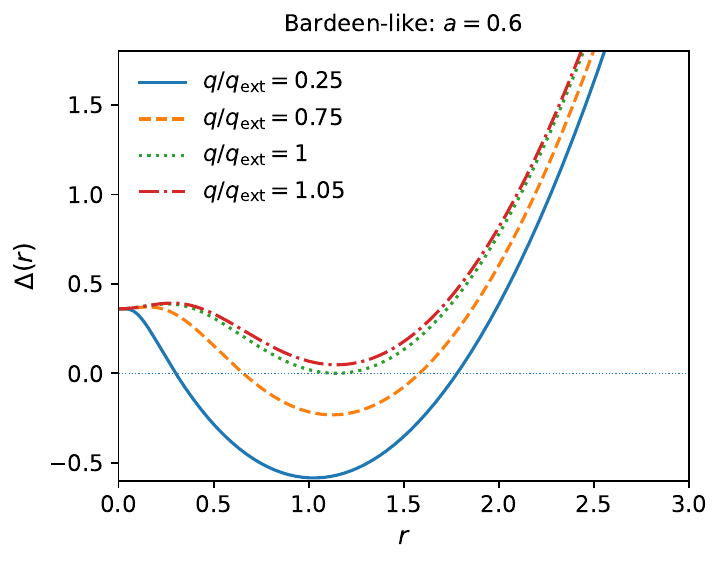}\\[1ex]
 \includegraphics[width=0.48\textwidth]{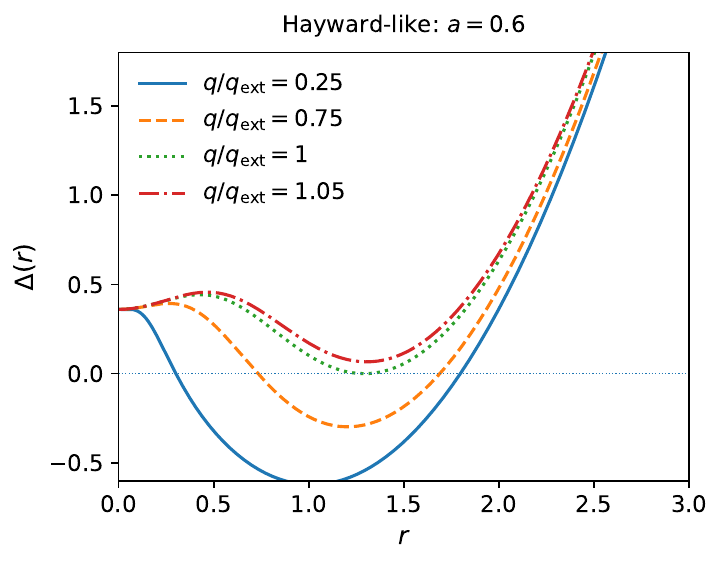}
 \caption{Radial functions at $a=0.6$ for Maxwellian, Bardeen-like and Hayward-like profiles, respectively. The four curves in each panel have $q/q_{\rm ext}=0.25,0.75,1,1.05$. The dotted extremal curve has a double root; the dash-dotted supercritical curve is horizonless.}
 \label{fig:delta}
\end{figure*}

At fixed $q=q_*/2$, Fig.~\ref{fig:horizon-thermo} follows the outer horizon, temperature and angular velocity as $a/a_{\rm ext}(q)$ increases. The extremal values are included as geometric endpoints, with $T_{\rm H}=0$. These endpoints are not used in the simple-horizon detector formulas. The outer horizon contracts and $\OmH$ rises along the displayed branches.

\begin{figure*}[t]\revcolor\centering
 \includegraphics[width=0.48\textwidth]{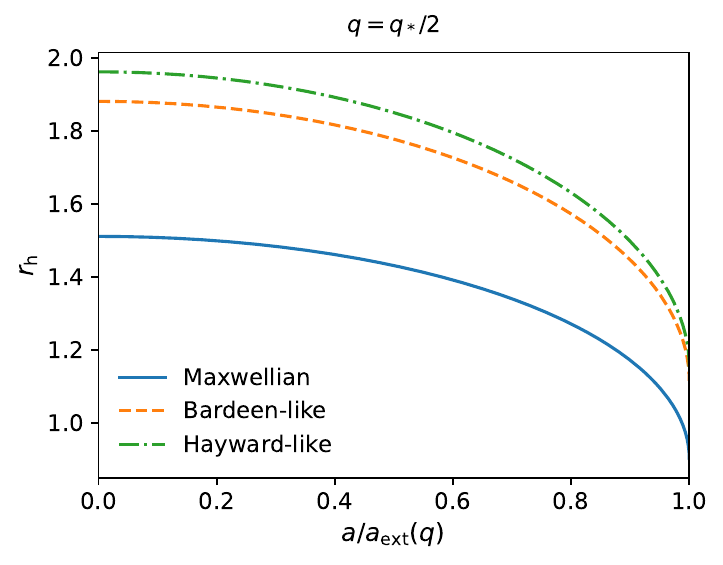}\hfill
 \includegraphics[width=0.48\textwidth]{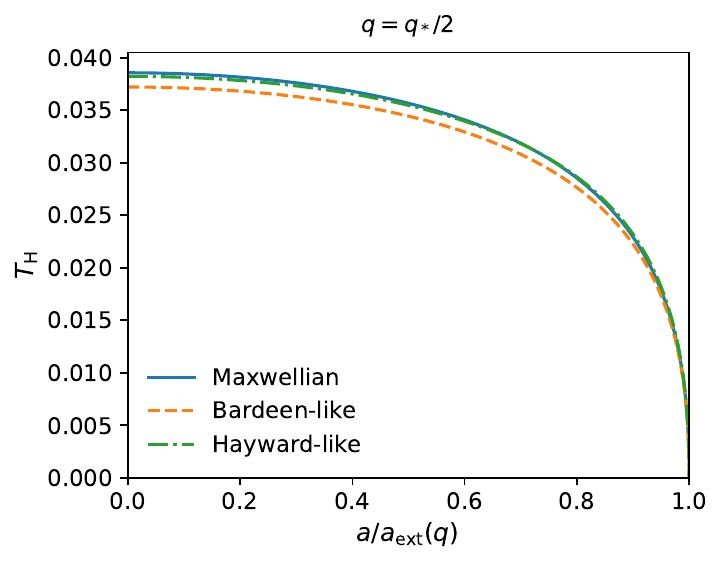}\\[1ex]
 \includegraphics[width=0.48\textwidth]{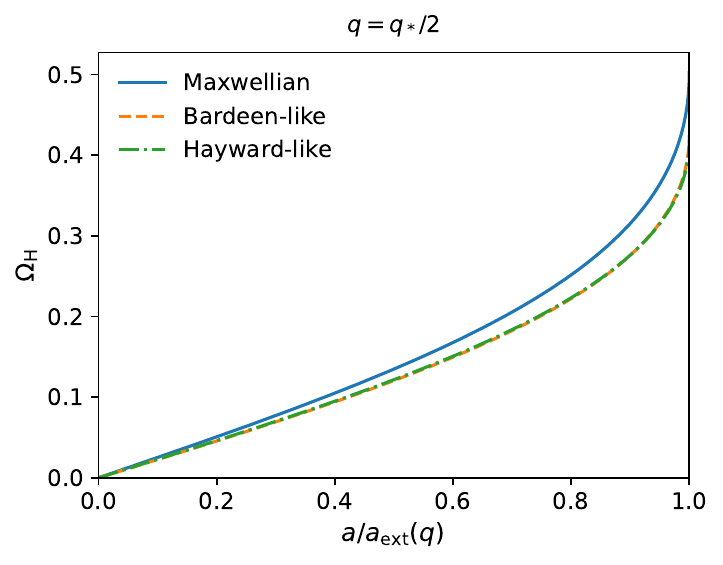}
 \caption{Outer radius (top left), Hawking temperature (top right) and horizon angular velocity (bottom) versus normalized spin, with $q=q_*/2$ separately in each model. The curves include the exact geometric extremal endpoint. All plotted variables are already dimensionless.}
 \label{fig:horizon-thermo}
\end{figure*}

Figure~\ref{fig:ergoregion} shows the corresponding horizon and stationary-limit surfaces at the representative parameters. Each surface is drawn as one continuous closed contour with the same coordinate limits and aspect ratio in all panels. The surfaces coincide at the poles and differ at the equator. These are coordinate cross sections, not isometric embeddings or projections of a black hole shadow.

\begin{figure*}[t]\revcolor\centering
 \includegraphics[width=0.32\textwidth]{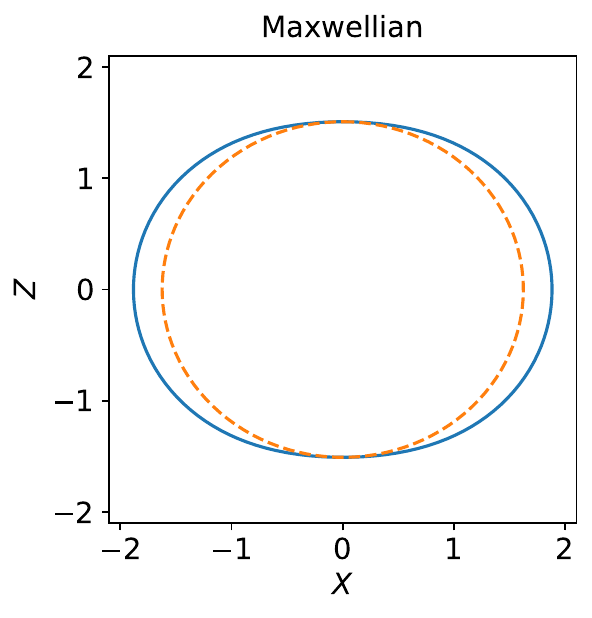}\hfill
 \includegraphics[width=0.32\textwidth]{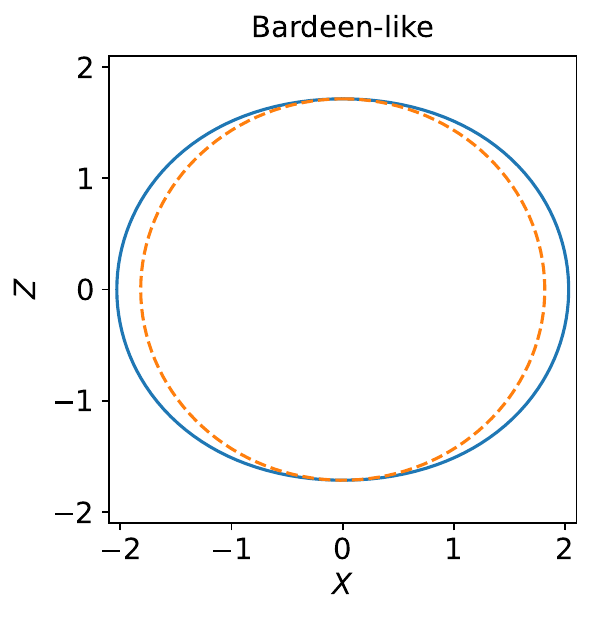}\hfill
 \includegraphics[width=0.32\textwidth]{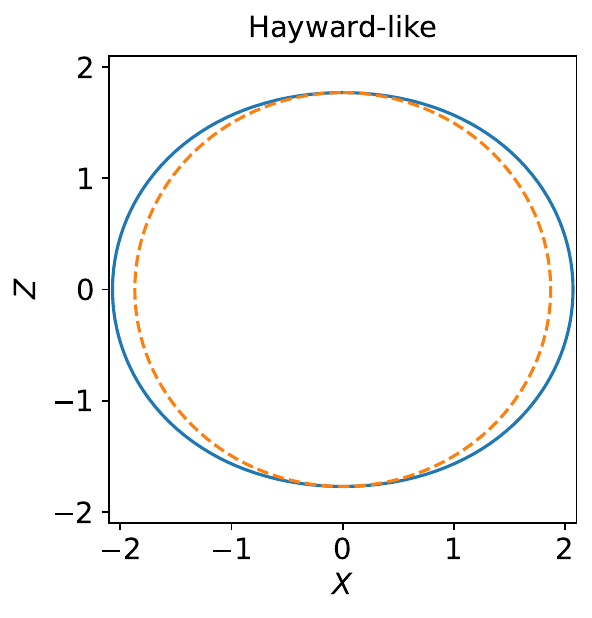}
 \caption{Meridional coordinate sections at the parameters in Table~\ref{tab:representative}. The solid contour is the stationary limit and the dashed contour is the event horizon. The dimensionless quasi-Cartesian coordinates are $X=\sqrt{r^2+a^2}\sin\theta$ and $Z=r\cos\theta$. All panels use the same scale.}
 \label{fig:ergoregion}
\end{figure*}

The charge dependence of the temperature is shown in Fig.~\ref{fig:temperature-charge}(a). At $a=0.6$ the three models begin at the Kerr value and cool to their own zero-temperature extremal endpoints. The Bardeen-like $\mu=3$ and $\mu=4$ comparison in panel (b) has the same qualitative behavior. This tests one additional mass-profile index; it does not establish a model-independent ranking of all finite-curvature geometries.

\begin{figure*}[t]\revcolor\centering
 \panel{0.48}{\textbf{(a)}\\\includegraphics[width=\linewidth]{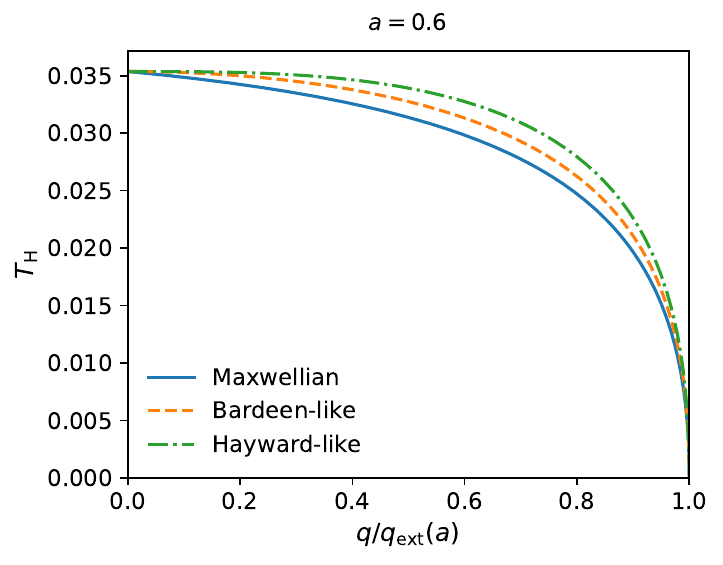}}\hfill
 \panel{0.48}{\textbf{(b)}\\\includegraphics[width=\linewidth]{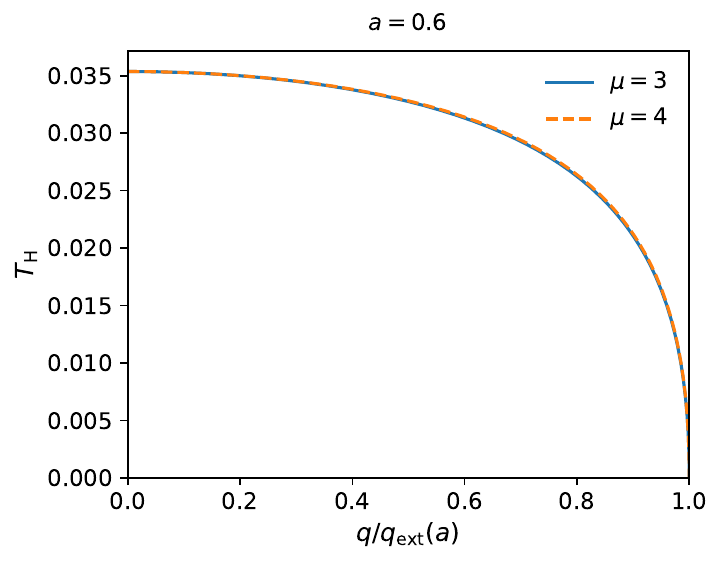}}
 \caption{Temperature versus model-normalized charge at $a=0.6$. (a) The three $\mu=3$ profiles. (b) The Bardeen-like profile at $\mu=3$ and $\mu=4$, each normalized by its own extremal charge. The endpoints are geometric limits, not extremal detector results.}
 \label{fig:temperature-charge}
\end{figure*}

\subsection{CQM index and leading local kernels}

Figure~\ref{fig:quantum-response}(a) displays the CQM coupling at $a=0.6$, $m=1$, $\omega=0.8$. Its growth as $\kap$ decreases is a phase-index enhancement, not an enhancement of radiation. The normalized leading spectrum is
\begin{equation}\label{eq:normalized-spectrum}
 \frac{\omega_0^2\mathscr R_{e,s}^{\rm loc}}{2\pi g^2}
 =\mathcal K_{\rm HBAR}.
\end{equation}
Panel (b) compares this kernel for the representative geometries and Kerr. Each curve approaches its own nonsuperradiant threshold from $\epsh>0$. Although the kernel tends to $1/(2\pi)$ as $\epsh\to0^+$, the associated ideal Bose occupation diverges; a stationary state at exactly $\epsh=0$ is not inferred.

At fixed $\omega=0.55$ and $m=1$, panel (c) shows the competition between the decreasing $\kap$ and the increasing $\OmH$. The Maxwellian curve is mildly nonmonotonic before its near-extremal suppression, while the Bardeen-like and Hayward-like kernels decrease along the displayed scan. These are formal leading kernels, not full finite-gate probabilities or globally normalized luminosities. None of the curves establishes the response of a fixed detector taken to extremality.

\begin{figure*}[t]\revcolor\centering
 \panel{0.48}{\textbf{(a)}\\\includegraphics[width=\linewidth]{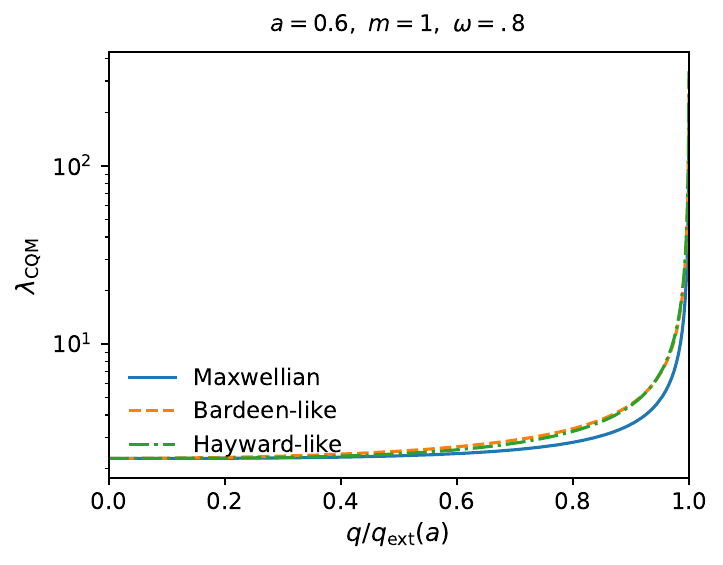}}\hfill
 \panel{0.48}{\textbf{(b)}\\\includegraphics[width=\linewidth]{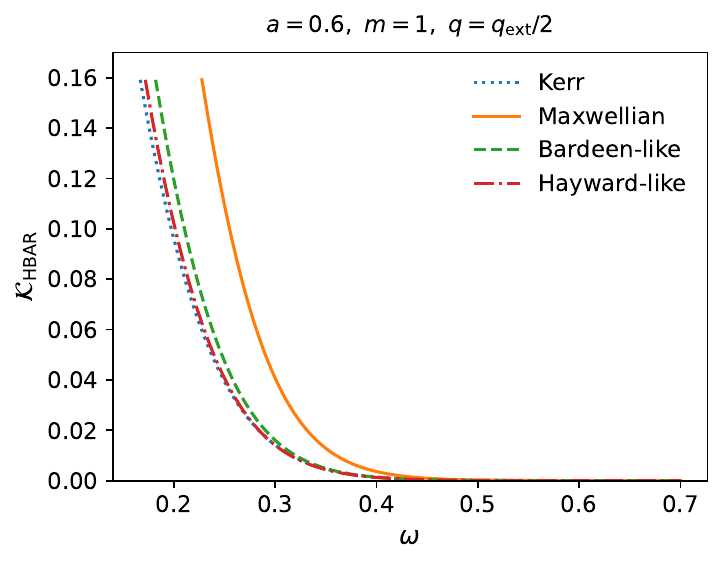}}\\[1ex]
 \panel{0.48}{\textbf{(c)}\\\includegraphics[width=\linewidth]{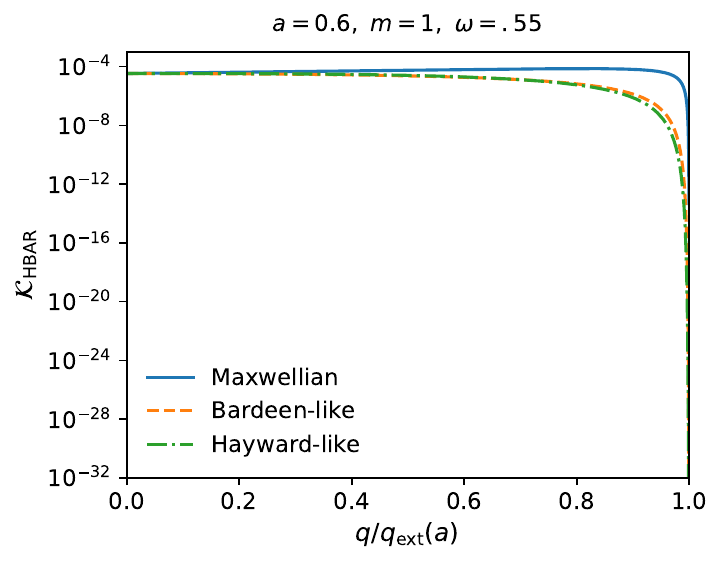}}
 \caption{Leading branch-resolved local quantities. (a) CQM coupling at $a=0.6$, $\omega=0.8$, $m=1$. (b) Local kernel for the representative geometries, restricted to $\epsh>0$. (c) Formal fixed-frequency kernel at $\omega=0.55$, $m=1$. Near-extremal points are not labeled detector-valid: a separate relative-error and switching analysis is required.}
 \label{fig:quantum-response}
\end{figure*}

\subsection{Complete finite-gate test of the outgoing branch}
\label{subsec:numerical-response}

Figure~\ref{fig:gap-dynamics}(a) uses the direct outgoing-branch calculation of Sec.~\ref{subsec:finite-gate}, including its exact exterior radial amplitude and worldline. For $T=0.03$, the gate gives the dominant finite-gap departure from $2\pi b_s$. The normalization by $2\pi b_s$ makes the three curves nearly coincide because their leading gate factor is common; this is not evidence that the mass profiles are identical. The increase with $\omega_0$ in Fig.~\ref{fig:gap-dynamics}(a) does not contradict the decrease of the orbital affinity in Fig.~\ref{fig:A}. The two figures describe different atomic motions and therefore different proper-time phases of the selected field mode. For stationary circular motion, the atom samples a harmonic phase and the orbital affinity decreases as $\mathcal A_{s,\varsigma}^{\rm orb}\sim 4\varpi_{s,\varsigma}/\omega_0$ at large gap, so it approaches zero as $\omega_0$ increases. For the horizon-crossing trajectory, the outgoing mode instead carries the logarithmic near-horizon phase. The corresponding affinity approaches the nonzero Hawking value $2\pi b_s$, while the finite-gate correction decreases in magnitude as the gap increases. Consequently, $\mathcal A_s^{\rm out}/(2\pi b_s)$ rises toward unity in Fig.~\ref{fig:gap-dynamics}(a), whereas $\mathcal A_{s,\varsigma}^{\rm orb}$ falls toward zero in Fig.~\ref{fig:A}.

To resolve the additional geometric term, panel (b) plots
\begin{equation}\label{eq:gate-subtracted-affinity}
 \frac{\omega_0}{4}
 (\mathcal A_s^{\rm out}-\mathcal A_s^{\rm gate})
 \longrightarrow E_s^{\rm geom},
\end{equation}
where $E_s^{\rm geom}=b_s\operatorname{Re}d_s^{\rm geom}+\operatorname{Im}d_s^{\rm geom}$. The analytic limits for the three profiles are $0.056428$, $0.033739$ and $0.086229$. Subtracting the exact pure-power gate isolates this smaller geometric effect; it is not subtraction of a measured experimental background. Table~\ref{tab:detector-numerical} compares the integrated affinity with the first-order formula at $\omega_0=300$.

\begin{figure*}[t]\revcolor\centering
 \panel{0.48}{\textbf{(a)}\\\includegraphics[width=\linewidth]{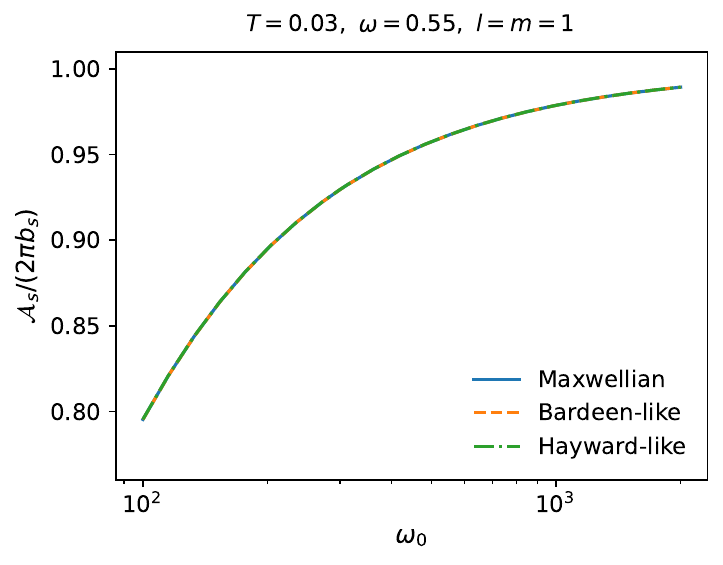}}\hfill
 \panel{0.48}{\textbf{(b)}\\\includegraphics[width=\linewidth]{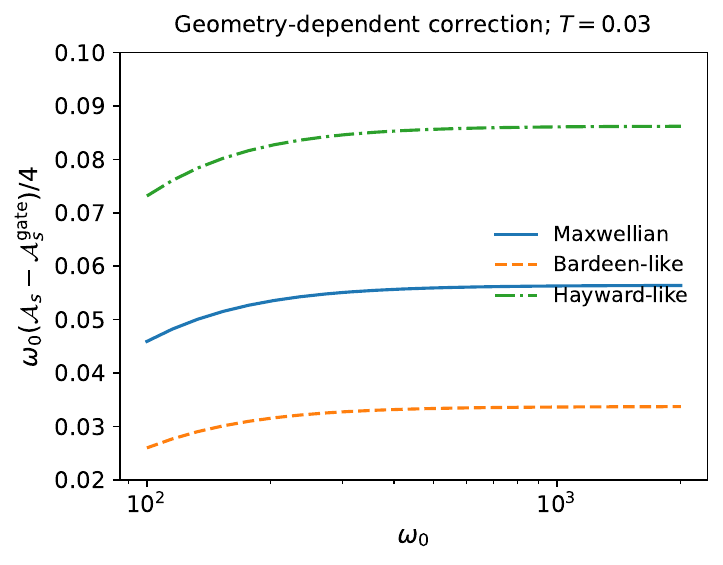}}
 \caption{Protocol-defined finite-gap response for the outgoing solution, with the parameters in Eq.~\eqref{eq:finite-gate-numerical-parameters}, $e=1$, $\ell=\mathcal Q=0$, and $T=0.03$. (a) Full branch affinity relative to the Hawking value. (b) Gate-subtracted affinity scaled to approach $E_s^{\rm geom}$. The calculation includes the full outgoing radial shape but not an unselected second branch or global Klein--Gordon normalization.}
 \label{fig:gap-dynamics}
\end{figure*}

\begin{table*}[t]\revcolor
\caption{Direct outgoing-branch finite-gate affinities at $\omega_0=300$, $T=0.03$. The first-order column uses Eq.~\eqref{eq:proper-time-affinity}, with the complete coefficient including the gate slope. The final column is $100|\mathcal A_{\rm NLO}/\mathcal A_{\rm num}-1|$. All quantities are dimensionless.}
\label{tab:detector-numerical}
\begin{ruledtabular}\begin{tabular}{lrrrr}
Profile & $2\pi b_s$ & $\mathcal A_{\rm num}$ & $\mathcal A_{\rm NLO}$ & Difference (\%)\\\hline
Maxwellian & 10.275860 & 9.552695 & 9.549744 & 0.0309\\
Bardeen-like & 11.239930 & 10.448552 & 10.445318 & 0.0310\\
Hayward-like & 11.152938 & 10.368380 & 10.365179 & 0.0309\\
\end{tabular}\end{ruledtabular}\end{table*}

The default calculation uses a twelfth-order outgoing series at $x=10^{-7}$, differential-equation relative tolerance $2\times10^{-12}$, 5000 smooth interpolation samples and a gate cutoff $z/T=42$. Refining these to order 14, tolerance $2\times10^{-13}$, 9000 samples and $z/T=50$ changes the affinities by less than $10^{-8}$ at the tested gaps $100$, $300$, $1000$ and $2000$. Quadrature-error estimates relative to each transition amplitude are also supplied in the notebook. These are numerical convergence checks for this nonextremal protocol, not uniform bounds for all gates, modes or parameter paths.

\subsection{Mode relaxation and the shrinking horizon scale}

Figure~\ref{fig:relax-validity}(a) solves Eq.~\eqref{eq:nbar-evolution} from an empty selected mode with $\omega=0.28$, $m=1$ and the leading outgoing Hawking rates. Setting the common weight $\mathcal W_s=1$ isolates the mathematical dependence on $\epsh/\kap$. The Maxwellian representative has the largest occupation because its threshold is closest to the chosen frequency. These normalized curves do not determine comparative laboratory or astrophysical relaxation times, which require the physical channel weights and injection rate.

Panel (b) displays $x_f/L_\Delta$ for the illustrative fixed radial window $x_f=0.05\rh$. Its growth near extremality shows that a fixed fractional window eventually leaves the linear-horizon region. This replaces a phase-plus-cutoff ``minimum atomic gap'' estimate: the plot is purely a geometric scale check and is neither a sufficient detector-validity criterion nor an error estimate for the direct finite-gate integration.

\begin{figure*}[t]\revcolor\centering
 \panel{0.48}{\textbf{(a)}\\\includegraphics[width=\linewidth]{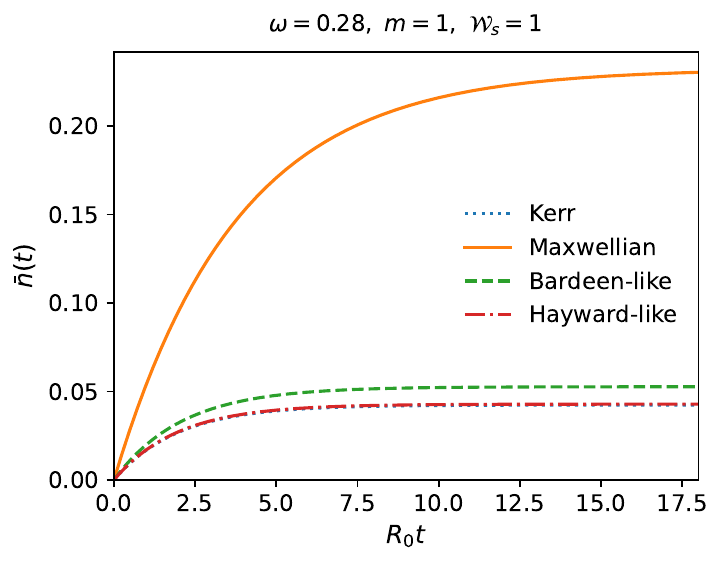}}\hfill
 \panel{0.48}{\textbf{(b)}\\\includegraphics[width=\linewidth]{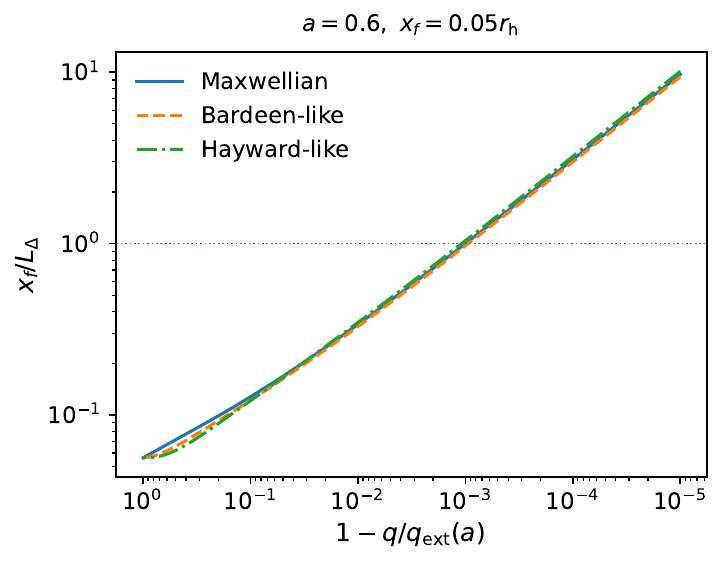}}
 \caption{(a) Empty-mode relaxation under ideal large-gap outgoing rates, with all common mode weights set to unity. (b) The ratio of a fixed radial window to $L_\Delta=2\Delta_1/|\Delta_2|$ at $a=0.6$. Extremality is approached to the right. The dotted line marks equality of the two lengths, not a physical minimum detector gap.}
 \label{fig:relax-validity}
\end{figure*}
\end{revision}

\begin{revision}
\section{Neutral scalar quasinormal modes}\label{sec:scalar-qnms}

The same separated scalar operator also defines an exterior resonance problem. Unlike the leading normalized HBAR kernel, QNMs depend on the full radial equation between the horizon and infinity. We restrict the numerical spectrum to a neutral massless test scalar, $\mu_\Phi=0$. These are not the coupled gravitational and electromagnetic perturbations of an unspecified rotating NLED source. The charged prescribed-field separation in Appendix~\ref{app:charged} is not used in the numerical spectrum.

\subsection{Radial potential and boundary conditions}\label{subsec:qnm-separation}

The exact angular and radial equations are Eqs.~\eqref{eq:angular} and \eqref{eq:radial-exact}; they need not be derived a second time. Define
\begin{align}\label{eq:K-lambda-definitions}
 A(r)&=r^2+a^2,&K&=A\omega-am,\nonumber\\
 \lambda_{lm}&=\Lambda_{\omega lm}+a^2\omega^2-2am\omega,
 &u&=\sqrt{A}\,R.
\end{align}
With $\dd r_*/\dd r=A/\Delta$, the radial equation becomes (\cite{Uktamjon:2024fjb})
\begin{equation}\label{eq:schrodinger-scalar}
 \frac{\dd^2u}{\dd r_*^2}+[\omega^2-U(r,\omega)]u=0,
\end{equation}
where
\begin{align}\label{eq:frequency-dependent-potential}
 U={}&\frac{2am\omega}{A}-\frac{a^2m^2}{A^2}\nonumber\\
 &+\frac{\Delta}{A^2}\left[
 \lambda_{lm}+\mu_\Phi^2r^2
 +\frac{(r\Delta)'}{A}-\frac{3r^2\Delta}{A^2}\right].
\end{align}
Both frame dragging and the spheroidal eigenvalue make this potential frequency dependent. With the convention $\e^{-\ii\omega t}$, the massless boundary conditions are
\begin{equation}\label{eq:qnm-boundary-conditions}
 u\sim\begin{cases}
 \e^{-\ii(\omega-m\OmH)r_*},&r_*\to-\infty,\\
 \e^{+\ii\omega r_*},&r_*\to+\infty.
 \end{cases}
\end{equation}
Damped modes have $\operatorname{Im}\omega<0$. For a massive scalar the outgoing infinity behavior instead includes the long-range phase
\begin{equation}\label{eq:massive-infinity}
 u\sim\e^{\ii k_\infty r_*}
 r^{\ii M_{\rm ADM}\mu_\Phi^2/k_\infty},
 \qquad k_\infty=\sqrt{\omega^2-\mu_\Phi^2},
\end{equation}
with the branch selected by outgoing analytic continuation. This is different from an exponentially decaying quasi-bound-state condition~\cite{KonoplyaZhidenko2013,XuEtAl2021}. The massive refinement does not affect the massless numerical results.

\subsection{Frequency-dependent WKB and independent collocation}
\label{subsec:qnm-wkb}

The third-order Schutz--Will--Iyer construction~\cite{SchutzWill1985,IyerWill1987} is solved as a complex-frequency problem. Let
\begin{equation}
 D_*\equiv\frac{\Delta}{A}\frac{\dd}{\dd r},\qquad
 U_j=\left.D_*^jU\right|_{r_p},\qquad\alpha_n=n+\tfrac12.
\end{equation}
We solve the peak condition and the quantization condition simultaneously:
\begin{align}\label{eq:implicit-wkb-system}
 \partial_rU(r_p,\omega)&=0,\nonumber\\
 \omega^2&=U_0+\mathcal L_2
 -\ii\alpha_n\sqrt{-2U_2}\,(1+\mathcal O_3).
\end{align}
The corrections $\mathcal L_2$ and $\mathcal O_3$ are displayed in Appendix~\ref{app:numerical}; the notation $\mathcal O_3$ here denotes a WKB coefficient, not an order symbol. The four real unknowns are the real and imaginary parts of $r_p$ and $\omega$. The square-root branch and resonance branch are followed from Schwarzschild through Kerr and then through the regular deformation. Taylor-jet differentiation supplies derivatives through order six without differencing noisy samples. WKB order changes are numerical indicators, not rigorous error bars~\cite{Konoplya2019WKBRecipes}.

At every frequency iteration, the angular eigenvalue is recalculated from a fixed-$m$, fixed-parity spherical-harmonic matrix,
\begin{equation}\label{eq:spheroidal-matrix}
 \sum_{j'}\left[j(j+1)\delta_{jj'}
 -c^2\langle jm|\cos^2\theta|j'm\rangle\right]b_{j'}
 =\Lambda_{lm}b_j,
\end{equation}
where $c^2=a^2(\omega^2-\mu_\Phi^2)$. The moderate spheroidicities used here allow continuation of the branch connected to $l(l+1)$; broader scans would require additional eigenvector tracking.

For an independent frequency calculation, we factor the exact radial boundary behavior and use Chebyshev collocation. With $M_{\rm ADM}=1$,
\begin{align}\label{eq:spectral-ansatz}
 y&=1-\frac{\rh}{r},\qquad
 p_{\rm in}=-\frac{\ii[(\rh^2+a^2)\omega-am]}{\Delta_1},\nonumber\\
 R&=\e^{\ii\omega r}r^{2\ii\omega-1}y^{p_{\rm in}}\Psi(y).
\end{align}
The remaining function is represented on Chebyshev--Lobatto points $y_j=[1-\cos(j\pi/N)]/2$. The endpoint relations follow from the horizon Frobenius expansion and the first outgoing inverse-radius coefficient at infinity, as given in Appendix~\ref{app:numerical}. The eigenvalue $\Lambda_{lm}(\omega)$ is updated in this calculation as well. A WKB root can seed the branch, but the final root solves the exact radial boundary-value matrix, not a refitted WKB equation.

\subsection{Reference backgrounds and convergence}\label{subsec:qnm-validation}

The $q=0$ limit is Kerr with unit ADM mass. We also replace $\Delta$ only in the validation code by
\begin{equation}\label{eq:Delta-KN-validation}
 \Delta_{\rm KN}=r^2-2r+a^2+Q_{\rm KN}^2.
\end{equation}
The auxiliary Kerr--Newman electric charge $Q_{\rm KN}$ is distinct from the regular-profile parameter $q$. The neutral massless Kerr--Newman equation admits an independent continued-fraction calculation~\cite{BertiKokkotas2005,KonoplyaZhidenko2013}. Table~\ref{tab:qnm-validation} compares that calculation with WKB3. At $a=0.6$ the continued-fraction Kerr result is
\begin{equation}\label{eq:kerr-baseline-spectral}
 \omega_{220}^{\rm Kerr}=0.6173637080-0.0912453958\ii.
\end{equation}
The corresponding scalar angular eigenvalue is $\Lambda_{22}=5.980794923+0.005815344\ii$, consistent with the spheroidal data in Ref.~\cite{BertiCardoso2006}. The Chebyshev Kerr limit agrees with the independently implemented continued fraction within the reported numerical precision.

\begin{table*}[t]\revcolor
\caption{Neutral massless fundamental $(l,m,n)=(2,2,0)$ modes in the unit-mass Kerr and auxiliary Kerr--Newman backgrounds. CF denotes the independently computed continued fraction. The final column is $100|\omega_{\rm WKB3}-\omega_{\rm CF}|/|\omega_{\rm CF}|$.}
\label{tab:qnm-validation}
\begin{ruledtabular}\begin{tabular}{lccccc}
Background & $a$ & $Q_{\rm KN}$ & $\omega_{\rm CF}$ & $\omega_{\rm WKB3}$ & Error (\%)\\\hline
Kerr & 0.0 & 0.0 & $0.48364387-0.09675878\ii$ & $0.48321103-0.09680485\ii$ & 0.0883\\
Kerr & 0.2 & 0.0 & $0.51712058-0.09638217\ii$ & $0.51668441-0.09643359\ii$ & 0.0835\\
Kerr & 0.4 & 0.0 & $0.55964695-0.09493115\ii$ & $0.55920440-0.09498487\ii$ & 0.0785\\
Kerr & 0.6 & 0.0 & $0.61736371-0.09124540\ii$ & $0.61690462-0.09130233\ii$ & 0.0741\\
Kerr & 0.8 & 0.0 & $0.70682338-0.08152026\ii$ & $0.70630729-0.08161773\ii$ & 0.0738\\
Kerr-Newman & 0.6 & 0.2 & $0.62474516-0.09099008\ii$ & $0.62428741-0.09104501\ii$ & 0.0730\\
Kerr-Newman & 0.6 & 0.4 & $0.64959834-0.08978574\ii$ & $0.64913992-0.08983384\ii$ & 0.0703\\
Kerr-Newman & 0.6 & 0.6 & $0.70470001-0.08501865\ii$ & $0.70420496-0.08505280\ii$ & 0.0699\\
\end{tabular}\end{ruledtabular}\end{table*}

For the regular fundamental branch, radial resolutions $N=30,40,50$ and angular bases of 6, 8, 12 and 16 functions are checked at the representative points. The $N=40$ to $N=50$ frequency changes are below $10^{-9}$ throughout the displayed $0\leq\eta\leq0.9$ scan. Representative resolution tables and normalized matrix residuals are supplied with the notebook. Agreement under these changes is convergence evidence, not a mathematical error certificate or a proof that no other resonance family exists.

\subsection{Scalar frequencies and the exterior-field comparison}
\label{subsec:qnm-results}

We use $a=0.6$, $l=m=2$, $n=0$ and $\eta=q/q_{\rm ext}(a)$. At this spin,
\begin{equation}\label{eq:qext-values}
 q_{\rm ext}=0.1390148387,\quad0.5536077321,\quad0.8385172309
\end{equation}
for $\nu=1,2,3$, respectively. Table~\ref{tab:regular-qnms} uses $\eta=1/2$ and reports the independent spectral frequencies as the primary numerical values. The complex WKB error is measured against those values, rather than inferred from an order-to-order spread.

\begin{table*}[t]\revcolor
\caption{Independent spectral and WKB3 fundamental scalar frequencies at $a=0.6$, $q/q_{\rm ext}=1/2$, $M=0$, $\mu=3$, and $(l,m,n)=(2,2,0)$. The horizon data are given in Table~\ref{tab:representative}. The last column is $100|\omega_{\rm WKB3}-\omega_{\rm sp}|/|\omega_{\rm sp}|$.}
\label{tab:regular-qnms}
\begin{ruledtabular}\begin{tabular}{lccc}
Profile & $\omega_{\rm sp}$ & $\omega_{\rm WKB3}$ & Error (\%)\\\hline
Maxwellian & $0.71449413-0.08534642\ii$ & $0.71402904-0.08539147\ii$ & 0.06494\\
Bardeen-like & $0.63740474-0.08710387\ii$ & $0.63687382-0.08712435\ii$ & 0.08259\\
Hayward-like & $0.62272953-0.08912225\ii$ & $0.62217180-0.08913520\ii$ & 0.08868\\
\end{tabular}\end{ruledtabular}\end{table*}

Relative to Eq.~\eqref{eq:kerr-baseline-spectral}, the spectral real-frequency shifts are $15.7331\%$, $3.2462\%$ and $0.8692\%$. The damping magnitudes decrease by $6.4650\%$, $4.5389\%$ and $2.3269\%$. This establishes the hierarchy for the stated normalization and parameters; it is not a comparison at equal dimensional magnetic charge. The rotating Bardeen/Hayward scalar problem has an independent literature~\cite{PengHuang2025}, but different parameter conventions must be matched before numerical shifts are compared.

Figure~\ref{fig:qnm-eta-scan} displays the continued spectral branch. The real frequency increases and the damping magnitude decreases as $\eta$ increases along this scan. The computed modes satisfy $\operatorname{Re}(\omega-m\OmH)>0$ and $\operatorname{Im}\omega<0$. These facts establish damping of the tracked scalar modes, not full mode stability or the absence of superradiant trapped families and zero-damped branches.

\begin{figure*}[t]\revcolor\centering
 \panel{0.48}{\textbf{(a)}\\\includegraphics[width=\linewidth]{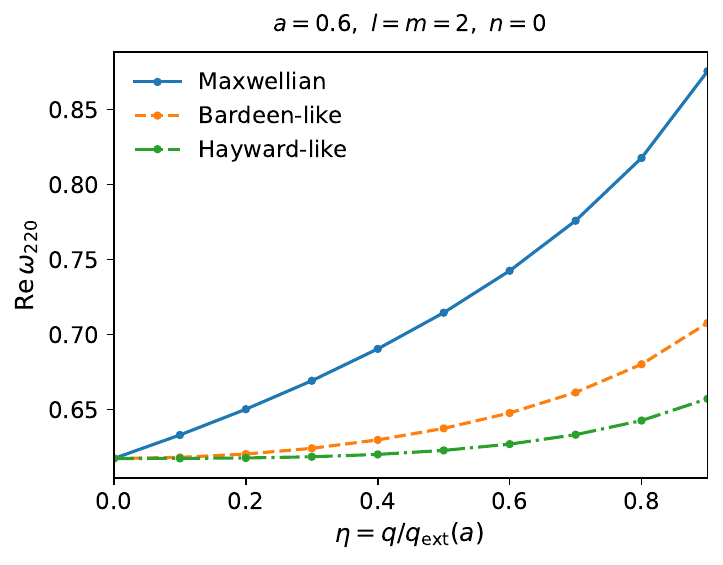}}\hfill
 \panel{0.48}{\textbf{(b)}\\\includegraphics[width=\linewidth]{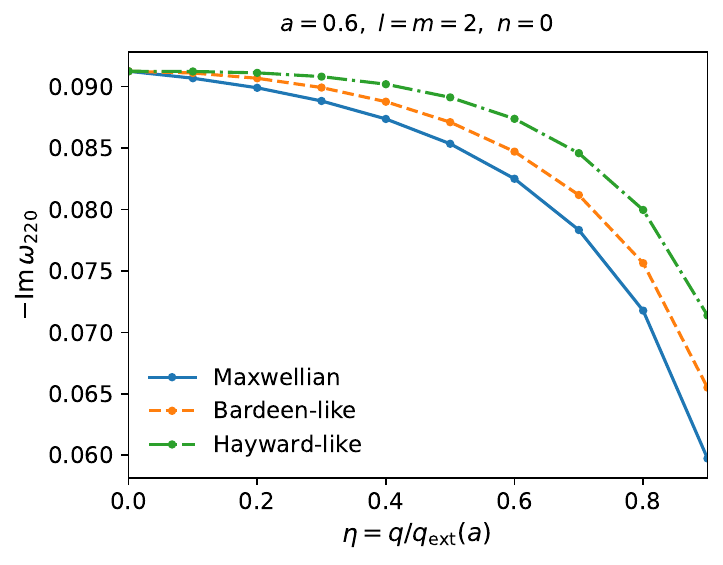}}
 \caption{Independently collocated fundamental neutral massless scalar frequencies for $M=0$, $\mu=3$, $a=0.6$, and $(l,m,n)=(2,2,0)$. (a) Real frequency. (b) Damping magnitude. Markers are computed spectral roots and lines guide the eye. The $\eta=0$ endpoint is Kerr. The notebook includes WKB comparisons and $N=40$/$N=50$ checks at every plotted point. No exactly extremal mode is computed.}
 \label{fig:qnm-eta-scan}
\end{figure*}

The WKB2--WKB3 changes at $\eta=1/2$ are $0.449\%$, $0.572\%$ and $0.623\%$, larger than the actual fundamental WKB3 errors in Table~\ref{tab:regular-qnms}; these spreads are not calibrated confidence intervals. The first overtone is substantially less accurate in WKB3. Independent spectral estimates are
\begin{equation}\label{eq:first-overtone-estimates}
 \omega_{221}^{\rm sp}\simeq
 \begin{cases}
 0.707536-0.257132\ii,&\nu=1,\\
 0.628360-0.262977\ii,&\nu=2,\\
 0.612614-0.269308\ii,&\nu=3.
 \end{cases}
\end{equation}
Their $N=40$/$N=50$ changes are below $10^{-7}$ at these points, while WKB3 differs by $3.52\%$, $3.51\%$ and $3.56\%$. This independently confirms that fundamental-mode calibration must not be transferred to overtones.

Restoring the physical length $\Mrem$ and the speed of light gives
\begin{equation}\label{eq:restore-qnm-dimensions}
 f^{\rm SI}=\frac{c\operatorname{Re}\omega}{2\pi\Mrem},
 \qquad \tau^{\rm SI}=\frac{\Mrem}{c[-\operatorname{Im}\omega]}.
\end{equation}
The scalar spectrum and the finite-gate response thus share the same mass scaling but represent distinct boundary-value problems. A QNM is not a normalizable thermal mode; substituting its complex frequency into a Bose occupation is not part of this calculation.
\end{revision}

\section{Application of the WKB approximation to the quasi-bound states and trapped modes }

It is clear from the expressions given in Eqs.(\ref{eq:frequency-dependent-potential},\ref{eq:charged-potential-correct}) that at infinity $U(r\to\infty)\to\mu^2_\Phi$ so a massive field can support quasi-bound states with $\omega^2<\mu^2_\Phi$ as the wave decays exponentially at large radii. The same algebraic WKB formula given in (\ref{eq:implicit-wkb-system}) can be used, with $\frac{dU}{dr_*}> 0$, to obtain the local spectrum in the well (see, \cite{2026arXiv260910880K} for details).

Since the WKB series is typically asymptotic, adding higher-order terms does not guarantee a better result. Pad\'e approximants allow the truncated series to be rearranged. One has to define a polynomial $P_\kappa(\epsilon)$ to employ Pad\'e approximants (\cite{Konoplya:2019hlu}):
\begin{eqnarray}
    P_\kappa(\epsilon)&=&U_0+A_2(\mathcal{K}^2)\epsilon^2+A_4(\mathcal{K}^2)\epsilon^4+A_6(\mathcal{K}^2)\epsilon^6...\\\nonumber
    &-&ik\sqrt{-2U_2}\left[\epsilon+A_3(\mathcal{K}^2)\epsilon^3+A_5(\mathcal{K}^2)\epsilon^5\right]\,.
\end{eqnarray}
A family of Pad\'e approximants $P_{\Tilde{n}/\Tilde{m}}(\epsilon)$  is introduced for the polynomial $P_\kappa(\epsilon)$ in a vicinity of $\epsilon=0$, where $\Tilde{n}+\Tilde{m}=k$:
\begin{eqnarray}
  P_{\Tilde{n}/\Tilde{m}}(\epsilon)=\frac{Q_0+Q_1\epsilon+...+Q_n\epsilon^n}{R_0+R_1\epsilon+...+R_n\epsilon^n}\,,  
\end{eqnarray}
such that
\begin{eqnarray}
   P_{\Tilde{n}/\Tilde{m}}-P_\kappa(\epsilon)=\mathcal{O}(\epsilon^{\kappa+1})\,. 
\end{eqnarray}
In our further calculation we use $P_{8/8}(1)$ as near-diagonal approximation $\Tilde{n}=\Tilde{m}$ is the most accurate. Since this method provides only an approximation to the real part of $\omega$, the imaginary part--which gives the decay rate of the mode--can only be obtained by also considering tunnelling through the potential barrier.

In the leading Gamow approximation, the imaginary part is treated as a small nonperturbative correction to the real level (\cite{Gamow:1928zz}), so the barrier action and the well action can be expressed as:
\begin{eqnarray}
    K(\omega_R^2)&=&\int^{r_{*2}}_{r_{*1}}\sqrt{U(r_*)-\omega_R^2}dr_*\,,\\\nonumber
    I(\omega_R^2)&=&\frac{1}{2}\int_{r_{*2}}^{r_{*3}}\frac{dr_*}{\sqrt{\omega_R^2-U(r_*)}}\,,
\end{eqnarray}
where $\omega_R$ is the real part of $\omega$ obtained from WKB method and $r_{*1}<r_{*2}<r_{*3}$ are the three real roots of the $U(r_*)=\omega_R^2$. Finally full expression for the quasi-bound-state frequencies. $\omega$ can be written as (\cite{2026arXiv260910880K}):
\begin{eqnarray}
    \omega=\omega_R-i\frac{\exp{\left[-2K(\omega_R^2)\right]}}{8\omega_RI(\omega_R^2)}+\mathcal{O}(\omega_R^3)\,.
\end{eqnarray}

\begin{table*}[t]
Frequencies of the fundamental quasi-bound state of a massive scalar field  at $a=0.6$, $M=0$, $\mu=3$, $\mu_\phi=0.3$ and $(l,m,n)=(1,1,0)$. To obtain the WKB real parts, the WKB equation was solved numerically, employing the Pad\'e approximant with $\Tilde{m}=\Tilde{n}=8$ at the local minimum. The imaginary parts are determined by applying the Gamow formula to a frequency-independent potential and the superradiant horizon-flux correction is not included.
\label{tab:regular-qnms1}
\begin{ruledtabular}\begin{tabular}{lcc}
Profile & $q/q_{ext}=0.1$ & $q/q_{ext}=1/2$ \\\hline
Maxwellian & $0.296174-4.40954\times10^{-7}\ii$ & $0.296237-7.89716\times10^{-8}\ii$ \\
Bardeen-like & $0.296166-5.89301\times10^{-7}\ii$ & $0.29616-3.78744\times10^{-7}\ii$ \\
Hayward-like & $0.296158-5.99208\times10^{-7}\ii$ & $0.26186-5.17618\times10^{-7}\ii$ \\
\end{tabular}\end{ruledtabular}\end{table*}

\begin{revision}

To be more informative we give some numerical values of the quasi-bound-state frequencies $\omega$ in table (\ref{tab:regular-qnms1}).

\section{Domain of validity and limitations}\label{sec:limitations}

The geometry is an effective off-shell Kerr metric. Its static seed has an Einstein--NLED interpretation, but the rotating reconstruction has a known consistency issue~\cite{Rodrigues:2017tfm,Toshmatov:2017note}. Separability, geodesics, horizon quantities and test-scalar propagation are statements about the prescribed metric. A self-consistent rotating matter source, gravitational perturbations and dynamical thermodynamic work terms are not determined here. The first-law defect in Eq.~\eqref{eq:first-law-defect} makes this distinction explicit.

The $\mu=3$ branch has finite but direction-dependent central curvature limits; it is not presented as a proof of smooth extension through the rotating center. The detector and QNM calculations are exterior calculations and cannot by themselves distinguish regularity behind the horizon. Dimensionless charge scans follow Eq.~\eqref{eq:physical-parameter-path}, not a fixed-$\alpha$ path in one microscopic theory. The $\mu=4$ comparison tests the exterior horizon trends for one additional branch.

The CQM reduction and proper-time expansion require a simple outer horizon. Their fixed-$b_s$ asymptotics are not uniform at extremality. The gate and higher smooth coefficients, the shrinking radial scale $L_\Delta$, and the relative error in the suppressed emission amplitude must all be controlled along any proposed near-extremal sequence. A hard outer cutoff cannot substitute for this analysis, and a fixed atom cannot be assigned an arbitrarily large gap while neglecting recoil.

The direct finite-gate calculation includes the full radial amplitude and equatorial worldline of a horizon-normalized outgoing solution. It does not normalize a global scattering or cavity state or calculate its second radial branch. The complete-channel affinity differs by Eq.~\eqref{eq:branch-affinity}. Angular nodes, switching nodes, frequency-degenerate sectors, superradiant gain, finite-size effects, strong coupling, overlapping atomic interactions and non-Markovian memory require separate treatments. A gate smooth on the exterior half-line is not an interaction switched smoothly through the horizon.

\end{revision}

\begin{revision}
\section{Conclusions}\label{sec:conclusions}

We have constructed a branch-resolved HBAR calculation for Kerr-like rotating finite-curvature metrics with nonlinear-electrodynamic mass profiles. The exact extremal charge--spin parametrization fixes the allowed domain on the $M=0$ finite-curvature branch. The dimensionful interpretation of a charge scan follows a correlated variation of the charge parameter and coupling.

The leading scalar CQM index depends only on $\epsh/\kap$, with $\epsh=\omega-m\OmH$. The finite detector response contains additional information. By expanding in proper time, we obtained a first affinity correction that includes the radial Frobenius coefficient, angular variation, complete first subleading trajectory and switching slope. The explicit cancellation in the radial contribution and the Schwarzschild limit show why a trajectory-phase multiplier alone is insufficient. Direct integration of the outgoing radial solution and an equatorial trajectory with a specified exponential gate verifies the finite-gap expansion over the reported nonextremal parameter range. It separates the dominant gate effect from the smaller geometry-dependent correction without interpreting either as a universal shifted Hawking temperature. We also analyzed forced equatorial circular atoms as a distinct detector motion. Although their proper acceleration can become large in the ultra-relativistic near-horizon regime, their transition rates depend on the orbital proper-time frequency and switching profile rather than obeying the universal infalling Hawking affinity.

The same distinction persists for an atomic cloud. Complete selected-channel probabilities determine a geometric stationary state and a nonnegative population entropy-production identity. An unselected radial branch changes the affinity through coherent interference, and cavity output must be distinguished from a change in mode occupation. The off-shell first-law defect prevents promotion of these identities to an unconditional horizon-area law.

Independent Chebyshev calculations validate the neutral-scalar WKB fundamental frequencies and quantify the larger first-overtone errors. These resonances probe the exterior radial operator, whereas the leading normalized outgoing HBAR kernel uses only two horizon parameters. They are complementary scalar calculations, not a gravitational-wave spectrum or a rule for thermally populating complex-frequency QNMs.

A complete observational application requires a globally normalized real-frequency mode problem, a specified detector and output geometry, controlled branch selection, and a self-consistent rotating source where horizon work terms and backreaction can be derived. The present results identify which local thermal statements survive finite protocol corrections and which further physical inputs such an application must supply.

\section*{Data and code availability}
No external observational data were used. 
\end{revision}

\begin{acknowledgments}
A.~{\"O}. and R.~C.~P. acknowledge networking support from COST Actions CA21106 (CosmicWISPers), CA22113 (THEORY-CHALLENGES), CA21136 (CosmoVerse), CA23130 (BridgeQG), and CA23115 (Relativistic Quantum Information), funded by COST (European Cooperation in Science and Technology).  A.~{\"O}. also acknowledges support from SCOAP3, Switzerland, and T{\"U}B{\.I}TAK, T{\"u}rkiye.
\end{acknowledgments}

\appendix

\section{Mass-profile derivatives and the second radial derivative}\label{app:derivatives}

For
\begin{equation}
 \rho=\frac{r^\mu}{(r^\nu+q^\nu)^{\mu/\nu}},
\end{equation}
logarithmic differentiation gives Eqs.~\eqref{eq:rho-derivatives}.  For $\mu=3$,
\begin{equation}\label{eq:D2-general-mu3}
 \Delta''=2-\frac{6q^\nu r^2
 \left[4q^\nu+(1-\nu)r^\nu\right]}
 {(r^\nu+q^\nu)^{3/\nu+2}}.
\end{equation}
The three subclasses are
\begin{align}
 \nu=1:\quad
 \Delta''&=2\left[1-\frac{12q^2r^2}{(q+r)^5}\right],\\
 \nu=2:\quad
 \Delta''&=2\left[1-\frac{3q^2r^2(4q^2-r^2)}
 {(q^2+r^2)^{7/2}}\right],\\
 \nu=3:\quad
 \Delta''&=2\left[1-\frac{6q^3r^2(2q^3-r^3)}
 {(q^3+r^3)^3}\right].
\end{align}
These identities were checked symbolically and numerically in the notebook.

To derive Eq.~\eqref{eq:extremal-param}, note that at $r=r_{\rm e}$,
\begin{equation}
 \rho_{\rm e}=(1+y)^{-\mu/\nu},
 \qquad
 r_{\rm e}\rho'_{\rm e}=\frac{\mu y}{1+y}\rho_{\rm e}.
\end{equation}
The equation $\Delta'=0$ gives
\begin{equation}
 r_{\rm e}=\rho_{\rm e}+r_{\rm e}\rho'_{\rm e}
 =\rho_{\rm e}\frac{1+(\mu+1)y}{1+y},
\end{equation}
which is the first of Eqs.~\eqref{eq:extremal-param}.  The definitions of $y$ and $\Delta=0$ then give $q_{\rm e}$ and $a_{\rm e}^2$.  Finally, $a_{\rm e}^2\geq0$ is equivalent to $(\mu-1)y\leq1$.

\section{Near-horizon geodesic coefficients}\label{app:geodesics}

Let
\begin{gather}
 H=\rh^2+a^2,
 \qquad
 \Delta_1=\Delta'(\rh),
 \qquad
 \Delta_2=\Delta''(\rh),\\
 P_0=H\widetilde e,
 \qquad
 P_1=2e\rh,\\
 B_0=\rh^2+(\ell-ae)^2+\mathcal Q,\\
 c_1=-4e\rh P_0+\Delta_1B_0,
 \qquad
 X=ae\sin^2\theta_{\rm h}-\ell.
\end{gather}
The radial potential expands as
\begin{equation}
 \mathcal R=P_0^2-c_1x+\order(x^2),
\end{equation}
which immediately gives Eq.~\eqref{eq:kcoeff}.  The polar equation implies
\begin{equation}
 \rev{\theta(x)=\theta_{\rm h}+\theta_1x+\order(x^2),\qquad
 \theta_1=\mp\frac{\sqrt{\Theta_\theta(\theta_{\rm h})}}{P_0}},
\end{equation}
so every smooth polar factor can be evaluated at $\theta_{\rm h}$ at leading endpoint order.  Dividing the $t$ and $\phi$ equations in Eq.~\eqref{eq:geodesics} by the ingoing radial equation and expanding to constant order gives
\begin{align}\label{eq:C-generic}
 C={}&\rev{\frac{2\rh}{\Delta_1}
 -\frac{H\Delta_2}{2\Delta_1^2}
 +\frac{HB_0}{2P_0^2}
 -\frac{aX}{P_0}},\\
 \alpha_\phi={}&\frac{2a\rh}{H\Delta_1}
 +\frac{X\Sigma_{\rm h}}
 {H^2\widetilde e\sin^2\theta_{\rm h}}.
 \label{eq:alpha-generic}
\end{align}
The $2\rh/\Delta_1$ term in $C$ follows from the linear expansion
\begin{equation}
 r^2+a^2=H+2\rh x+\order(x^2)
\end{equation}
in the singular factor $(r^2+a^2)\mathcal P/(\Delta\sqrt{\mathcal R})$.  Omitting this derivative misses a genuine constant contribution to $\dd t/\dd r$.  The cancellation of $\Delta_2$ and $c_1$ from $\alpha_\phi$ follows from forming $\dot{\widetilde\phi}=\dot\phi-\OmH\dot t$.  The notebook compares Eqs.~\eqref{eq:C-generic} and \eqref{eq:alpha-generic} against direct numerical limits of the exact first-order geodesics at a generic nonequatorial crossing.

\begin{revision}
The sign in $\theta_1$ is opposite to the sign chosen in $\Sigma\dot\theta=\pm\sqrt{\Theta_\theta}$. The proper-time Jacobian through the next order is
\begin{align}\label{eq:k1-generic}
 -\frac{\dd\tau}{\dd x}&=k+k_1x+\order(x^2),\\
 k_1&=k\left[\frac{2\rh-2a^2\sin\theta_{\rm h}\cos\theta_{\rm h}\,\theta_1}
 {\Sigma_{\rm h}}+\frac{c_1}{2P_0^2}\right].\nonumber
\end{align}
The simplification of $C$ in Eq.~\eqref{eq:C-generic} uses $c_1=-2P_0P_1+\Delta_1B_0$. These coefficients were also checked against the exact nonequatorial Carter equations. The numerical finite-gate experiment specializes to $e=1$, $\ell=\mathcal Q=0$, and $\theta_{\rm h}=\pi/2$.
\end{revision}

\section{Gamma-function endpoint identity}\label{app:gamma}

For $a>0$, $s>0$, and regulator $\epsilon>0$,
\begin{equation}
 \int_0^\infty\dd x\,x^{-\ii a}\e^{-(\epsilon+\ii s)x}
 =\Gamma(1-\ii a)(\epsilon+\ii s)^{-1+\ii a}.
\end{equation}
Taking $\epsilon\to0^+$ and using
\begin{equation}
 |\Gamma(1+\ii a)|^2=\frac{\pi a}{\sinh(\pi a)}
\end{equation}
gives
\begin{equation}\label{eq:gamma-modulus}
 \left|\int_0^\infty\dd x\,x^{-\ii a}\e^{-\ii sx}\right|^2
 =\frac{2\pi a}{s^2(\e^{2\pi a}-1)}.
\end{equation}
Reversing the sign of the oscillatory phase gives the absorption factor $\e^{2\pi a}$.  Equation~\eqref{eq:gamma-modulus} is the origin of the local kernels in Eq.~\eqref{eq:probabilities-exact}; it was verified numerically in the accompanying notebook for several values of $a$ and $s$.

\begin{revision}
\section{Numerical coefficients and boundary relations}
\label{app:numerical}

For the WKB convention in Eq.~\eqref{eq:implicit-wkb-system}, with $\alpha_n=n+1/2$, the coefficients implemented in the notebook are
\begin{align}\label{eq:wkb-L2}
 \mathcal L_2={}&\frac{U_4}{8U_2}\left(\frac14+\alpha_n^2\right)
 -\frac1{288}\left(\frac{U_3}{U_2}\right)^2
 (7+60\alpha_n^2),\\
 \mathcal O_3={}&\frac{1}{-2U_2}\bigg[
 \frac{5}{6912}\left(\frac{U_3}{U_2}\right)^4(77+188\alpha_n^2)
 \nonumber\\
 &-\frac{U_3^2U_4}{384U_2^3}(51+100\alpha_n^2)\nonumber\\
 &+\frac{1}{2304}\left(\frac{U_4}{U_2}\right)^2(67+68\alpha_n^2)
 \nonumber\\
 &+\frac{U_3U_5}{288U_2^2}(19+28\alpha_n^2)
 -\frac{U_6}{288U_2}(5+4\alpha_n^2)\bigg].
 \label{eq:wkb-O3}
\end{align}
They are the third-order barrier expressions with the complex peak and frequency solved together~\cite{IyerWill1987,Konoplya2019WKBRecipes}. Taylor jets implement $D_*^j$, not ordinary radial derivatives, in $U_j$.

For the independent massless collocation ansatz~\eqref{eq:spectral-ansatz}, let $c_R^{\rm in}$ be Eq.~\eqref{eq:radial-first-Frobenius} evaluated at $p=p_{\rm in}$ and $\mu_\Phi=0$. The horizon endpoint relation is
\begin{equation}\label{eq:spectral-horizon-bc}
 \Psi_y(0)=\rh\left[c_R^{\rm in}-\ii\omega
 -\frac{2\ii\omega-1-p_{\rm in}}{\rh}\right]\Psi(0).
\end{equation}
At infinity write $\Delta=r^2-2r+d_0+\order(r^{-1})$, where $d_0=a^2+2\mu q$ for $\nu=1$ and $d_0=a^2$ for $\nu=2,3$. The first coefficient in
$R=\e^{\ii\omega r}r^{2\ii\omega-1}(1+c_\infty/r+\cdots)$ is
\begin{equation}\label{eq:spectral-cinf}
 c_\infty=\frac{(8-2d_0+a^2)\omega^2-\Lambda_{lm}}{2\ii\omega},
\end{equation}
so that
\begin{equation}\label{eq:spectral-infinity-bc}
 \Psi_y(1)=\left(-\frac{c_\infty}{\rh}-p_{\rm in}\right)\Psi(1).
\end{equation}
These endpoint equations are imposed together with the exact interior radial equation. An interior normalization removes the arbitrary amplitude. The remaining complex residual determines $\omega$; the normalization does not alter the QNM boundary conditions.

For the real-frequency detector calculation, set $R=x^pV$ with $p=\ii b_s/2$, $t_x=\ln x$, and $J(x)=\Delta/x$. The regularized radial equation is
\begin{equation}\label{eq:detector-radial-regular}
 V_{t_xt_x}+B(x)V_{t_x}+C(x)V=0,
\end{equation}
where
\begin{align}
 B(x)&=2p+xJ'/J,\nonumber\\
 C(x)&=p^2+pxJ'/J+K^2/J^2-xW/J.\label{eq:detector-BC}
\end{align}
The indicial identity sets $C(0)=0$. If $V=\sum_nc_nx^n$, $c_0=1$, its coefficients obey
\begin{equation}\label{eq:detector-recurrence}
 c_n=-\frac{\sum_{j=1}^n[(n-j)B_j+C_j]c_{n-j}}
 {n(n+2p)}.
\end{equation}
This recurrence initializes the numerical outgoing branch. Near-horizon series also remove the cancelling logarithms from the Carter equations; away from the endpoint the untruncated radial and geodesic equations are used. These numerical regularizations do not replace the physical gate by a regulator.
\end{revision}

\begin{revision}
\section{Charged scalar on a prescribed aligned magnetic background}
\label{app:charged}

For completeness, the charged-scalar equations can be formulated consistently as a separate test-field problem. This appendix prescribes both the metric and an aligned gauge potential; it does not assert that they solve a common rotating Einstein--NLED system. Distinguish the mass-profile parameter $q$, the gauge-potential parameter $P$, and the scalar electric charge $e_s$. In the static-seed convention of Ref.~\cite{Toshmatov:2017zpr}, the dimensional magnetic charge is $Q_m=q_{\rm dim}^2/\sqrt{2\alpha}$, illustrating why it must not simply be identified with the dimensionless profile parameter.

Take
\begin{equation}\label{eq:magnetic-potential}
 A_\mu\dd x^\mu=-\frac{P\cos\theta}{\Sigma}
 \left[a\,\dd t-(r^2+a^2)\dd\phi\right],
\end{equation}
and $D_\mu=\nabla_\mu-\ii e_sA_\mu$, with $p_m=e_sP$. The charged Klein--Gordon equation is $(D_\mu D^\mu-\mu_\Phi^2)\Phi=0$. In the displayed gauge, separation is exact in $a$ for arbitrary radial $\Delta$:
\begin{align}\label{eq:charged-angular-correct}
 0={}&\frac{1}{\sin\theta}\frac{\dd}{\dd\theta}
 \left(\sin\theta\frac{\dd S}{\dd\theta}\right)\nonumber\\
 &+\bigg[a^2(\omega^2-\mu_\Phi^2)\cos^2\theta
 -2a\omega p_m\cos\theta\nonumber\\
 &\hspace{8mm}-\frac{(m-p_m\cos\theta)^2}{\sin^2\theta}
 +\Lambda\bigg]S,\\
 0={}&\frac{\dd}{\dd r}\left(\Delta\frac{\dd R}{\dd r}\right)
 +\bigg[\frac{[(r^2+a^2)\omega-am]^2}{\Delta}
 \nonumber\\
 &\hspace{7mm}-\mu_\Phi^2r^2-a^2\omega^2+2am\omega-\Lambda\bigg]R.
 \label{eq:charged-radial-correct}
\end{align}
The magnetic parameter enters through the angular eigenproblem. 

The displayed potential has polar gauge strings. Regular north and south patches may be chosen as $A^{N}=A-P\dd\phi$ and $A^{S}=A+P\dd\phi$. Their azimuthal quantum numbers are $m_N=m-p_m$ and $m_S=m+p_m$, with $m_N,m_S\in\mathbb Z$. Single-valued patch transition functions require $2p_m\in\mathbb Z$ in these units. At $a=0$ the regular angular eigenfunctions are monopole harmonics with
\begin{equation}\label{eq:monopole-limit}
 \Lambda=j(j+1)-p_m^2,\qquad j\geq|p_m|,
\end{equation}
not the neutral $l(l+1)$ spectrum. No numerical charged spectrum or neutral-basis approximation is used in the main analysis.

For $M=0$, $\mu=3$, define $F(r)=(r^\nu+q^\nu)^{3/\nu}$, $f=1-2r^2/F$ and $\Delta_0=r^2f$. To first order in $a$, the substitutions $u_1=rR$ and $\dd r_*/\dd r=f^{-1}$ give
\begin{equation}
 \frac{\dd^2u_1}{\dd r_*^2}+[\omega^2-V]u_1=0,
\end{equation}
with
\begin{align}\label{eq:charged-potential-correct}
 V={}&f\left[\mu_\Phi^2+\frac{\Lambda}{r^2}
 +\frac{2(r^\nu-2q^\nu)}{(r^\nu+q^\nu)^{1+3/\nu}}\right]
 \nonumber\\
 &+\frac{4am\omega}{F(r)}+\order(a^2).
\end{align}
Here $\Lambda$ must be taken from the charged angular problem at the same rotation order. The angular barrier has a positive sign. The neutral Schwarzschild limit is
\begin{equation}\label{eq:charged-Schwarzschild-check}
 V=\left(1-\frac2r\right)
 \left[\mu_\Phi^2+\frac{l(l+1)}{r^2}+\frac{2}{r^3}\right],
\end{equation}
which is checked in the symbolic verification. The construction concerns a charged test field on prescribed backgrounds; an exact rotating NLED completion and a charged-detector HBAR calculation are not assumed.
\end{revision}

\revcolor\bibliography{mybib}\color{black}

\end{document}